\documentclass[longauth]{aa}  

\usepackage{graphicx}
\usepackage{txfonts}
\usepackage[colorlinks=True,allcolors=blue,breaklinks=True]{hyperref}
\usepackage{natbib,twoopt}
\usepackage{todonotes}
\usepackage{placeins}
\bibpunct{(}{)}{;}{a}{}{,} %% natbib format for A&A and ApJ
\makeatletter
\newcommandtwoopt{\citeads}[3][][]{\href{https://ui.adsabs.harvard.edu/\#abs/#3}%
        {\def\hyper@linkstart##1##2{}%
                \let\hyper@linkend\@empty\citealp[#1][#2]{#3}}}
\newcommandtwoopt{\citepads}[3][][]{\href{https://ui.adsabs.harvard.edu/\#abs/#3}%
        {\def\hyper@linkstart##1##2{}%
                \let\hyper@linkend\@empty\citep[#1][#2]{#3}}}
\newcommandtwoopt{\citetads}[3][][]{\href{https://ui.adsabs.harvard.edu/\#abs/#3}%
        {\def\hyper@linkstart##1##2{}%
                \let\hyper@linkend\@empty\citet[#1][#2]{#3}}}
\newcommandtwoopt{\citeyearads}[3][][]%
{\href{https://ui.adsabs.harvard.edu/\#abs/#3}
        {\def\hyper@linkstart##1##2{}%
                \let\hyper@linkend\@empty\citeyear[#1][#2]{#3}}}
\makeatother

\graphicspath{{pic/}}

\newcommand{\msun}{M$_\odot$}
\newcommand{\msunyr}{\msun\,yr$^{-1}$}

\begin{document} 

%%%%%%%%%%%%%%%%%%%%%%%%%%%%%%%%
%          META-DATA
%%%%%%%%%%%%%%%%%%%%%%%%%%%%%%%%

   \title{The VLT/SPHERE view of the \textsc{Atomium} cool evolved star sample}

   \subtitle{I. Overview: Sample characterization through polarization analysis}

   \author{M.~Montargès\inst{\ref{Inst:LESIA}}
          \and
          E.~Cannon\inst{\ref{Inst:Leuven}}
          \and
          A.~de~Koter\inst{\ref{Inst:Amsterdam},\ref{Inst:Leuven}}
          \and
          T.~Khouri\inst{\ref{Inst:Chalmers}}
          \and
          E.~Lagadec\inst{\ref{Inst:Nice}}
          \and
          P.~Kervella\inst{\ref{Inst:LESIA}}
          \and
          L.~Decin\inst{\ref{Inst:Leuven},\ref{Inst:Leeds_Chem}}
          \and
          I.~McDonald\inst{\ref{Inst:Manchester},\ref{Inst:OpenU}}
          \and
          W.~Homan\inst{\ref{Inst:ULB},\ref{Inst:Leuven}}
          \and
          L.~B.~F.~M.~Waters\inst{\ref{Inst:SRON},\ref{Inst:Nijmegen}}
          \and
          R.~Sahai\inst{\ref{Inst:CalTech}}
          \and
          C.~A.~Gottlieb\inst{\ref{Inst:CfA}}
          \and
          J.~Malfait\inst{\ref{Inst:Leuven}}
          \and
          S.~Maes\inst{\ref{Inst:Leuven}}
          \and
          B.~Pimpanuwat\inst{\ref{Inst:Thailand}}
          \and
          M.~Jeste\inst{\ref{Inst:MPIfR}}
          \and
          T.~Danilovich\inst{\ref{Inst:Leuven}}
          \and
          F.~De~Ceuster\inst{\ref{Inst:Leuven},\ref{Inst:UCL}}
          \and
          M.~Van~de~Sande\inst{\ref{Inst:Leeds_Astro},\ref{Inst:Leuven}}
          \and
          D.~Gobrecht\inst{\ref{Inst:Gothenburg}}
          \and
          S.~H.~J.~Wallström\inst{\ref{Inst:Leuven}}
          \and
          K.~T.~Wong\inst{\ref{Inst:IRAM}}
          \and
          I.~El~Mellah\inst{\ref{Inst:IPAG}}
          \and
          J.~Bolte\inst{\ref{Inst:Leuven}}
          \and
          F.~Herpin\inst{\ref{Inst:Bordeaux}}
          \and
          A.~M.~S.~Richards\inst{\ref{Inst:Manchester}}
          \and
          A.~Baudry\inst{\ref{Inst:Bordeaux}}
          \and
          S.~Etoka\inst{\ref{Inst:Manchester}}
          \and
          M.~D.~Gray\inst{\ref{Inst:Manchester},\ref{Inst:Thailand}}
          \and
          T.~J.~Millar\inst{\ref{Inst:Belfast}}
          \and
          K.~M.~Menten\inst{\ref{Inst:MPIfR}}
          \and
           H.~S.~P.~Müller\inst{\ref{Inst:Koln}}
          \and
          J.~M.~C.~Plane\inst{\ref{Inst:Leeds_Chem}}
          \and
          J.~Yates\inst{\ref{Inst:UCL}}
          \and
          A.~Zijlstra\inst{\ref{Inst:Manchester}}
          }

   \institute{LESIA, Observatoire de Paris, Université PSL, CNRS, Sorbonne Université, Université Paris Cité, 5 place Jules Janssen, 92195 Meudon, France,
              \email{\href{mailto:miguel.montarges@observatoiredeparis.psl.eu}{miguel.montarges@observatoiredeparis.psl.eu}}\label{Inst:LESIA}
              \and
              Institute of Astronomy, KU Leuven, Celestijnenlaan 200D, 3001, Leuven, Belgium\label{Inst:Leuven}
              \and
              University of Amsterdam, Anton Pannekoek Institute for Astronomy, 1090 GE, Amsterdam, The Netherlands\label{Inst:Amsterdam}
              \and
              Department of Space, Earth and Environment, Chalmers University of Technology, Onsala Space Observatory, 439 92, Onsala, Sweden\label{Inst:Chalmers}
              \and
              Université Côte d'Azur, Laboratoire Lagrange, Observatoire de la Côte d'Azur, 06304, Nice Cedex 4, France\label{Inst:Nice}
              \and
              University of Leeds, School of Chemistry, Leeds, LS2 9JT, UK\label{Inst:Leeds_Chem}
              \and
              Jodrell Bank Centre for Astrophysics, Department of Physics and Astronomy, University of Manchester, Manchester, M13 9PL, UK\label{Inst:Manchester}
              \and
              Open University, Walton Hall, Milton Keynes, MK7 6AA, UK\label{Inst:OpenU}
              \and        
              Institut d'Astronomie et d'Astrophysique, Université Libre de Bruxelles (ULB), CP 226, 1060, Brussels, Belgium\label{Inst:ULB}
              \and
              SRON Netherlands Institute for Space Research, 3584 CA, Utrecht, The Netherlands\label{Inst:SRON}
              \and
              Radboud University, Institute for Mathematics, Astrophysics and Particle Physics (IMAPP), Nijmegen, The Netherlands\label{Inst:Nijmegen}
              \and
              California Institute of Technology, Jet Propulsion Laboratory, Pasadena, CA, 91109, USA\label{Inst:CalTech}
              \and
              Harvard-Smithsonian Center for Astrophysics, 60 Garden Street, Cambridge, MA, 02138, USA\label{Inst:CfA}
              \and
              National Astronomical Research Institute of Thailand, Chiangmai, 50180, Thailand\label{Inst:Thailand}
              \and
              Max-Planck-Institut für Radioastronomie, 53121, Bonn, Germany\label{Inst:MPIfR}
              \and
              University College London, Department of Physics and Astronomy, London, WC1E 6BT, UK\label{Inst:UCL}
              \and
              School of Physics and Astronomy, University of Leeds, Leeds LS2 9JT, UK\label{Inst:Leeds_Astro}
              \and
              Department of Chemistry and Molecular Biology, University of Gothenburg, Kemigården 4, 412 96 Gothenburg, Sweden\label{Inst:Gothenburg}
              \and
              Institut de Radioastronomie Millimétrique, 300 rue de la Piscine, 38406, Saint Martin d'Hères, France\label{Inst:IRAM}
              \and
              Univ. Grenoble Alpes, CNRS, IPAG, 38000 Grenoble, France\label{Inst:IPAG}
              \and
              Université de Bordeaux, Laboratoire d'Astrophysique de Bordeaux, 33615, Pessac, France\label{Inst:Bordeaux}
              \and
              Astrophysics Research Centre, School of Mathematics and Physics, Queen's University Belfast, University Road, Belfast, BT7 1NN, UK\label{Inst:Belfast}
              \and
              Universität zu Köln, I. Physikalisches Institut, 50937, Köln, Germany\label{Inst:Koln}
             }
         
   \date{Received  8\textsuperscript{th} November 2022; accepted 21\textsuperscript{st} December 2022}

%%%%%%%%%%%%%%%%%%%%%%%%%%%%%%%%
%          ABSTRACT
%%%%%%%%%%%%%%%%%%%%%%%%%%%%%%%%

% \abstract{}{}{}{}{} 
% 5 {} token are mandatory
 
  \abstract
   {Low- and intermediate-mass asymptotic giant stars and massive red supergiant stars are important contributors to the chemical enrichment of the Universe. They are among the most efficient dust factories of the Galaxy, harboring chemically rich circumstellar environments. Yet, the processes that lead to dust formation or the large-scale shaping of the mass loss still escape attempts at modeling.}
   {Through the \textsc{Atomium}  project, we aim to present a consistent view of a sample of 17 nearby cool evolved stars. Our goals are to unveil the dust-nucleation sites and morphologies of the circumstellar envelope of such stars and to probe ambient environments with various conditions. This will further enhance our understanding of the roles of stellar convection and pulsations, and that of companions in shaping the dusty circumstellar medium.}
   {
     %The \textsc{Atomium} project is centered around an ALMA large program executed through cycle 6. 
     Here we present and analyze VLT/SPHERE-ZIMPOL polarimetric maps obtained in the visible ($645 - 820$\,nm) of 14 out of the 17 \textsc{Atomium} sources. They were obtained contemporaneously with the ALMA high spatial resolution data. To help interpret the polarized signal, we produced synthetic maps of light scattering by dust, through 3D radiative transfer simulations with the \texttt{RADMC3D} code.}
   {
        %We marginally spatially resolve the visible stellar photosphere $\pi^1$ Gru, and significantly resolve R\,Hya. 
        The degree of linear polarization (DoLP) observed by ZIMPOL spreads across several optical filters. We infer that it primarily probes dust located just outside of the point spread function of the central source, and in or near the plane of the sky. The polarized signal is mainly produced by structures with a total optical depth close to unity in the line of sight, and it represents only a fraction of the total circumstellar dust. The maximum DoLP ranges from 0.03-0.38 depending on the source, fractions that can be reproduced by our 3D pilot models for grains composed of olivine, melilite, corundum, enstatite, or forsterite.  The spatial structure of the DoLP shows a diverse set of shapes, including clumps, arcs, and full envelopes. Only for three sources do we note a correlation between the ALMA CO $\varv=0, \ J=2-1$ and SiO $\varv=0, \ J=5-4$ lines, which trace the gas density, and the DoLP, which traces the dust. 
   %Among the sample, W\,Aql stands out with a strong DoLP level, detected over a comparatively wide spatial extent.  This is likely not related to its S-type nature, as the other S-type star in our sample ($\pi^{1}$\,Gru) does not show such a prominent polarization signal.
        }
   {The clumpiness of the DoLP and the lack of a consistent correlation between the gas and the dust location show that, in the inner environment, dust formation occurs at very specific sites. This has potential consequences for the derived mass-loss rates and dust-to-gas ratio in the inner region of the circumstellar environment. Except for $\pi^1$~Gru and perhaps GY~Aql, we do not detect interactions between the circumstellar wind and the hypothesized companions that shape the wind at larger scales. This suggests that the orbits of any other companions are tilted out of the plane of the sky.}

   \keywords{stars: AGB and post-AGB -- supergiants -- stars: mass-loss -- stars: imaging -- circumstellar matter-- stars: evolution}

   \maketitle
%
%-------------------------------------------------------------------

%%%%%%%%%%%%%%%%%%%%%%%%%%%%%%%%
%       INTRODUCTION
%%%%%%%%%%%%%%%%%%%%%%%%%%%%%%%%

\section{Introduction}

Cool evolved stars are among the most active chemical sites in the Universe, with more than 100 molecules and 15 dust species identified in their environment \citepads{2021ARA&A..59..337D}. Their outflowing winds contribute $\sim$ 85\% of the gas (metals) and $\sim$ 35\% of the dust enrichment of the interstellar medium \citepads{2005pcim.book.....T}. As the wind cools by moving farther away from the star, various chemical processes occur, including uni-, bi- and ter-molecular gas-phase reactions as well as cluster and grain formation (see, e.g., \citeads{2022A&A...658A.167G}). Initially proposed by \citetads{1962MNRAS.124..417H}, the concept of the low- and intermediate-mass asymptotic giant branch (AGB) stellar wind driven through radiation pressure on dust is now well established (e.g., \citeads{2019A&A...623A.158H}, and references therein). However, the driving mechanism of the massive red supergiant (RSG) star wind is still debated. Several scenarios have been proposed: radiation pressure on molecular lines \citepads{2007A&A...469..671J}; chromospheric Alfv\'{e}n wave dissipation \citepads{2000ApJ...528..965A,2010ApJ...723.1210A}; and photospheric turbulent pressure \citepads{2021A&A...646A.180K}. The latter model allows for predictions of the mass-loss rate as a function of the photospheric turbulent velocity. These processes would lift the material from the photosphere, allowing dust to condense farther from the star. Once sufficient dust particles have formed, as for AGB stars, the wind is expected to be dust driven. So, the interplay between the gaseous and the dusty components of the circumstellar environment is at the core of the driving and chemical processes in the wind of cool evolved stars.

High angular resolution spectro-imaging observations have proven instrumental in unveiling the 3D structure, dynamics, and chemistry of the wind of cool evolved stars in the past few decades (e.g., \citeads{1987ApJ...313..400H,2012Natur.490..232M,2018A&A...616A..34H,2019MNRAS.485.2417M}). Recent achievements have even succeeded in imaging the photosphere of a sample of nearby giant (e.g., \citeads{2018Natur.553..310P} and \citeads{2016A&A...591A..70K}) and supergiant stars (e.g., \citeads{2017Natur.548..310O} and \citeads{2021Natur.594..365M}). In order to achieve a significant step forward in understanding the dust nucleation from gas-phase precursors, we began the \textsc{Atomium} project\footnote{\url{https://fys.kuleuven.be/ster/research-projects/aerosol/atomium/atomium}}: ALMA Tracing the Origins of Molecules formIng dUst in oxygen-rich M-type stars. At the core of this endeavor is an executed large program (2018.1.00659.L, PI L. Decin \& co-PI C. Gottlieb; see \citeads{2022A&A...660A..94G}) during the cycle 6 of the Atacama Large Millimeter Array (ALMA). The project focuses on O-rich (C/O $\lesssim 1$) AGB and RSG stars. The ALMA data provide essential information on the spatial distribution of the molecules considered the most likely candidates to form the first dust grains (gaseous TiO, TiO$_2$, AlO, AlOH, and SiO). From the observations we obtained multichannel molecular line emission maps, including the CO $\varv=0$ $J=2-1$, SiO $\varv=0$ $J=6-5$, $J=5-4$, and HCN $\varv=0$ $J=3-2$ transitions.

These maps reveal the complexity of the circumstellar environment of AGB stars, which is possibly linked to undetected (sub)stellar companions. The shaping mechanism(s) of the wind of AGB and post-AGB stars still represent(s) a missing piece in our understanding of their mass loss, and ultimately the formation of the structures seen in planetary nebulae \citepads{2020Sci...369.1497D}. By combining spatially resolved observations of the circumstellar environment through several molecular transitions, constraints are placed on the masses of the new companions for both $\pi^1$~Gru ($\lesssim 1.1$~M$_\odot$; \citeads{2020A&A...644A..61H}) and R~Hya ($\gtrsim 0.65$~M$_\odot$; \citeads{2021A&A...651A..82H}). For the S-type star W~Aql, observations of halide molecules show that its AlF abundance exceeds the solar fluorine abundance, confirming that fluorine synthesized {in situ} is already brought to the surface of the AGB star and expelled in its circumstellar environment \citepads{2021A&A...655A..80D}. In most cases, the ALMA observations do not retrieve spatially resolved information about the dust, because the continuum maps are dominated by the central source. For this reason, we obtained contemporaneous observations of the \textsc{Atomium} sources with the SPHERE-ZIMPOL instrument installed at the Very Large Telescope (VLT), which can localize dust through polarized scattering in the visible, at an angular resolution that matches the most extended configurations of ALMA (i.e., a beam size of $\sim$ 20~mas). 

SPHERE is the European Southern Observatory's SPectropolarimetric High-contrast Exoplanet REsearch instrument \citepads{2019A&A...631A.155B}. It is equipped with a high performance adaptive optics (AO) system that compensates for most of the atmospheric turbulence in the visible and the near-infrared. We used its ZIMPOL (Zurich IMaging POLarimeter; \citeads{2018A&A...619A...9S}) subunit, which operates in the visible. Although primarily designed to detect exoplanets, ZIMPOL has already been used to image stellar photospheres and pinpoint dust around nearby cool evolved stars. This is the case for the  AGB star L$_2$~Puppis during the science verification time, when its circumstellar disk, both in the intensity images and polarized frames, was unveiled \citepads{2015A&A...578A..77K}. Later on, warm spots were observed on the photosphere of the closest AGB star R~Dor \citepads{2016A&A...591A..70K}. As for the well-studied binary star Mira ($o$~Ceti), ZIMPOL traced the circumstellar environments of both the primary and the secondary stars \citepads{2018A&A...620A..75K}. \citetads{2016A&A...589A..91O,2017A&A...597A..20O} observed the AGB star W~Hya with ZIMPOL. They conclude that H$\alpha$-emitting warm gas is coexisting with relatively small ($0.4 - 0.5~\mu$m) dust grains (Al$_2$O$_3$, Mg$_2$SiO$_4$, or MgSiO$_3$) within a thin layer at two to three stellar radii. Moreover, by comparing several observation epochs, they saw the evolution of clumpy dust structures in the range 1.4 to 2.0~R$_\star$, on a timescale of 10 months. IK~Tau was observed with ZIMPOL by \citetads{2019A&A...628A.132A}; dust clumps were also observed surrounding this high mass-loss rate AGB star, mostly in the range $3.5 - 25$~R$_\star$, but also as far out as 73~R$_\star$. ZIMPOL was also used on RSGs such as Antares, the closest star of this class, to characterize a recently ejected dust cloud \citepads{2021MNRAS.502..369C}. The prototypical RSG Betelgeuse was observed by \citetads{2016A&A...585A..28K}, who resolved its photosphere and found a dust cloud in polarization. More recently, the exquisite angular resolution of SPHERE-ZIMPOL was used to interpret the Great Dimming of this same star \citepads{2021Natur.594..365M}: the historic drop in its brightness was caused by a cool photospheric patch that triggered dust nucleation in a preexisting gas cloud in the line of sight (see also \citeads{2022ApJ...936...18D}). 

In this paper we present ZIMPOL observations of 14 out of the 17 stars of the \textsc{Atomium} sample to provide a complete overview and to identify potential similarities and trends in the properties of the polarized light. In Sect.~\ref{Sect:TargetsObs} we present the sample selection, the observations, and the data-reduction procedure. In Sect.~\ref{Sect:Data_Analysis} we analyze the intensity and polarized images. Section~\ref{Sect:Simulations} is dedicated to the setup and analysis of numerical 3D radiative transfer dust simulations. Section~\ref{Sect:Discussion} discusses the simulations in the context of the ZIMPOL stellar sample. Concluding remarks are presented in Sect.~\ref{Sect:Conclusion}.

%%%%%%%%%%%%%%%%%%%%%%%%%%%%%%%%%%%%%%%%%%%%%
%    OBSERVATIONS AND DATA REDUCTION
%%%%%%%%%%%%%%%%%%%%%%%%%%%%%%%%%%%%%%%%%%%%%

\section{Observations}\label{Sect:TargetsObs}

\subsection{The \textsc{Atomium} sample}

The sources of the \textsc{Atomium} sample were chosen from oxygen-rich cool evolved stars, representative of various pulsational types and mass-loss rates (Table~\ref{Tab:ATOMIUM_caract}). Most distances come from the newly released \textit{Gaia} DR3 \citep{GaiaDR3}. However, in the case of GY~Aql and U~Her we used the values of \citet{2022arXiv220903906A}, who show that the \textit{Gaia} estimations for these two stars are unreliable. 

In order to ensure the observability of the sources, both with ALMA and the VLT, the \textsc{Atomium} sources were chosen from stars observable during the Atacaman night in June-August, when the large ALMA configurations (C43-8/C43-9) were scheduled. This ensured that the same spatial scales ($\sim$ 25-30~mas) were observed quasi-simultaneously with ALMA and VLT/SPHERE-ZIMPOL thanks to the Target of Opportunity (ToO; see Sect.~\ref{SubSect:Obs}) triggers. The ZIMPOL observations were carried out less than 10 days after the corresponding ALMA large configuration observations. This is well below the characteristic evolution time of 4.5~yr in a region of 100~mas at 200~pc, with a radially expanding wind at $\sim 20$~km\,s$^{-1}$. Additionally, for proper observations with VLT/SPHERE, we only chose sources brighter than R = 11\,mag, and with a photospheric angular diameter of at least 3~mas, to allow the inner circumstellar environment to be spatially resolved. Further description of the \textsc{Atomium} sample is available in \citet{2020Sci...369.1497D} and \citet{2022A&A...660A..94G}.

\begin{table*}
        \centering
        \caption{Main characteristics of the \textsc{Atomium} sources.}
        \label{Tab:ATOMIUM_caract}
        \begin{tabular}{l l l l l l l}
                \hline\hline
                \noalign{\smallskip}
                Target & Stellar type & Distance & Angular diameter & R mag\tablefootmark{a} & T\textsubscript{eff} & Mass-loss rate\tablefootmark{d} \\
                & & (pc) & (mas) & & (K) & (\msunyr) \\
                \hline
                \noalign{\smallskip}
                \object{S Pav} & AGB O-rich & $174 \pm 19$ (1) & 11.61\tablefootmark{b} &  $7.60 \pm 0.03$ & 3100 (4) & $8 \times 10^{-8}$ (5)\\
                \object{T Mic} & AGB O-rich     &  $186 \pm 8$ (1) & 9.26\tablefootmark{b} & $7.30 \pm 0.03$ & 3300 (4) & $8 \times 10^{-8}$ (5)\\
                \object{U Del} & AGB O-rich & $335 \pm 11$ (1) & $7.90 \pm 0.50$ (2) & $6.31 \pm 0.04$ & 3000 (19) & $1.5 \times 10^{-7}$  (5)\\
                \object{V PsA} & AGB O-rich & $304 \pm 11$ (1) & 11.4\tablefootmark{c} & $8.03 \pm 0.04$ & 2400 (5) & $3 \times 10^{-7}$ (5)\\
                \object{SV Aqr} & AGB O-rich & $431 \pm 13$ (1) & 5.7\tablefootmark{c} & $9.20 \pm 0.03$ & 3400 (4) & $3 \times 10^{-7}$ (5)\\
                \object{R Hya} & AGB O-rich & $148 \pm 10$ (1) & $23.7 \pm 1.0$ (2) & $5.66 \pm 0.04$ & 2100 (6) & $4 \times 10^{-7}$ (6)\\
                \object{U Her} & AGB O-rich & $270 \pm 20$ (20) & $11.2 \pm 0.6$ (2) & $8.83 \pm 0.03$ & 2100 (19) & $5.9 \times 10^{-7}$ (7)\\
                \object{$\pi^1$ Gru} & AGB S-type & $162\pm 12$ (1) & $18.37 \pm 0.18$ (3) & $5.66 \pm 0.04$ & 2300 (6) & $7.7 \times 10^{-7}$ (8)\\
                \object{R Aql} & AGB O-rich & $234 \pm 9$ (1) & $10.9 \pm 0.3$ (2) & $7.52 \pm 0.03$ & 2800 (19) & $1.1 \times 10^{-6}$ (7)\\
                \object{W Aql} & AGB S-type & $374 \pm 19$ (1) & $11.6 \pm 1.8$ (2) & $10.09 \pm 0.04$ & 2800 (6) & $3 \times 10^{-6}$ (9)\\
                \object{GY Aql} & AGB O-rich & $340 \pm 30$ (20) & 20.46\tablefootmark{b} & $10.44 \pm 0.04$ & 3100 (4) & $4.1 \times 10^{-6}$ (10)\\
%               \object{GY Aql} & AGB & $678 \pm 77$ (1) & 11.1\tablefootmark{b} & 3100 (4) & $4.1 \times 10^{-6}$ (10)\\
                \object{AH Sco} & RSG & $2260 \pm 190$ (13) & $5.81 \pm 0.15$ (14) &  $7.13 \pm 0.04$ & $3682 \pm 190$ (14) & \textit{Unknown}\\
                \object{KW Sgr} & RSG & $2400 \pm 300$ (14) & $3.91 \pm 0.25$ (14) & $8.81 \pm 0.04$ & $3720 \pm 183$ (14) & \textit{Unknown}\\
                \object{VX Sgr} & RSG & $1560 \pm 110$ (15) & $8.82 \pm 0.50$ (16) & $8.94 \pm 0.04$ & 3150 (17) & [$1 - 6$] $\times 10^{-5}$ (6, 17, 18) \\
                \hline
                \noalign{\smallskip}
                RW Sco & AGB O-rich& $578 \pm 33$ (1) & 4.87\tablefootmark{b} & 13.13 & 3300 (4) & $2.1 \times 10^{-7}$ (12)\\
                IRC-10529 & AGB OH/IR & 620 (6) & 6.47\tablefootmark{b} & 19.17 & 2700 (6) & $4.5 \times 10^{-6}$ (6)\\
                IRC+10011 & AGB OH/IR & 740 (11) & 6.53\tablefootmark{b} & 18.68 & 2700 (6) & $1.9 \times 10^{-5}$ (6)\\           
                \hline
        \end{tabular}
        \tablebib{(1)~\citet{GaiaDR3};
%               (1b)~\citetads{2018A&A...616A...1G};
%               (2)~\citetads{1997A&A...323L..49P};
%               (3)~\citetads{2002MNRAS.334..498Z};
%               (3)~\citetads{2007A&A...472..547V};
                (2)~\citetads{2005A&A...431..773R};
                (3)~\citetads{2018Natur.553..310P};
                (4)~\citetads{2008A&A...482..883M};
                (5)~\citetads{2002A&A...391.1053O};
                (6)~\citetads{2010A&A...523A..18D};
                (7)~\citetads{1995ApJ...445..872Y};
                (8)~\citetads{2017A&A...605A..28D};
                (9)~\citetads{2017A&A...605A.126R};
                (10)~\citetads{1993A&AS...99..291L};
                (11)~\citetads{2001MNRAS.326..490O};
                (12)~\citetads{1999A&AS..140..197G};
                (13)~\citetads{2008ApJ...681.1574C};
                (14)~\citetads{2013A&A...554A..76A};
                (15)~\citetads{2018ApJ...859...14X};
                (16)~\citetads{2010A&A...511A..51C};
                (17)~\citetads{2017MNRAS.466.1963L};
                (18)~\citetads{1986MNRAS.220..513C,2011A&A...526A.156M,2018AJ....155..212G,2020A&A...644A.139G};
                (19)~\citetads{2020Sci...369.1497D};
                (20)~\citetads{2022arXiv220903906A}.
        }
        \tablefoot{The stars listed in the second part of the table were not observed with ZIMPOL due to their faint magnitude in the R band.\\
\tablefoottext{a}{The R magnitudes were obtained from the USNO-B catalog \citepads{2003AJ....125..984M}.}
                \tablefoottext{b}{These angular diameters were derived from the bolometric luminosity, effective temperature and distance \citepads{2020Sci...369.1497D}.}
        \tablefoottext{c}{These angular diameters were derived using the magnitude-color relation of \citetads{1999AJ....117..521V}.}
        \tablefoottext{d}{Mass-loss rates from references (5), (6), (7), (8), (9), (10), and (12) are derived from the 1D modeling of their CO rotational line emission. For VX~Sgr, in (17), the authors model the thermal emission from dust from the optical to the far-infrared, and assume a gas-to-dust ratio of 200 from the litterature. (18) compiles various mass-loss diagnostics of VX~Sgr: H$_2$O and OH maser emissions, infrared excess, and mid- and far-infrared photometry and imaging.}        
        }       
\end{table*}

\subsection{Data acquisition and data reduction}\label{SubSect:Obs}

Observations of the \textsc{Atomium} sources were performed with VLT/SPHERE-ZIMPOL between 2019 April 7 and 2019 October 1. We used ESO's ToO-soft mode to have the observations executed within 10 days of the corresponding ALMA observations of the \textsc{Atomium} sources in the extended array configuration. It allowed a quasi-simultaneous view of the dust (ZIMPOL) and gas (ALMA) components of the inner and intermediate circumstellar environments of these stars. The detailed log of the observations is presented in Table \ref{Tab:ObsLog}. Several filter configurations were used to adapt to the R band magnitude of the sources. For bright sources, we used three narrow band filters: CntH$\alpha$ (centered at 644.9\,nm), Cnt748 (747.4\,nm), and Cnt820 (817.3\,nm). Intermediate sources were observed with the N\_R (645.9\,nm) and N\_I (816.8\,nm) filters. Finally, relatively faint sources were observed with the VBB (735.4\,nm) broadband filter only. The complete characteristics of ZIMPOL's filters are available on ESO's dedicated website\footnote{\url{https://www.eso.org/sci/facilities/paranal/instruments/sphere/inst/filters.html}}. Each target observation was followed by the observation of a reference spatially unresolved star (usually a K giant), used as proxy of the point spread function (PSF). We chose stars with R magnitudes close to the \textsc{Atomium} sources in order to have a similar setup for the AO system. 

The data were reduced through the SPHERE DataCenter \citepads{2017sf2a.conf..347D,2018A&A...615A..92G}. Custom Python routines were applied to derive the polarimetric observables (Stokes $I$, polarized flux, degree of linear polarization (DoLP), and polarization angle; see \citeads{2015A&A...578A..77K} for more details). However, \citetads{2017A&A...607A..90E} state that the polarized flux (and the DoLP) are affected by a systematic bias when the signal-to-noise ratio (S/N) is weak. This comes from the squaring of Stokes $U$ and Stokes $Q$ at low values when computing the polarized flux $P = \sqrt{U^2 + Q^2}$. The authors suggest instead using the locally defined azimuthal and radial $U_\phi$ and $Q_\phi$ parameters:
\begin{eqnarray}
        Q_\phi &=& -(Q\cos 2\phi + U\sin 2\phi)\\
        U_\phi &=& -Q\sin 2\phi + U\cos 2\phi
,\end{eqnarray}
\noindent where $\phi$ is the polar angle between north and the point of interest, measured from the north over east. The polarized flux is then directly $P = |Q_\phi|$ under the assumption that the dust features are optically thin, and that a single scattering occurs that leaves the electric vector of the radiation field aligned tangentially with respect to the
central source. We verified that $|U_\phi|$ shows only instrumental noise for our sources. The DoLP definition remains unchanged: $\mathrm{DoLP} = P/I$. The flux calibration was performed with the PSFs calibrators, using the method outlined in \citetads{2021MNRAS.502..369C}.

%%%%%%%%%%%%%%%%%%%%%%%%%%%%%%%%
%        DATA ANALYSIS
%%%%%%%%%%%%%%%%%%%%%%%%%%%%%%%%

\section{Data analysis}\label{Sect:Data_Analysis}

\subsection{Intensity images}

\begin{figure*}
        \centering
        \includegraphics[width=.99\columnwidth]{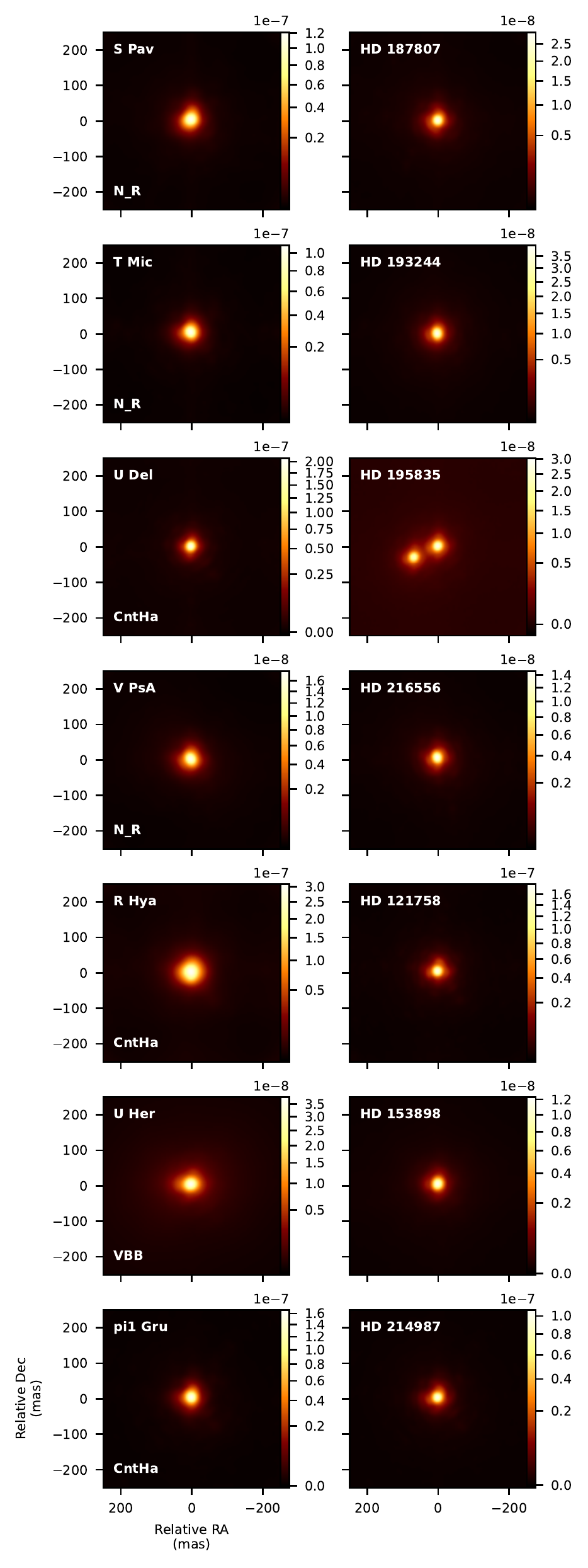}~
        \includegraphics[width=.99\columnwidth]{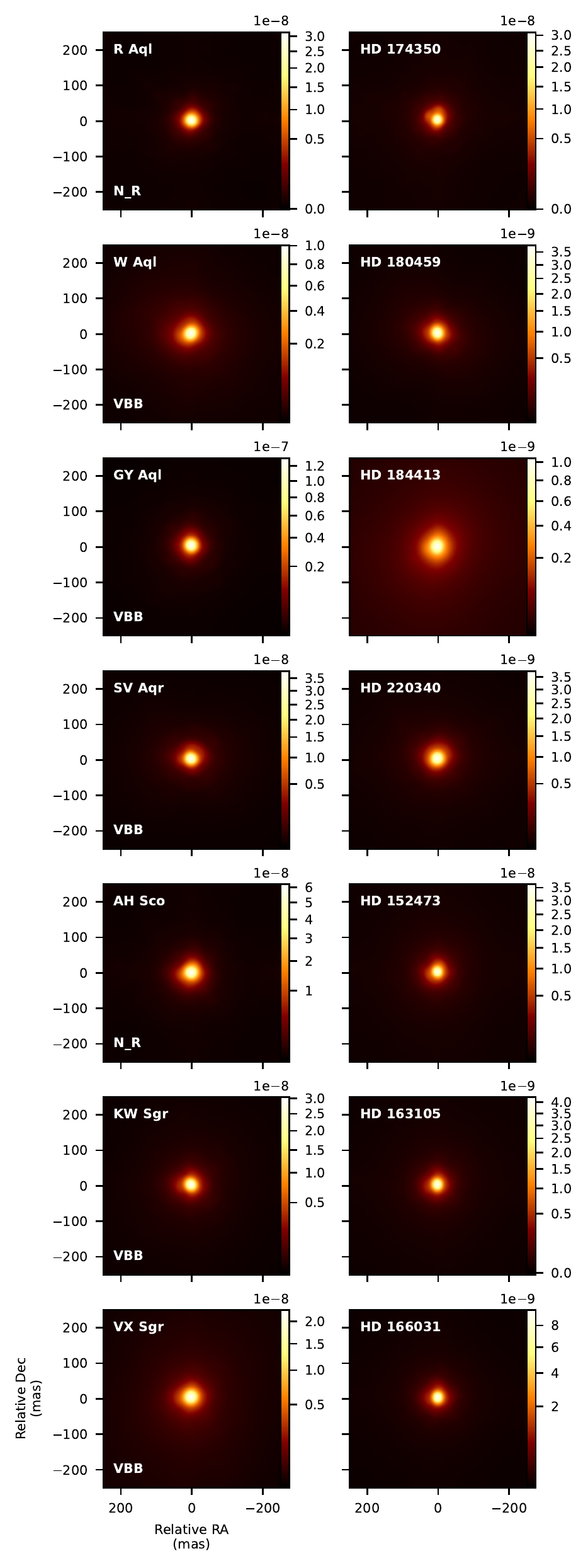}
        \caption{Intensity images of the 14 \textsc{Atomium} sources observed with SPHERE-ZIMPOL. The first and third columns correspond to the science sources, and columns two and four correspond to their respective PSF references. The filter used is indicated in the bottom-left corner of each science source. The color scale represents W m$^{-2}$ $\mu$m$^{-1}$ arcsec$^{-2}$.}
        \label{Fig:ZIMPOL_intensity}
\end{figure*}

The intensity images of the \textsc{Atomium} sources are shown in Fig.~\ref{Fig:ZIMPOL_intensity} for the inner 500~mas. Besides the  central star, few features are visible. We note that U~Del's PSF calibrator, namely HD~195835, is actually a binary system, which was suggested, after our observations were executed, by its high \textit{Gaia} Early Data Release 3 (EDR3) renormalized unit weight error (RUWE) of 13.7 from \citet{2022A&A...657A...7K}. We note that it is currently listed as an interferometric calibrator in \citetads{2014ASPC..485..223B,2019MNRAS.490.3158C}. The angular separation between the two centroids in the CntH$\alpha$ filter is 74.2~mas. With a parallax of $2.66 \pm 0.23$~mas \citep{GaiaDR3}, this translates into a separation of $27.8 \pm 2.4$~au. From now on, and in subsequent \textsc{Atomium}-ZIMPOL papers, we will use R~Hya's PSF reference (HD~121758, observed three days earlier with the same filter) instead if we need to deconvolve the intensity images.

To characterize the observed extension of the \textsc{Atomium} sources and their PSF references, we fitted the intensity images with a circular 2D Gaussian. The resulting derived full widths at half maximum (FWHMs) are presented in Table~\ref{Tab:FitIntensity}. Their comparability to the 20~mas theoretical angular resolution of an 8.2~m telescope in the R band is a good proxy of the quality of the AO correction (i.e., the Strehl ratio). In most cases (except U~Del, GY~Aql, and SV~Aqr), the size of the PSF calibrator is smaller than the scientific source, and in the range 25-30~mas. ~In the case of U~Del (for which we fitted a two component PSF on HD~195835) and SV~Aqr, the PSF reference is slightly larger than the main source, suggesting a slight evolution of Earth's atmospheric conditions (hence the AO correction). SV~Aqr's PSF calibrator (HD~220340) can still be used because the FWHM increase is relatively small. For U~Del we already decided to use HD~121758 in future studies, as stated in the previous paragraph. GY~Aql's PSF reference (HD~184413) has been observed under very different atmospheric conditions than its main source, leading to a FWHM 1.7 times larger than GY~Aql. Since HD~184413 has an angular diameter of $0.150 \pm 0.003$~mas (\citeads{2014ASPC..485..223B}, to be compared with GY~Aql's diameter of 20.46~mas; see Table~\ref{Tab:ATOMIUM_caract}), we can conclude that the AO correction was poor during this observation. Therefore, we replace GY~Aql's PSF calibrator with W~Aql's HD~180459, the closest observed in time with the VBB filter.

\begin{table}
        \centering
        \caption{FWHMs of the observed sources resulting from the fit of the ZIMPOL intensity images by a 2D circularly symmetric Gaussian.}\label{Tab:FitIntensity}
        
        \begin{tabular}{l l l}
                \hline\hline
                \noalign{\smallskip}
                Star & Filter & FWHM (mas)\\
                \hline
                \noalign{\smallskip}
                S Pav & N\_R & 34.3 \\
                \object{HD 187807} & N\_R & 26.9 \\
                \noalign{\smallskip}
                T Mic & N\_R & 31.7 \\
                HD 193244 & N\_R & 26.9 \\
                \noalign{\smallskip}
                U Del & CntH$\alpha$ & 24.4 \\
                \object{HD 195835}a & CntH$\alpha$ & 28.1 \\
                HD 195835b & CntH$\alpha$ & 28.6 \\
                \noalign{\smallskip}
                V PsA & N\_R & 33.1 \\
                \object{HD 216556} & N\_R & 26.5 \\
                \noalign{\smallskip}
                R Hya & CntH$\alpha$ & 44.3 \\
                \object{HD 121758} & CntH$\alpha$ & 26.0 \\
                \noalign{\smallskip}
                U Her & VBB & 37.0 \\
                \object{HD 153898} & VBB & 26.6 \\
                \noalign{\smallskip}
                $\pi^1$ Gru & CntH$\alpha$ & 30.7 \\
                \object{HD 214987} & CntH$\alpha$ & 26.4 \\
                \noalign{\smallskip}
                R Aql & N\_R & 30.0 \\
                \object{HD 174350} & N\_R & 29.3 \\
                \noalign{\smallskip}
                W Aql & VBB & 37.7 \\
                \object{HD 180459} & VBB & 33.0 \\
                \noalign{\smallskip}
                GY Aql & VBB & 31.3 \\
                \object{HD 184413} & VBB & 52.0 \\
                \noalign{\smallskip}
                SV Aqr & VBB & 31.4 \\
                \object{HD 220340} & VBB & 36.0 \\
                \noalign{\smallskip}
                AH Sco & N\_R & 36.6 \\
                \object{HD 152473} & N\_R & 30.4 \\
                \noalign{\smallskip}
                KW Sgr & VBB & 31.3 \\
                \object{HD 163105} & VBB & 29.4 \\
                \noalign{\smallskip}
                VX Sgr & VBB & 39.3 \\
                \object{HD 166031} & VBB & 28.2 \\
                \hline
        \end{tabular}
	\tablefoot{Each named \textsc{Atomium} source is followed by its PSF reference star (designated by its HD number), observed in the same sequence.}
\end{table}

It is difficult to determine whether an individual \textsc{Atomium} source is spatially resolved when its FWHM is larger than the one of its PSF reference. Indeed a significantly larger FWHM can be caused by fluctuations of the atmospheric turbulence. However, an inspection of Table~\ref{Tab:ObsLog} shows that a significant jump in the seeing value for the main source was only observed in VX~Sgr. It indicates a strong degradation of the atmospheric turbulence during the observation sequence. Therefore, taking into account the angular diameters of Table~\ref{Tab:ATOMIUM_caract}, we can consider that ZIMPOL slightly spatially resolves the photospheres of $\pi^1$~Gru, and significantly R~Hya.

\begin{figure}
        \centering
        \includegraphics[width=\columnwidth]{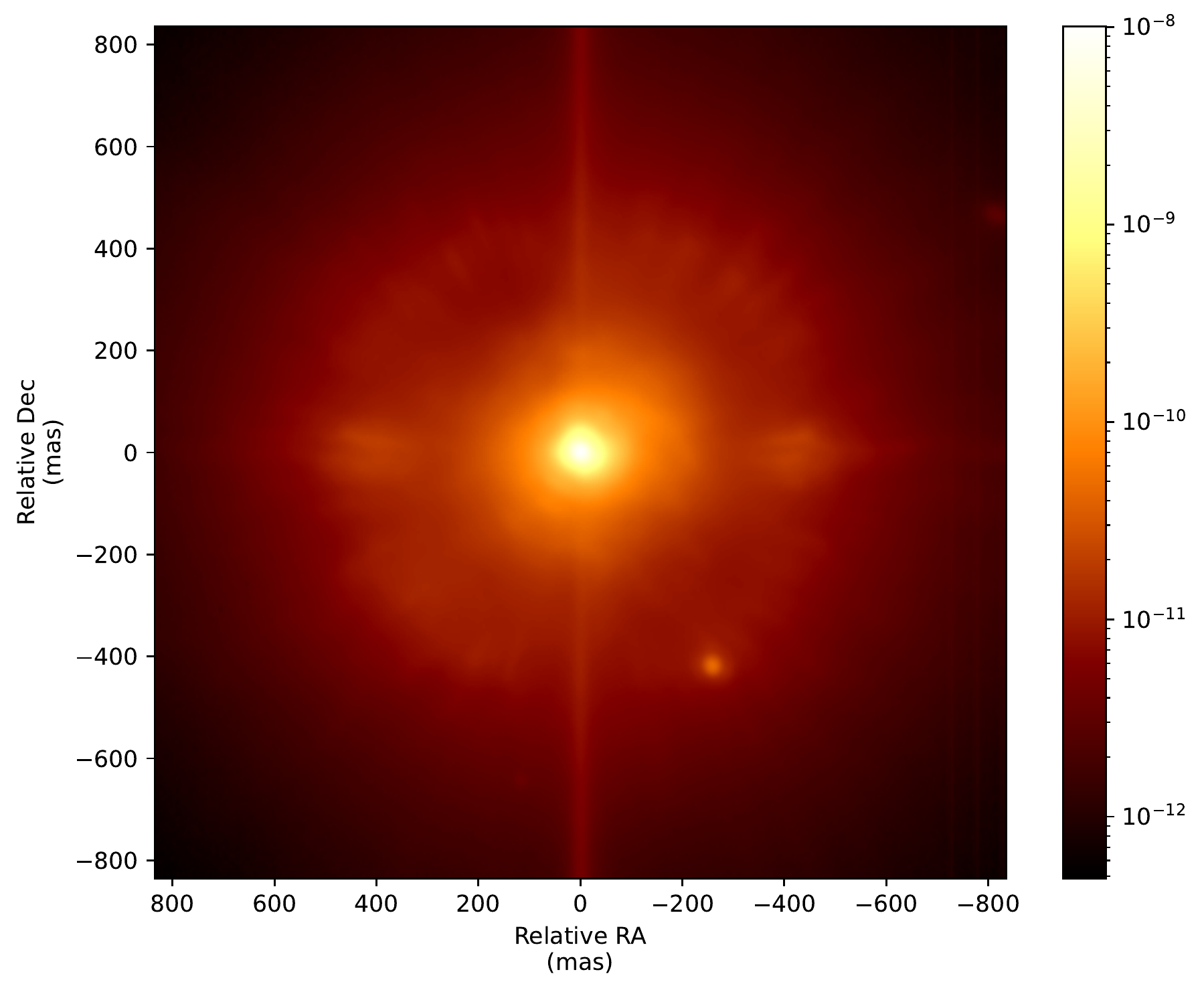}
        \caption{Full intensity image of W Aql in the VBB filter. The logarithmic color scale represents W m$^{-2}$ $\mu$m$^{-1}$ arcsec$^{-2}$.}\label{Fig:W_Aql_intensity}
\end{figure}

Regarding the intensity images, we note that W~Aql's known companion \citepads{1965VeBam..27..164H,2011A&A...531A.148R,2013A&A...549A..69M,2015A&A...574A..23D} is visible in the full field of view of the ZIMPOL intensity image (Fig.~\ref{Fig:W_Aql_intensity}) at ($-270.00 \pm        0.18$~mas; $-417.60 \pm 0.18$~mas) in (RA; Dec). The study of this system will be the subject of a dedicated forthcoming paper (Danilovich et al. in prep.). We checked for previously undetected companions around other \textsc{Atomium} sources in the ZIMPOL field of view and could find none.

\subsection{Degree of linear polarization}

\begin{figure*}
        \centering
        \includegraphics[width=2\columnwidth]{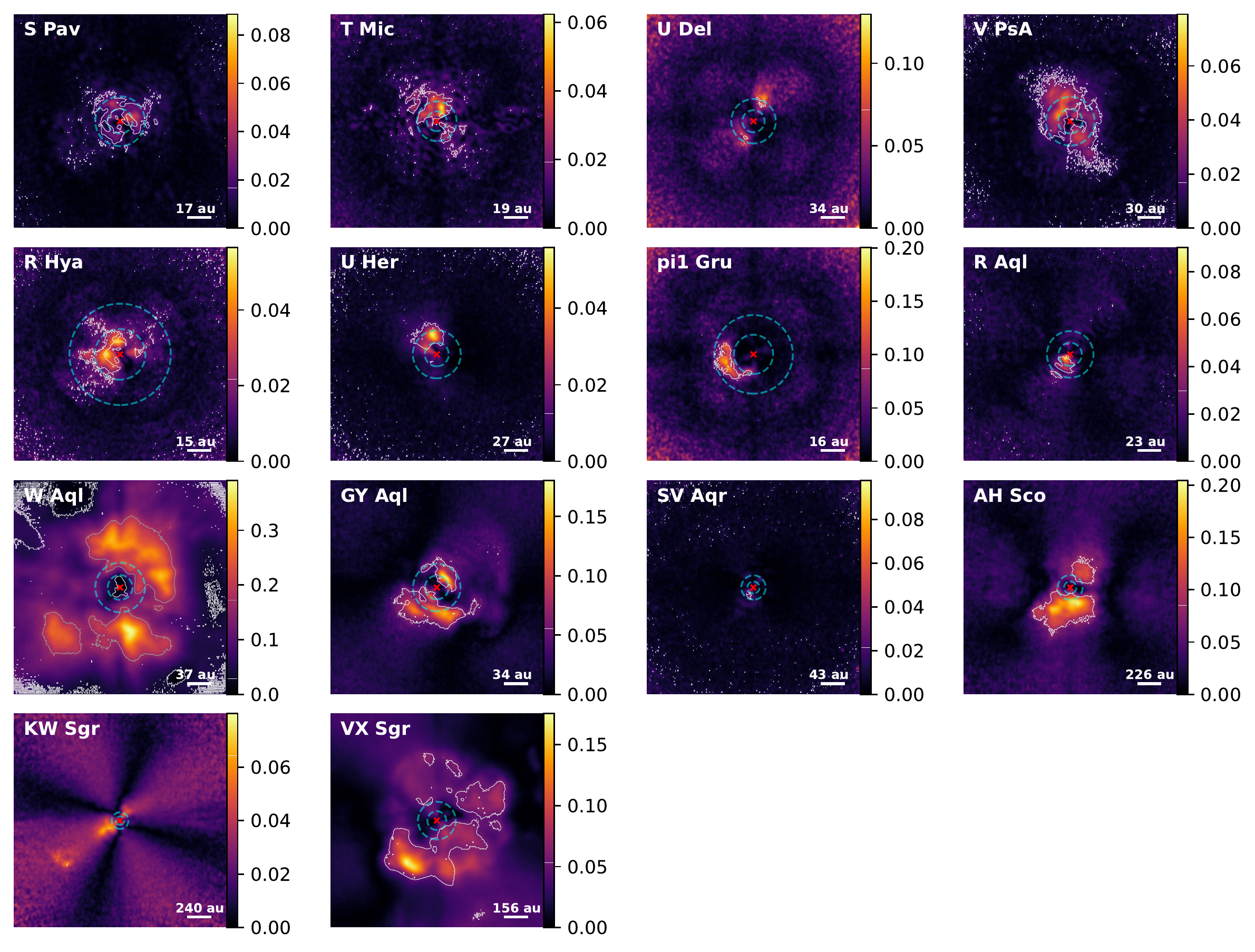}
        \caption{DoLP observed with VLT/SPHERE-ZIMPOL for the \textsc{Atomium} sources. For each star the image taken with the filter at the shortest available wavelength is shown. The white contours correspond to the $5\sigma$ (and $30\sigma$ for W~Aql) level, the red cross indicates the star position, the dashed cyan circles correspond to distances of 10 and 20 R$_\star$ from the star center. The field of view is 1 arcsec x 1 arcsec for each source. The write ruler at the bottom of the image defines the size scale. North is up, and east is to the left. We note that the color scale has been adapted to each source to highlight the polarization signal.}
        \label{Fig:ZIMPOL_DoLP_all}
\end{figure*}

The DoLP of each \textsc{Atomium} source is represented in Fig.~\ref{Fig:ZIMPOL_DoLP_all}. A $5\sigma$ signal or higher is retrieved for all stars, except for SV Aqr and KW Sgr. \citetads{2019A&A...631A.155B} estimate the instrumental polarization to be $\sim 0.4\%$, well below what is observed on our \textsc{Atomium} sources. For comparison, the DoLP of the PSF sources is presented in Fig.~\ref{Fig:ZIMPOL_DoLP_PSF}. No significant polarization is present in any of them. 

Some instrumental artifacts are visible in the images of both the main sources and the PSF calibrators. First, in all PSF sources except HD~180459, HD~184413, HD~166031, and in only the \textsc{Atomium} sources S~Pav, T~Mic, U~Del, V~PsA, R~Hya, and $\pi^1$~Gru, the DoLP shows the imprint of the AO correction ring as well as a cross oriented north-south and east-west (see Figs.~\ref{Fig:ZIMPOL_DoLP_all} and \ref{Fig:ZIMPOL_DoLP_original}). It is\ visible as a brighter area in the intensity images (see example on W Aql in Fig.~\ref{Fig:W_Aql_intensity}) and as lower DoLP regions. Secondly, SV~Aqr and AH~Sco show some unusual low DoLP patterns (a cross for the former and two inverted brackets to the east and west for the latter, Fig.~\ref{Fig:ZIMPOL_DoLP_all}). Figure~\ref{Fig:ZIMPOL_DoLP_original} represents the original DoLP, with a lower S/N, derived  from the computation of the polarized flux directly from Stokes parameters $U$ and $Q$. It shows that these isolated artifacts are introduced from the $U_\phi$/$Q_\phi$ calculation. Since the improvement in S/N exceeds the degradation caused by the artifacts for all other sources, we continue with the $U_\phi$/$Q_\phi$ data processing.

In Fig.~\ref{Fig:DoLP_all_filters} we plot the DoLP for all observed filters for stars that have multi-filter observations. The DoLP we observe spreads across several wavelengths. In line with results from \citetads{2016A&A...585A..28K} and \citetads{2021MNRAS.502..369C} for RSGs, we attribute this observation to circumstellar dust scattering the light anisotropically. Table~\ref{Table:MaxDOLP} shows the maximum value of the DoLP in the inner 400~mas (inside the adaptive optics correction ring), for the different sources across filters. Both Fig.~\ref{Fig:DoLP_all_filters} and Table~\ref{Table:MaxDOLP} show a similar trend: the DoLP decreases for the longest wavelengths, compared to the shortest ones. This could be due to decreased instrumental sensitivity (although to our knowledge none has been reported) but it is also consistent with the trend obtained from the radiative transfer simulation discussed in Sect.~\ref{Sect:Simulations}. In the following, because of the very narrow field of view of SPHERE-ZIMPOL, we will not consider additional scattering on the line of sight from the outer circumstellar environment that could lower the DoLP.

The only feature that is present in all the DoLP images is the dark central area corresponding to the star (the DoLP is the polarized flux divided by the total intensity, this implies a close to zero central disk corresponding to the bright star intensity image). The DoLP in the circumstellar environment displays very diverse shapes, from clumps (U~Her and U~Del) and arcs (S~Pav, AH~Sco, $\pi^1$~Gru, and GY~Aql) to full envelopes (W~Aql and V~PsA). These signals are compared to the ALMA observations in Sect.~\ref{SubSect:ALMA}. In Sect.~\ref{SubSect:IndivTargets} we detail the signal detected on specific sources we deem most interesting.

\begin{table*}
        \centering
        \caption{Maximum DoLP observed on the \textsc{Atomium} sources for the different ZIMPOL filters in the inner arcsec$^2$.}\label{Table:MaxDOLP}
        \begin{tabular}{lllllll}
                \hline\hline
                \noalign{\smallskip}
                & \multicolumn{6}{c}{Filter names \& central wavelengths (nm)}\\
                Source &  CntHa &  N\_R &  VBB &  Cnt748 &  N\_I &  Cnt820 \\
                 & 644.9 & 645.9 & 735.4 & 747.4 & 816.8 & 817.3 \\
                \hline
                \noalign{\smallskip}
                S Pav &     - & 0.06 &   - &    0.05 &   - &    0.06 \\
                T Mic &     - & 0.06 &   - &    0.06 &   - &    0.05 \\
                U Del &   0.12 &   - &   - &    0.13 &   - &    0.09 \\
                V PsA &     - & 0.06 &   - &      - & 0.04 &      - \\
                SV Aqr &     - &   - & 0.04 &      - &   - &      - \\
                R Hya &   0.06 &   - &   - &    0.04 &   - &    0.03 \\
                U Her &     - &   - & 0.06 &      - &   - &      - \\
                $\pi^1$ Gru &   0.20 &   - &   - &    0.13 &   - &    0.12 \\
                R Aql &     - & 0.09 &   - &    0.10 &   - &    0.06 \\
                W Aql &     - &   - & 0.39 &      - &   - &      - \\
                GY Aql &     - &   - & 0.18 &      - &   - &      - \\
                AH Sco &     - & 0.20 &   - &    0.21 &   - &    0.16 \\
                KW Sgr &     - &   - & 0.07 &      - &   - &      - \\
                VX Sgr &     - &   - & 0.18 &      - &   - &      - \\
                \hline
        \end{tabular}
	\tablefoot{ We consider the uncertainty to be 0.004 on each measurement, from the instrumental polarization.}
\end{table*}

%%%%%%%%%%%%%%%%%%%%%%%%%%%%%%%%
%      TOY SIMULATIONS
%%%%%%%%%%%%%%%%%%%%%%%%%%%%%%%%

\section{Polarization signature modeling using \texttt{RADMC3D}}\label{Sect:Simulations}

In order to evaluate the effect of the dust properties and distribution on the DoLP measured by ZIMPOL, we ran 3D dust radiative transfer simulations. We used the publicly available code \texttt{RADMC3D}\footnote{\url{https://www.ita.uni-heidelberg.de/~dullemond/software/radmc-3d/}} \citepads{2012ascl.soft02015D}. The DoLP in the vicinity of the \textsc{Atomium} sources shows very diverse values (between a few percent and $\sim$ 40\%) and morphologies. It is beyond the scope of the present study to attempt to reproduce in detail the DoLP for each individual star. Instead, we aim to characterize the effects of the various circumstellar parameters on the polarized signal.  We computed the simplest model possible: a spherical dust clump placed around a typical AGB star from the \textsc{Atomium} sample. Using this common model, we interpret the observations of the \textsc{Atomium} sample as a whole in Sect.~\ref{Sect:Discussion}.

\subsection{RADMC3D general setup}

In the simulations, the central star was modeled using \texttt{PHOENIX} atmospheres \citepads{2013A&A...553A...6H} with an effective temperature of 2800~K, $\log g = 0.0$, solar metallicity, located at 200~pc with an angular diameter of 16~mas (R = 344 R$_\odot$). We chose this set of parameters to be representative of the \textsc{Atomium} sample (see Table~\ref{Tab:ATOMIUM_caract}). Some of the \textsc{Atomium} sources have an effective temperature that differs by up to 600~K from the value we chose. However, since the DoLP is the ratio between the polarized flux from scattering and the total intensity at a given pixel, the temperature dependence is negligible. The main consequence will be a change in the dust onset radius, which should be evaluated individually, for instance by combining the SPHERE-ZIMPOL and ALMA observations in future papers.

The numerical grid was built using spherical coordinates centered on the star center. According to the \texttt{RADMC3D} manual, the star has to be outside the grid. Therefore, we set its extension from the mean radius between the stellar photosphere and the location of the edge of the clump closest to the photosphere, to 1.5 times the radial coordinate of the outer edge of the clump. For the longitudinal direction ($\theta$) the grid explored the range (0, $\pi$), and for the azimuthal direction ($\phi$) it ranged between 0 and 2$\pi$. The grid had (20, 40, 40) initial cells, linearly spread, in the (r, $\theta$, $\phi$) directions. If dust were present in a cell, the cell size was refined up to three times using an adaptive mesh refinement. In the following we discuss the clump position in a Cartesian coordinate system (RA, Dec, $z$) with RA (resp. Dec) the relative right ascension (resp. declination) with respect to the star center, and $z$ the position on the line of sight with respect to the star center.

We used the full anisotropic scattering functionality of \texttt{RADMC3D} to produce the three Stokes parameters ($I$, $Q$, $U$), which were then combined to produce the ZIMPOL observables. Because we were only interested in the dust induced polarization, we did not consider gas in these simulations. We did not use any smooth stellar wind either (dust or gas), because the features observed on our sources (Fig.~\ref{Fig:ZIMPOL_DoLP_all}) are not consistent with such an outflow. The dust properties were taken from the Jena Database of Optical Constants for Cosmic Dust\footnote{\url{https://www.astro.uni-jena.de/Laboratory/OCDB/}}. The following five dust species -- commonly found in the environment of cool evolved stars -- were input individually in separate simulations with the \texttt{RADMC3D} code: glassy MgFeSiO$_4$ (olivine, \citeads{1994A&A...292..641J,1995A&A...300..503D}), glassy Ca$_2$Al$_2$SiO$_7$ (melilite, \citeads{1998A&A...333..188M}), Al$_2$O$_3$ (corundum, \citeads{2013A&A...553A..81Z}), amorphous MgSiO$_3$ (enstatite, \citeads{2003A&A...401...57J}), or amorphous Mg$_2$SiO$_4$ (forsterite, \citeads{2003A&A...401...57J}). The \texttt{RADMC3D} code provides a Python interface to transform the optical constants to optical opacities as a function of wavelength (using Mie theory) when the appropriate specific mass, and chosen grain-size range and distribution are provided. We set 30 grain sizes that were logarithmically spread over the interval. We used randomly oriented dust grains. According to the \texttt{RADMC3D} manual, if all grains are randomly oriented and without any preferential helicity, for each particle shape there are equal numbers       of particles with that shape and with its mirror copy shape.

The \texttt{RADMC3D} simulation computes the dust temperature (assuming radiative equilibrium) and traces the light scattering before producing images and/or spectra at the requested wavelength.

As state above, we used a spherical clump with homogeneous density distribution (the simplest geometric setup) in order to assess the DoLP that can be recovered from the simulations. We ran several parameter studies (changing the clump characteristics and dust properties) whose results are assessed below, after an investigation of the PSF convolution effect.

\subsection{Effect of the PSF convolution on the polarization}\label{SubSect:PSF_conv}

For these preliminary tests we used a clump of 4~au in radius, with a constant dust density of 10$^{-17}$ g~cm$^{-3}$ (a value in the range derived for clumps offset cool evolved stars; e.g., \citeads{2021MNRAS.502..369C,2021Natur.594..365M}). The clump was located at 10~au from the star center, exactly to the east in the plane of the sky through the center of the star. For this setup, the dust consisted only of olivine, with grain sizes ranging between 0.01 and 0.1~$\mu$m.

\begin{figure}
        \centering
        \includegraphics[width=\columnwidth]{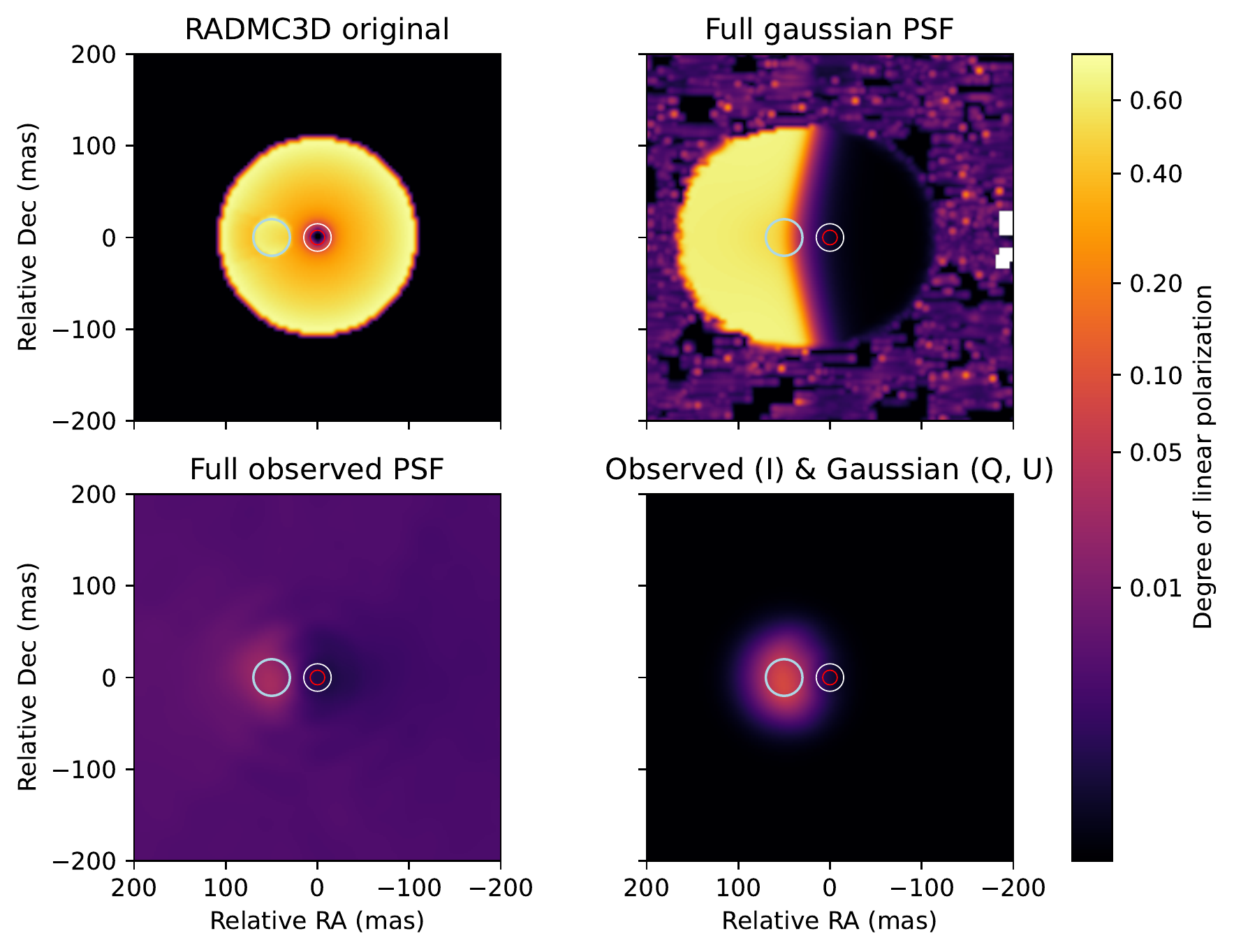}
        \caption{Example of DoLP maps obtained with \texttt{RADMC3D} with different PSF convolutions. All the other parameters of the simulation remain the same. The observation wavelength is 644.9~nm. The color scale is a power law with an exponent of 0.25. The red circle indicates the stellar angular diameter, the white circle corresponds to the PSF reference's FWHM, and the light blue circle corresponds to the physical localization of the dust clump. \textit{Top left:} Original \texttt{RADMC3D} DoLP. \textit{Top right:} Resultant DoLP after convolution of Stokes I, Q, U signals with a synthetic Gaussian beam of 30~mas FWHM. \textit{Bottom left:} Resultant DoLP after convolution of Stokes I, Q, U with an observed PSF calibrator by ZIMPOL (HD 220340). \textit{Bottom right:} Resultant DoLP after convolution of the Stokes I with the observed PSF reference, and Stokes Q and U with the synthetic Gaussian.}\label{Fig:RADMC3D_PSF}
\end{figure}

\texttt{RADMC3D} produces images in the three Stokes parameters (I, Q and U). To allow a meaningful comparison with observational data, these images needed to be convolved with the PSF. The accompanying Python module \texttt{radmc3dPy} provides tools to perform the PSF convolution by a 2D Gaussian or an Airy pattern corresponding to a circular aperture with a central obscuration (caused by the secondary mirror). Once the convolution was performed on the Stokes parameters, the DoLP could be derived. The synthetic PSF references of \texttt{RADMC3D} produce very sharp DoLP features with a level that is always higher than the actual data, regardless of the input parameters (Fig.~\ref{Fig:RADMC3D_PSF}, top right). The DoLP also spreads beyond the actual clump extension. The resulting maps are unrealistic. Another possibility is to use the observed PSF calibrator from ZIMPOL in intensity, to be convolved with each one of the three Stokes parameters (Fig.~\ref{Fig:RADMC3D_PSF}, bottom left). This produces maps with a low S/N. The asymptotic value of the DoLP reaches more than 1\%, which is inconsistent with the observed instrumental polarization of 0.4\% \citepads{2019A&A...631A.155B}.

Because of the relatively sharp edges of its kernel and the absence of instrumental noise, the convolution with a synthetic PSF reference preserves a strong DoLP and cause it to spread well beyond the clump physical extension. In contrast to this, the convolution of Stokes I, Q, and U by the observed PSF calibrator destroys most of the polarized signal due to the asymptotic noise level and AO correction features it contains.

To derive a polarized signal consistent with the ZIMPOL observations, we used a procedure inspired by the way the Stokes parameters are obtained from the ZIMPOL reduced frames. Indeed, the ZIMPOL pipeline delivers Stokes $+Q$ and Stokes $+U$, as well as their opposite values $- Q$ and $-U$, leading to the following polarization processing \citepads{2015A&A...578A..77K}: 
\begin{eqnarray}
        I &=& \sqrt{I_Q^2 + I_U^2}\\
        Q &=& \frac{+Q - (-Q)}{2}\\
        U &=& \frac{+U - (-U)}{2}.
\end{eqnarray} 
\noindent This implies a lower systematic bias for the final $Q$ and $U$ than for $I$, since they are derived from the subtraction of +$Q$/-$Q$ and +$U$/-$U$. \texttt{RADMC3D} only gives us the $Q$ and $U$ frames directly. The convolution of these maps was performed using the following procedure: we convolved the $Q$ and $U$ frames with a synthetic Gaussian with a FWHM of 30~mas (corresponding to the averaged values of the observed PSF references; see Table \ref{Tab:FitIntensity}), and the $I$ frame with the observed PSF intensity. This results in the DoLP map shown in the bottom-right panel of Fig.~\ref{Fig:RADMC3D_PSF}, which shows a signal contained at the clump position, which is consistent with the observations for a large range of the input parameters. Henceforth, we use this ``mixed'' convolution approach.

\subsection{Parameter studies}\label{SubSect:RADMC3D_ParamStudies}

We performed several simulations for different characteristics of the dust clump, in order to determine their effect on the polarized signal. In the following, all parameters were left untouched, but for the one being probed. Only the maximum DoLP of the model (after PSF convolution) is discussed, but the information is extracted from images such as the one presented in the bottom-right panel of Fig.~\ref{Fig:RADMC3D_PSF}.

\begin{figure}
        \centering
        \includegraphics[width=\columnwidth]{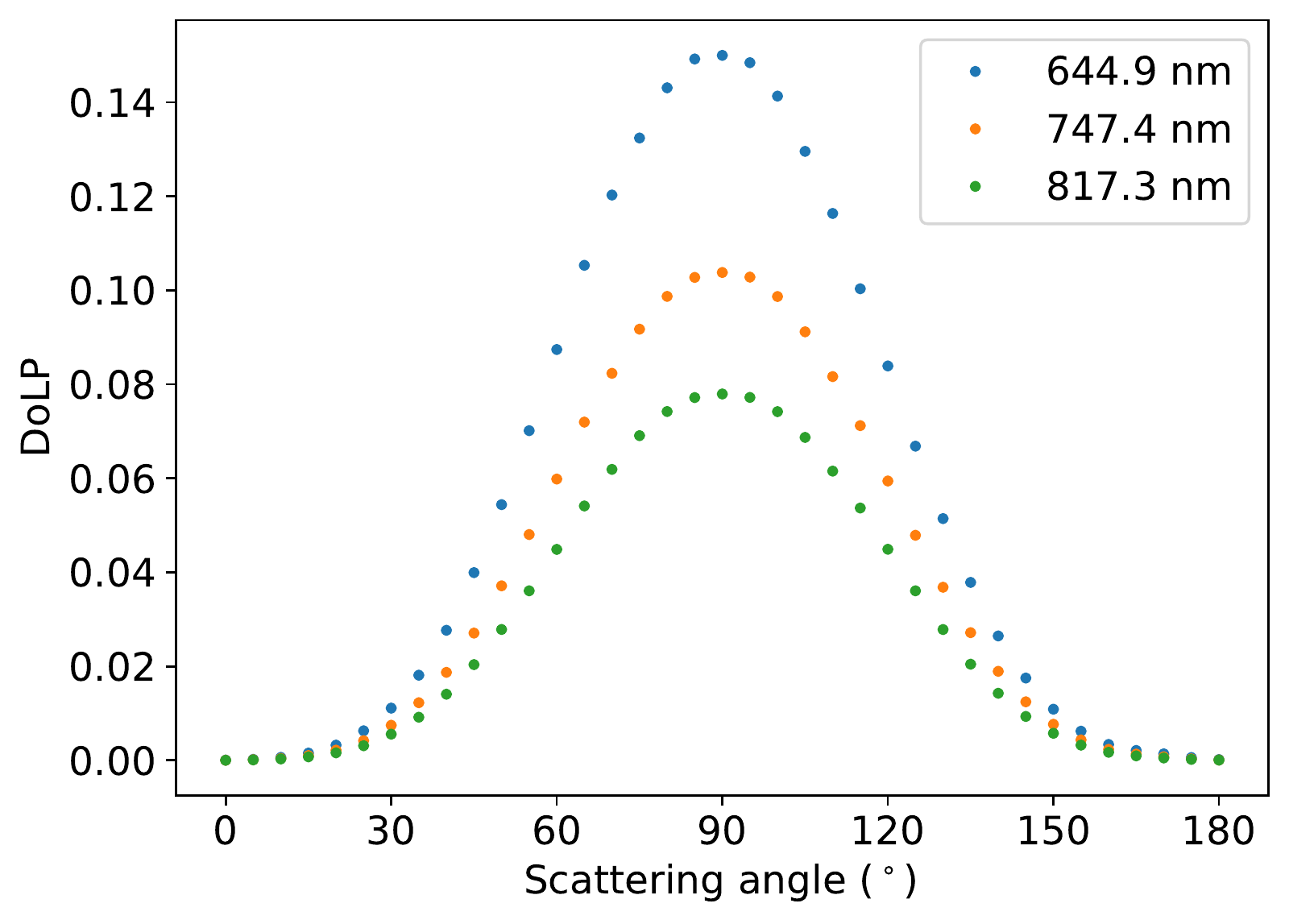}
        \caption{Maximum DoLP for a dust clump at 15~au from the model AGB star as a function of the scattering angle. Each color represents a different ZIMPOL filter.}\label{Fig:RADMC3D_ScatAngle}
\end{figure}

\paragraph{Scattering angle.} As in Sect.~\ref{SubSect:PSF_conv}, we set the spherical clump radius to 4~au, and its homogeneous density is set to 10$^{-17}$ g~cm$^{-3}$. The dust was composed of olivine, with the dust grain size ranging from 0.01 to 0.1~$\mu$m. We put the clump center position at 15~au ($\sim 9$~R$_\star$) from the star center. Fig.~\ref{Fig:RADMC3D_ScatAngle} shows how the maximum DoLP from the clump changes, as we explored the full range of possible scattering angles (from 0$^\circ$ directly in front of the star to 180$^\circ$ directly behind it). As expected, the maximum DoLP is reached for a scattering angle of 90$^\circ$, when the clump is in the plane of the sky. The wavelength dependence will be discussed in the dust properties paragraph.

\begin{figure}
        \centering
        \includegraphics[width=\columnwidth]{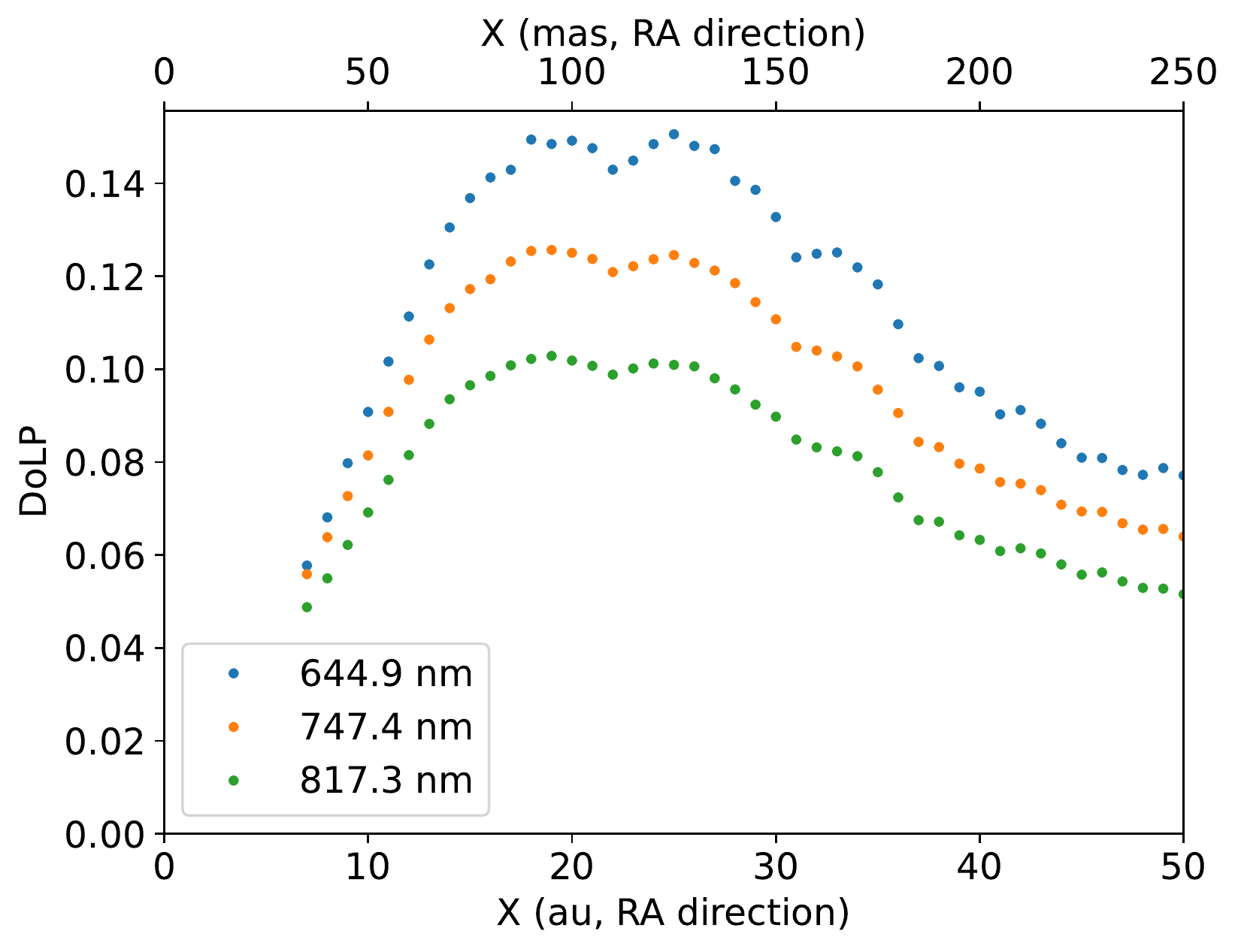}
        \caption{Maximum DoLP for a dust clump placed at distances (X) between 7 and 50~au from the star center.}\label{Fig:RADMC3D_pos}
\end{figure}

\paragraph{Clump distance from the star.} In a next step, we set the clump in the plane of the sky, exactly to the east of the star (null relative declination). The clump center position was placed in the range from 7 to 50~au from the star center (see Fig.~\ref{Fig:RADMC3D_pos}). The DoLP increased when we moved the clump away from the star intensity halo. It reached a plateau at $0.12 \pm 0.03$ (depending on the wavelength) when the clump center was at 17~au from the star center (i.e., the inner edge of the clump was at 13~au from the center). This translates to 85 and 65~mas, respectively, at 200~pc. The DoLP decreased after its center passed the 25~au position, implying that its inner edge reached the 21~au distance from the star center (i.e., at 125~mas and 105~mas, respectively, at 200~pc). In this regime, the polarized flux decreases because from the clump, the star appears within a smaller solid angle. Simultaneously, the total scattered light decreases in the same way, causing a constant DoLP in the clump. However, in the observed images of the PSF reference stars (used in the convolution), the total intensity reaches a floor due to the instrumental noise. This causes the decreasing polarized flux to be divided by a roughly constant intensity, hence a decreasing DoLP with increasing distance. Consequently, the sensitivity to scattered dust will decreases over the field of view, until reaching the AO ring limit at $\sim 360$~mas (= 72~au at 200~pc), after which the AO decreased performance prevents any detection. To summarize, our SPHERE observations have the highest sensitivity to the dust polarized signal just outside the PSF halo (from 85 to 125~mas).

\begin{figure}
        \centering
        \includegraphics[width=\columnwidth]{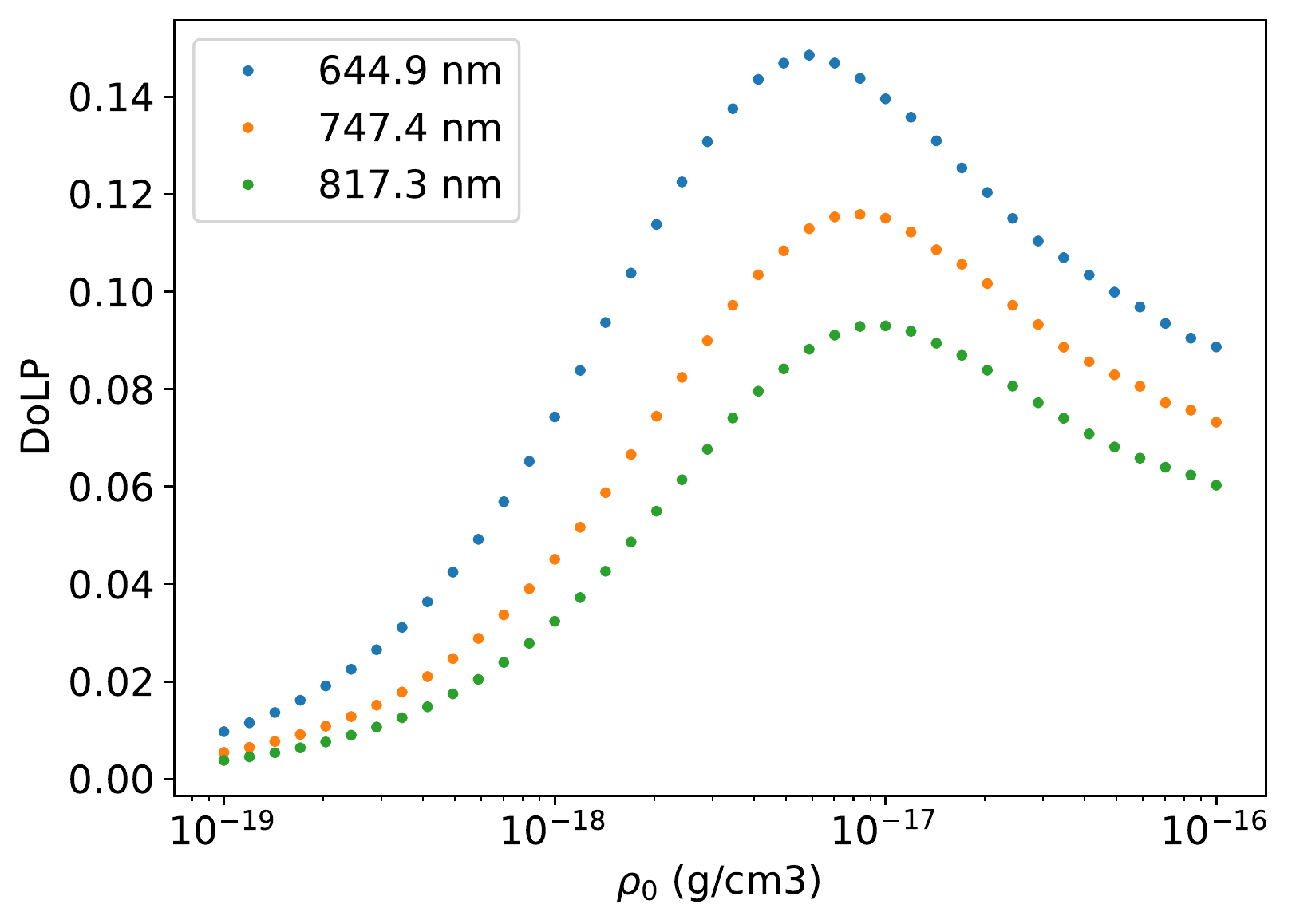}
        \caption{Maximum DoLP for a dust clump placed at 15~au from the star center as a function of the clump density.}\label{Fig:RADMC3D_density}
\end{figure}

\paragraph{Dust density.} The clump was set at 15~au from the star, center to center, in the plane of the sky, exactly in the eastern direction. We kept a grain size distribution in the range $0.01-0.1~\mu$m.  Only the clump dust density was allowed to vary, from 10$^{-19}$ to $10^{-16}$~g~cm$^{-3}$, using logarithmically spaced values. The simulations explored the effect of dust optical depth on the polarized signal. Fig.~\ref{Fig:RADMC3D_density} shows that the DoLP increases as the clump contains more and more dust, until approaching $8\times10^{-18} - 10^{-17}$~g~cm$^{-3}$ (depending on the wavelength). At these densities, we find that the clump is optically thick. In this new regime, less and less dust contributes to the emergent polarized signal through light scattering, implying that the maximum DoLP decreases when the clump density increases.

\begin{figure*}
        \centering
        \includegraphics[width=2\columnwidth]{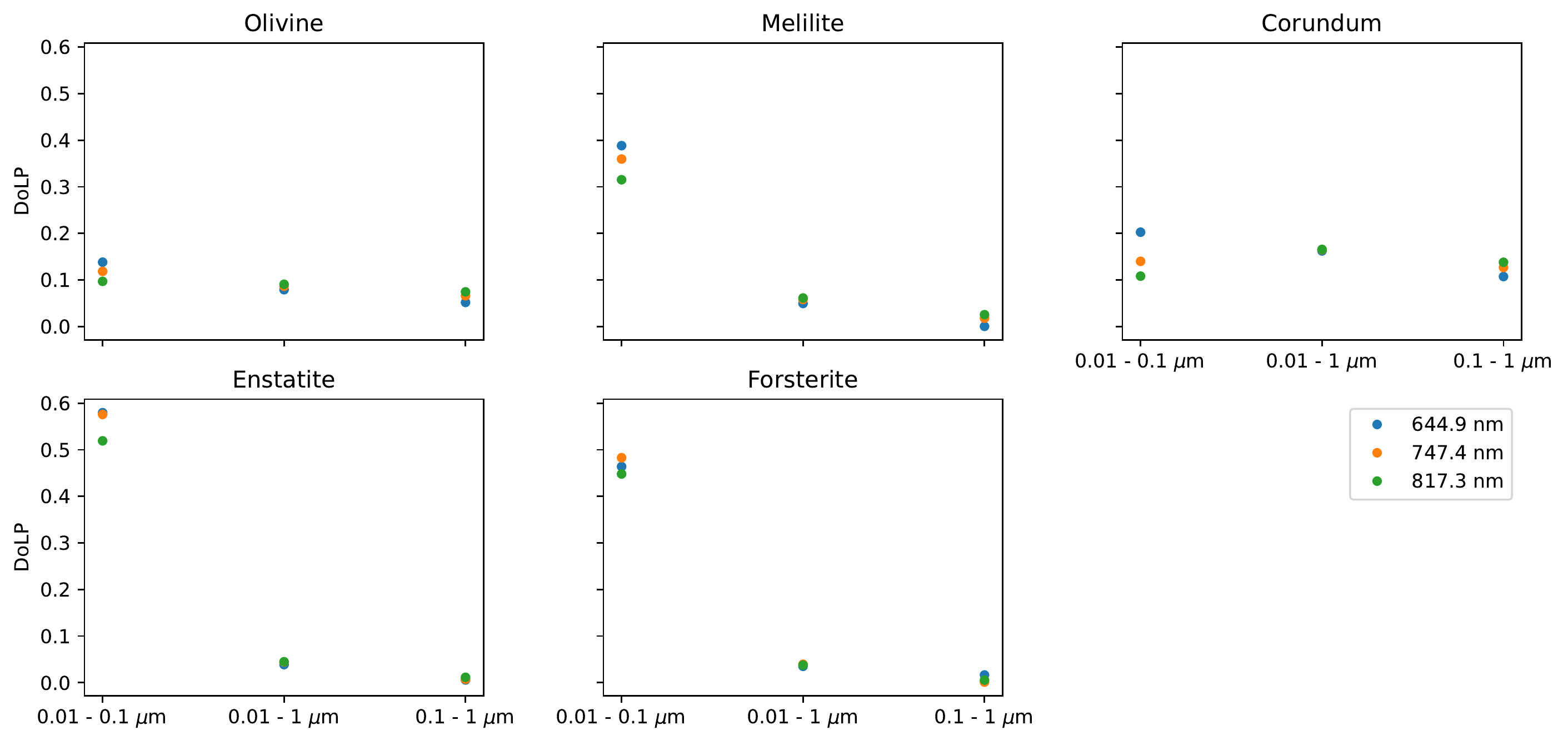}
        \caption{Maximum DoLP from a dust clump as a function of the dust composition (in the different panels) and grain size ranges (x-axis). The colors correspond to the observed wavelengths given in the legend.}\label{Fig:RADMC3D_dust}
\end{figure*}

\paragraph{Dust properties.} In this final setup, we used a dust density of $10^{-17}$~g~cm$^{-3}$ and we explored the various dust compositions described earlier: olivine, melilite, corundum, enstatite, and forsterite, for three grain size ranges: 0.01 to 0.1~$\mu$m, 0.1 to 1~$\mu$m, and 0.01 to 1~$\mu$m. In Fig.~\ref{Fig:RADMC3D_dust}, we observe very different trends depending on whether large grains (0.1 to 1~$\mu$m) are included. In the small grain regime (0.01 to 0.1~$\mu$m), that is, where the wavelength is smaller than the observed wavelength, enstatite and forsterite produce the stronger DoLP (up to 45-60\%) at any wavelength. For the three other species, the DoLP is maximized at the shorter wavelength (644.9~nm). For large grains, the DoLP remains relatively constant and weak over the observed wavelength range. For this reason, in the previous parameter studies we opted for the small grain-size distribution. Additionally, the ZIMPOL observations were obtained close to the stars, where we expect dust nucleation to start. For this smallest grain sizes on the order of 0.01~$\mu$m considered in this study, we note that finite size effects like large surface-to-volume ratios and atomic segregation can impact the structure of the grain. As a consequence these small size grain properties, including their optical constants, could be different from those in the Jena database. Overall, the wavelength dependence of the DoLP appears related to the grain size, at a given dust composition. With a broader wavelength baseline (e.g., extending to the infrared), it would be possible to discriminate this dust parameter. Regarding dust composition, our observations do not allow us do draw decisive conclusions: when taking the other parameters studied in the previous paragraphs into account, it appears that olivine and corundum, or forsterite and enstatite, produce rather similar signals. However, we only tested five representative dust species among many others that could be present, which prevents us from establishing the composition of the dust in the star sample. We also decided to keep the grains spherical. Indeed, we have no measurement of the magnetic field around the \textsc{Atomium} sources, which would be the only way to privilege a specific grain orientation. In this case the nonspherical shape of the grains would enhance the polarized signal.

%%%%%%%%%%%%%%%%%%%%%%%%%%%%%%%%
%      DISCUSSION
%%%%%%%%%%%%%%%%%%%%%%%%%%%%%%%%

\section{Discussion}\label{Sect:Discussion}

\subsection{The degree of linear polarization and its link with the dust distribution}\label{SubSect:DiscDustPolar}

Figure~\ref{Fig:DoLP_all_filters} and Table \ref{Table:MaxDOLP} establish that the observed polarization can be traced through several filters, thereby confirming that the signal arises from anisotropic scattering on dust grains, and not scattering off molecules. Moreover, for V~PsA, R~Hya, and $\pi^1$~Gru we observe a decrease of the DoLP with increasing wavelength hinting at the presence of small grains in the circumstellar environment (see Sect.~\ref{SubSect:RADMC3D_ParamStudies}). For the other stars observed with several filters (S~Pav, T~Mic, U~Del, R~Aql, and AH~Sco), the situation is not as clear, with the maximum DoLP remaining constant between two or three filters, which might imply that a wider range of grain sizes is present in their circumstellar dust (Fig.~\ref{Fig:RADMC3D_dust}). Only with polarized imaging extending toward the near-infrared can this be clarified.

From Fig.~\ref{Fig:RADMC3D_ScatAngle}, it is clear that any DoLP observed from the \textsc{Atomium} sources has the highest probability of originating from a 90$^\circ$ scattering angle, that is, from dust located in the plane of the sky through the center of the star. This does not mean that dust is only present in the plane of the sky, or is only present in the location where the DoLP is significant: there could be dust features elsewhere, unable to produce a significant DoLP. For those sources where the spatial extent of the DoLP -- originating most probably from the plane of the sky (see Sect.~\ref{SubSect:RADMC3D_ParamStudies}) -- is limited (clumps, arcs), it seems reasonable to assume that this is also the case in the line-of-sight direction. Consequently, for those targets this would imply that the observed dust features are confined close to the plane of the sky. Significant exceptions are W~Aql and VX~Sgr, which show nearly complete shells spatially extended in the plane of the sky. These may also extend outside this plane along both directions in the line of sight. 

\begin{figure}
        \centering
        \includegraphics[width=\columnwidth]{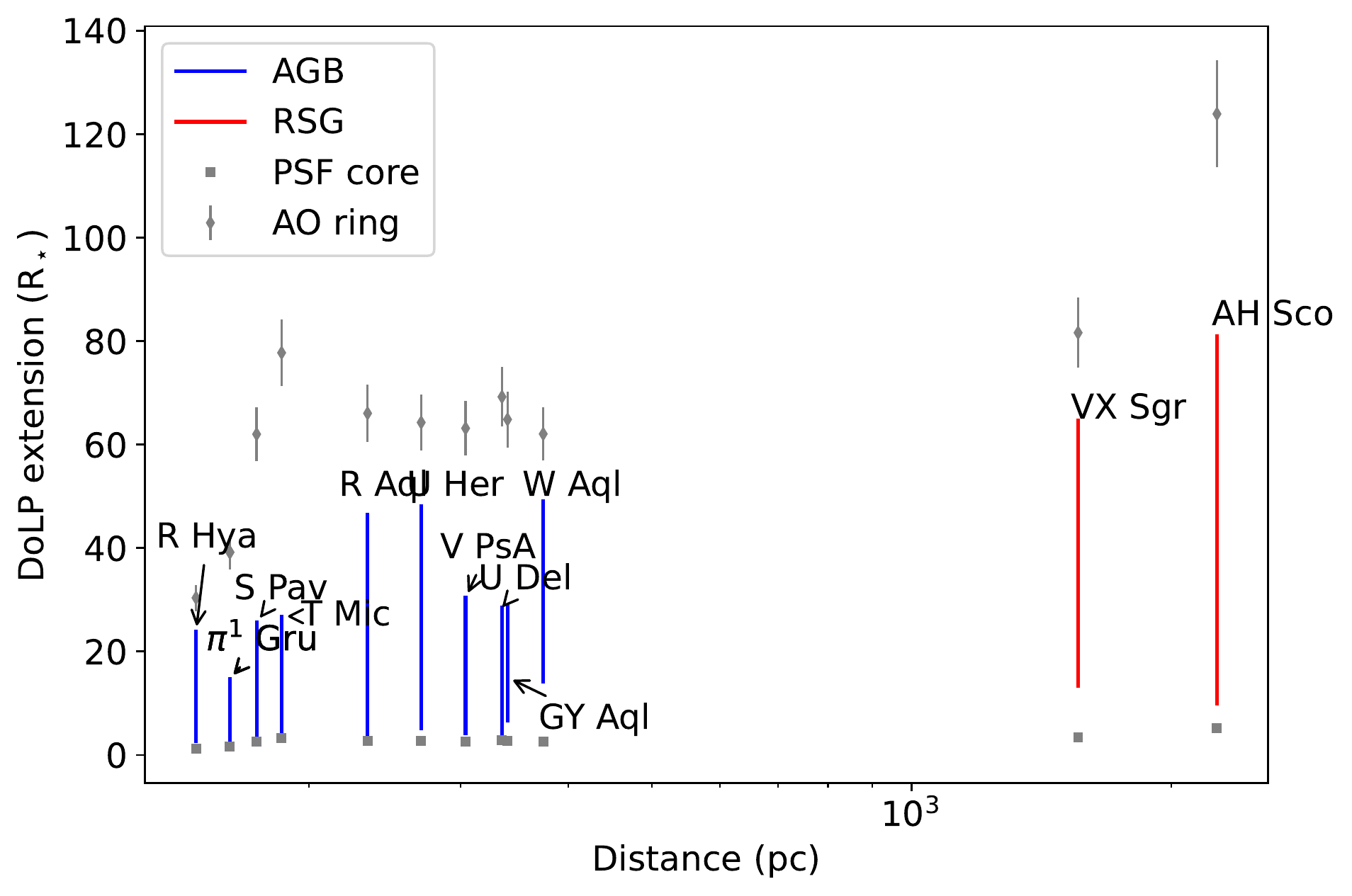}
        \caption{DoLP radial extension in R$_\star$, taken as the radial range over which the cumulative DoLP lies between 10\% and 95\% of its maximum, as a function of the star distance. The RSGs are in red, and the AGB stars are in blue. The gray squares correspond to the PSF calibrator half width at half maximum (in units of R$_\star$) at the distance of each star. The gray diamonds with an error bar correspond to the position and extent of the AO ring of SPHERE. We used the filter at the shortest wavelength observed for each star.}\label{Fig:DoLP_extension}
\end{figure}

In Fig.~\ref{Fig:DoLP_extension} we show the radial extension of the DoLP as a function of the distance of the star from Earth. In order to compute the radial extension, we derived the cumulative DoLP as a function of the radial distance in the image by summing the DoLP through radial bins. To avoid the areas dominated by noise, we defined the DoLP extension as the radial range where 10 and 95\% of the maximum cumulative DoLP are reached. We excluded SV~Aqr and KW~Sgr from this analysis because they do not display a DoLP above the $5\sigma$ limit. It appears that the inner extension of the DoLP is limited by the PSF core, which creates a weak DoLP zone at the center of the image (any polarized signal from this region is rendered insignificant by the high intensity of the central star core). The only exceptions appear to be W~Aql and VX~Sgr. In the case of VX~Sgr, this is most probably due to the adverse seeing conditions during the observations (Table~\ref{Tab:ObsLog}). For W~Aql, the greater inner extension of the DoLP appears physical and not an instrumental artifact. However, we stress that this does not imply that there is no dust closer to the central star: dust could be present without efficiently producing a polarized signal due to its location, grain shape or size, density, etc. In most cases (with the exception of W~Aql and VX~Sgr), the AO correction ring limit that is visible in the intensity images (and represented in Fig.~\ref{Fig:DoLP_extension}) is farther than the largest radial distance at which a significant DoLP is observed. This implies that the upper limit of the DoLP extension of the polarized signal is due to the dust configuration around the star, and is not an instrumental artifact.

To summarize, in most cases the inner extension of the DoLP of our sample is instrumental and comes from the PSF core. In contrast, the outer limitation is mostly physical. Between the PSF core and the AO correction ring, the DoLP provides a reliable estimation of the dust distribution close to (or in) the plane of the sky. However, we saw in Sect.~\ref{SubSect:RADMC3D_ParamStudies} and Fig.~\ref{Fig:RADMC3D_pos} that the DoLP decreases when the dust is farther away from the star, regardless of the position of the AO correction ring of ZIMPOL.

\begin{figure}
        \centering
        \includegraphics[width=\columnwidth]{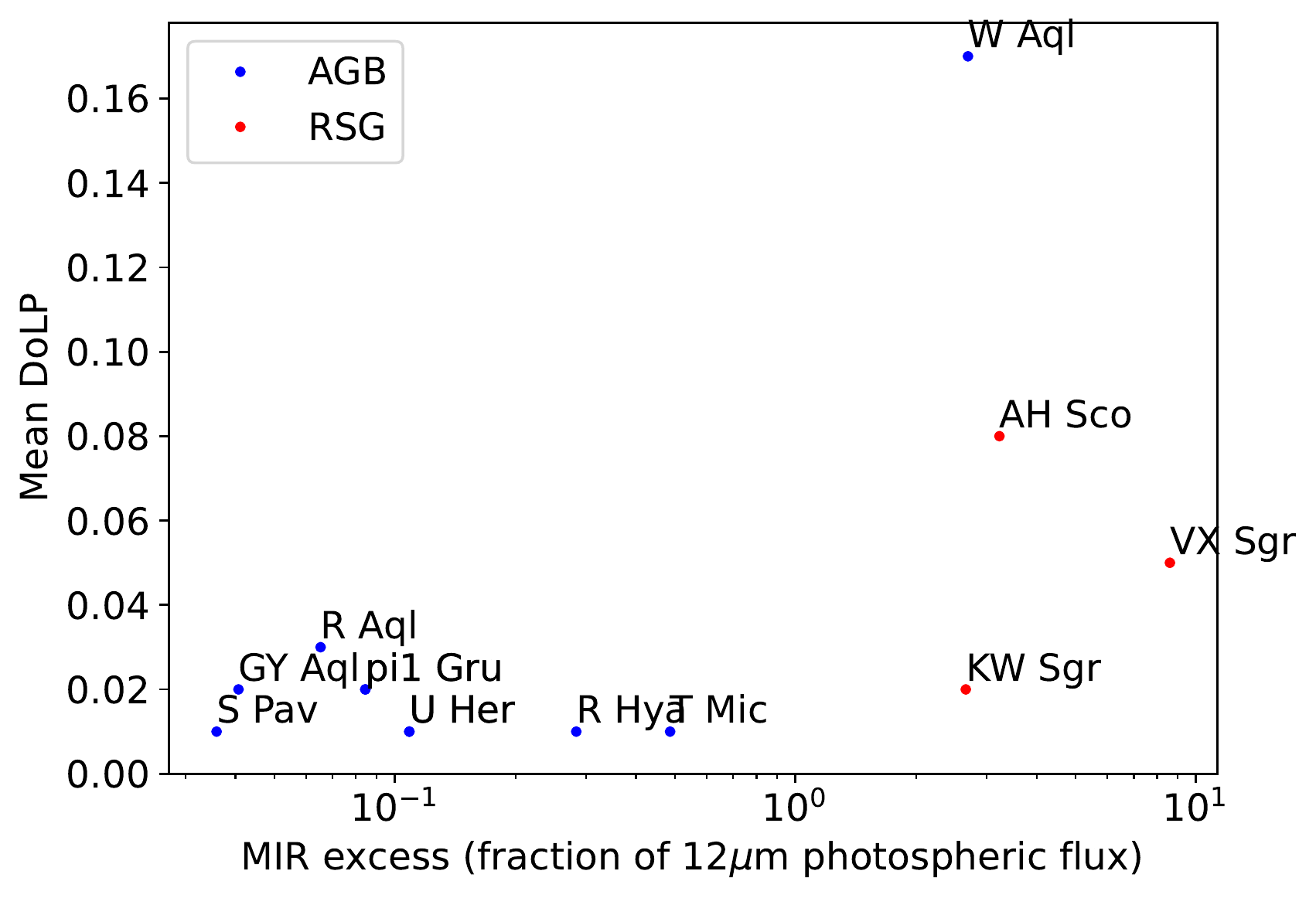}
        \caption{Average DoLP in the inner 50~R$_\star$ for each source, as a function of the MIR excess expressed as the fraction of the 12 $\mu$m photospheric blackbody flux. The RSG stars are in red, and the AGB stars are in blue. We used the filter at the shortest wavelength observed for each star.}\label{Fig:DoLP_MIR_excess}
\end{figure}

We compared the dust detected via polarization close to the star with the spatially unresolved detection of the thermal emission of warm dust for our sources. In Fig.~\ref{Fig:DoLP_MIR_excess}, we plot the average DoLP in the inner 50~R$_\star$ for each source as a function of the mid-infrared (MIR) excess, that is, the difference between the Infrared Astronomical Satellite (IRAS) observations at 12~$\mu$m \citepads{1988iras....7.....H} and the blackbody emission of each star derived from its effective temperature (Table~\ref{Tab:ATOMIUM_caract}). The IRAS 12~$\mu$m signal probes the thermal emission of warm dust. We expect the narrow field of view of SPHERE-ZIMPOL to closely match the expected extent of the 12\,$\mu$m excess. First we note that for U~Del, V~PsA, and SV~Aqr the MIR excess is negative. This is unphysical and could arise from erroneous estimates of their angular diameter or their effective temperature, or from variability of their photometry. Most stars are in a single group, whatever their MIR excess is: they show a mean DoLP of 0.01 to 0.04. This indicates that whatever their warm dust content is, the observed dust in polarization (i.e., in the plane of the sky) in the inner and intermediate circumstellar environments is similar (and relatively low in density, producing an optical depth close to unity) among these sources. This may imply that the ZIMPOL data are catching newly formed dust. In this inner region, dust likely forms only in favorable (density, temperature, pressure) areas. This could explain the clumpy shapes that are observed in most ZIMPOL DoLP maps. Such hypotheses were tested by \citetads{2018A&A...619A..47L} who used the CO5BOLD convective photospheres \citepads{2017A&A...600A.137F} as inner boundaries for their \textsc{Darwin} simulations. They successfully reproduce wind velocities and mass-loss rates for typical AGB stars, as well as density variations along the wind. 
%Farther out on other stars, dust is observed in more spatially extended features, including full clumpy envelopes \citepads{2011A&A...531A.117K,2014A&A...568A..17O}. This means that at greater distances from the photosphere, dust can nucleate in larger regions owing to more favorable ambient conditions. 
A notable outlier in the \textsc{Atomium} sample is KW~Sgr, a distant RSG (Table~\ref{Tab:ATOMIUM_caract} and Fig.~\ref{Fig:DoLP_extension}) for which the dust-onset region is not probed by the ZIMPOL data, because it is smaller than our angular resolution. For this star, the circumstellar dust is not located in a place or does not have a density suitable to produce a detectable polarized signal. The two RSGs AH~Sco and VX~Sgr produce a stronger averaged DoLP, but owing to their greater distance (Table~\ref{Tab:ATOMIUM_caract} and Fig.~\ref{Fig:DoLP_extension}) we are probing a region farther away in their environment than for AGB stars. Combined with their high MIR excess, the probability of detecting dust in a favorable position for producing a strong DoLP is higher. The real outlier is W~Aql, which has the highest averaged DoLP (see Fig.~\ref{Fig:DoLP_MIR_excess}). We conclude that W~Aql (one of two S-type stars in our sample, the other is $\pi^1$~Gru) has produced circumstellar dust that is very efficient at producing a polarized signal, possibly owing to a peculiar dust species or grain size.

\subsection{Comparison with ALMA data}\label{SubSect:ALMA}

\begin{figure*}
        \centering
        \includegraphics[width=0.95\columnwidth]{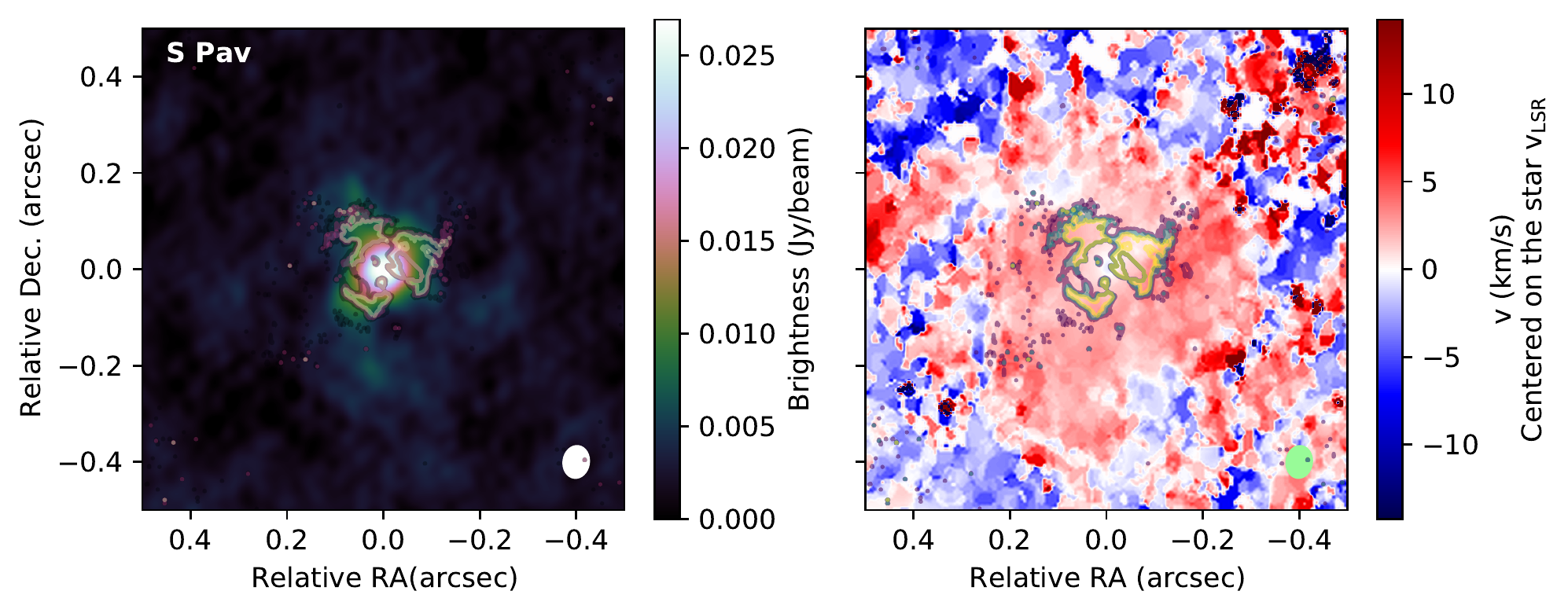}\hfill
        \includegraphics[width=0.95\columnwidth]{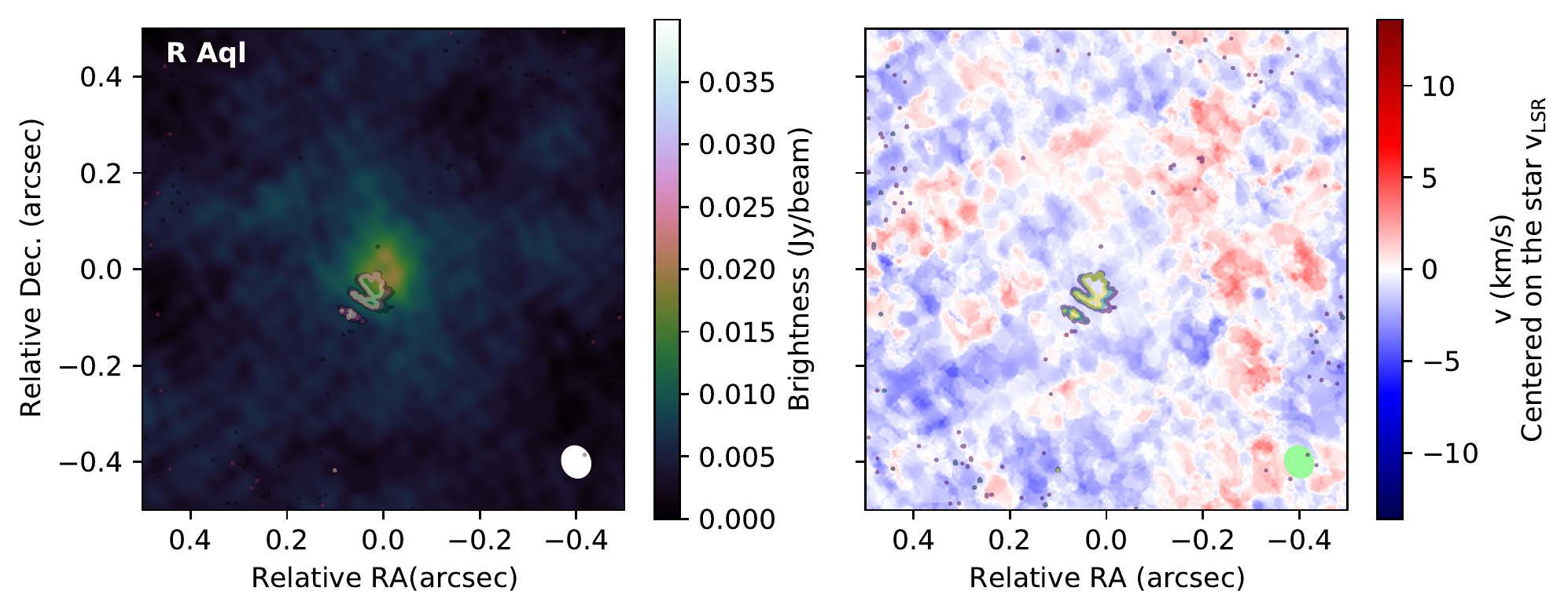}
        
        \includegraphics[width=0.95\columnwidth]{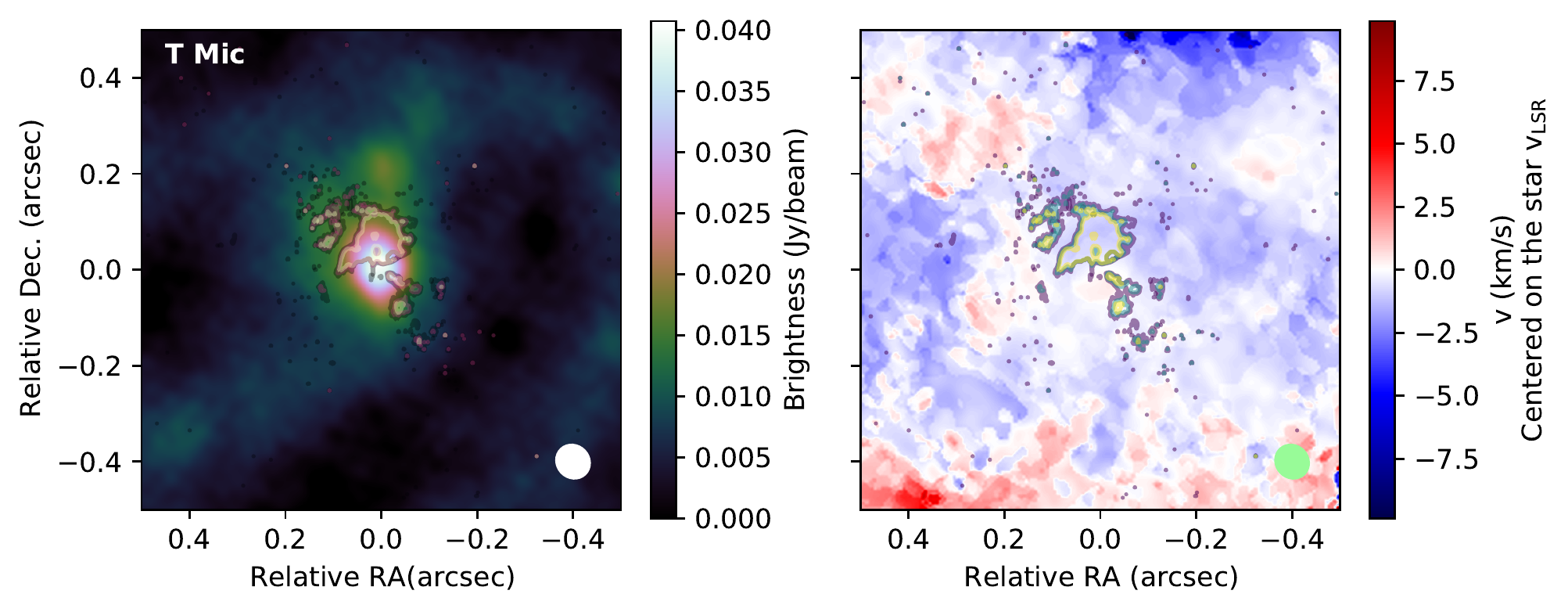}\hfill
        \includegraphics[width=0.95\columnwidth]{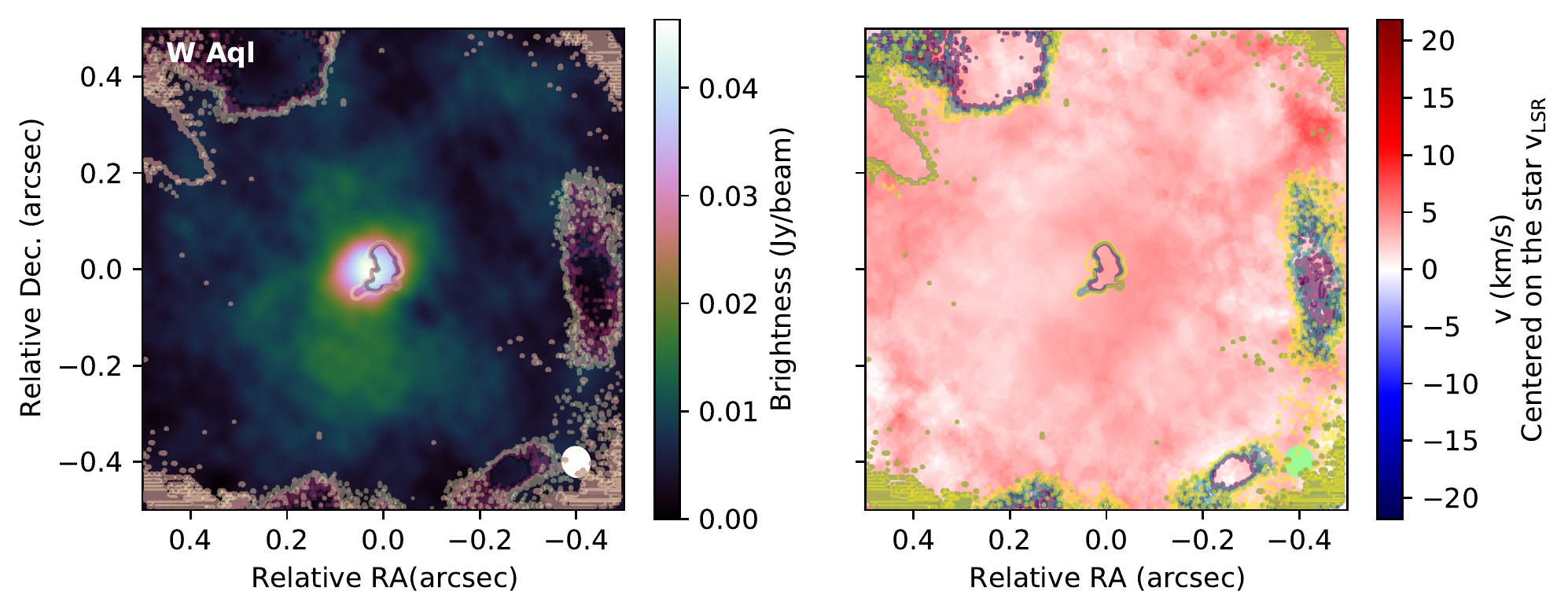}
        
        \includegraphics[width=0.95\columnwidth]{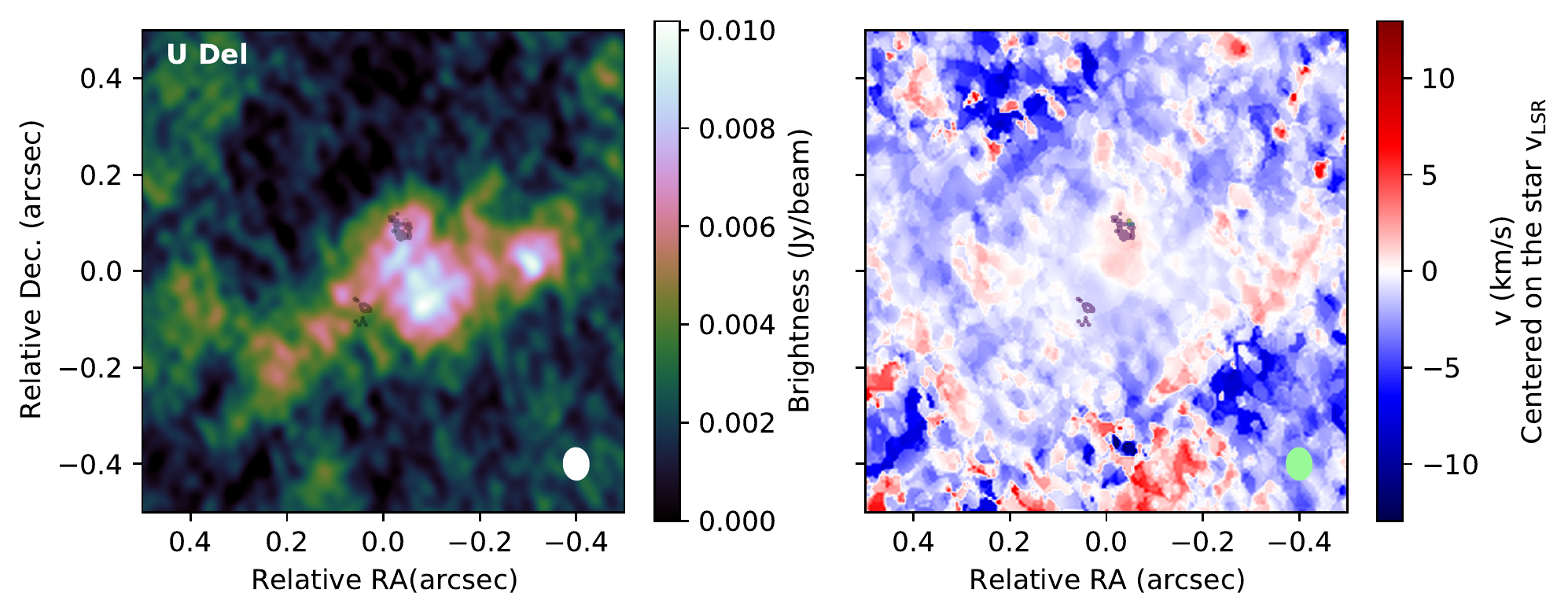}\hfill
        \includegraphics[width=0.956\columnwidth]{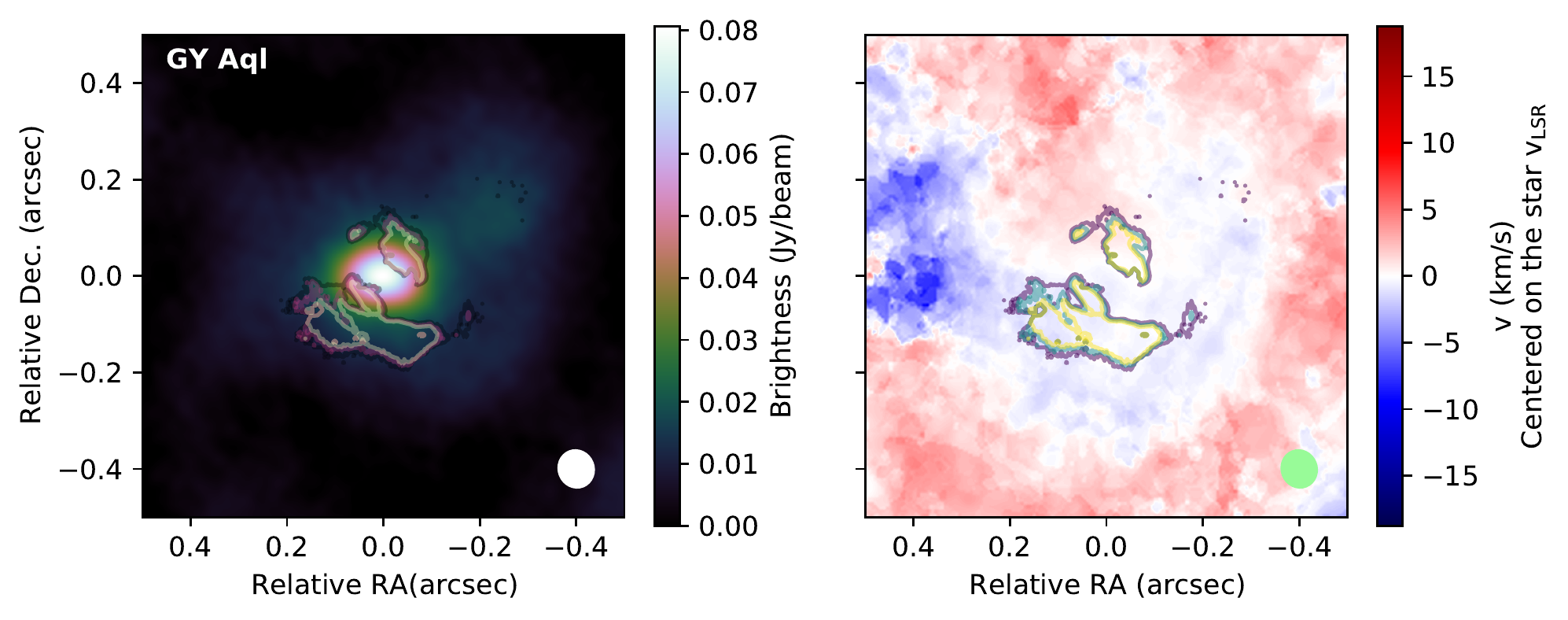}
        
        \includegraphics[width=0.95\columnwidth]{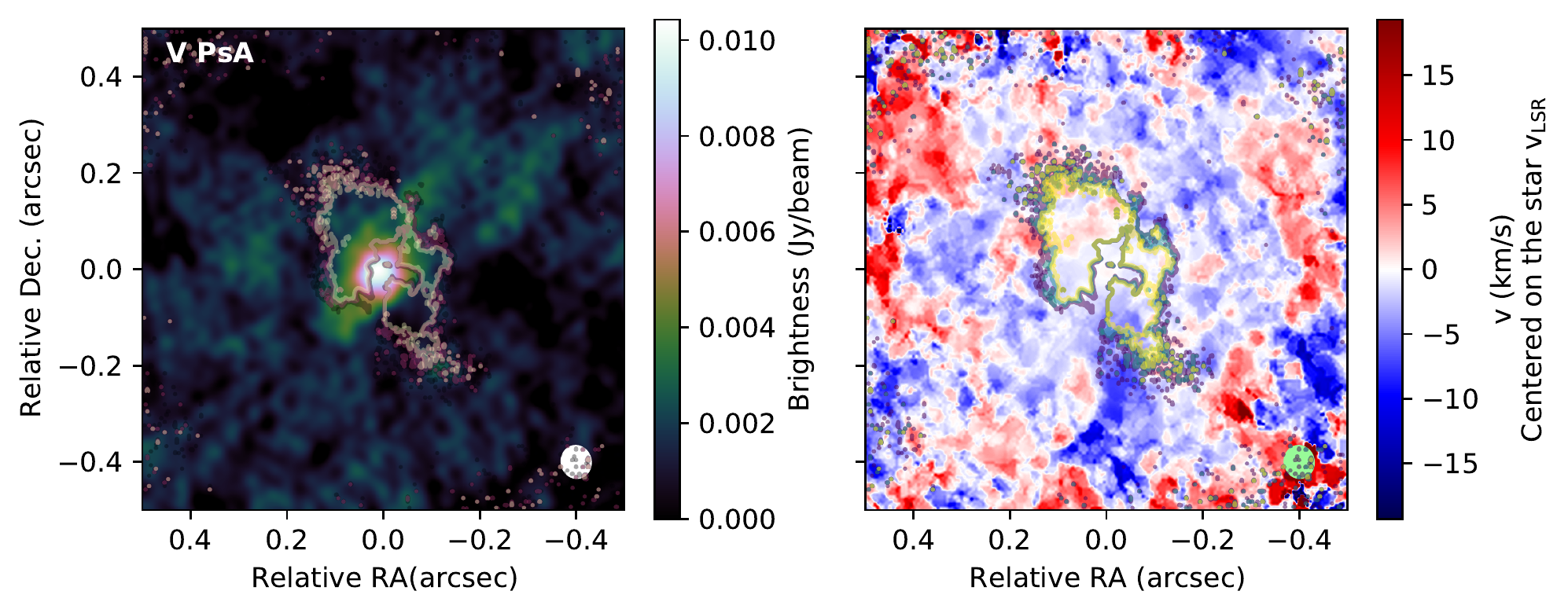}\hfill
        \includegraphics[width=0.95\columnwidth]{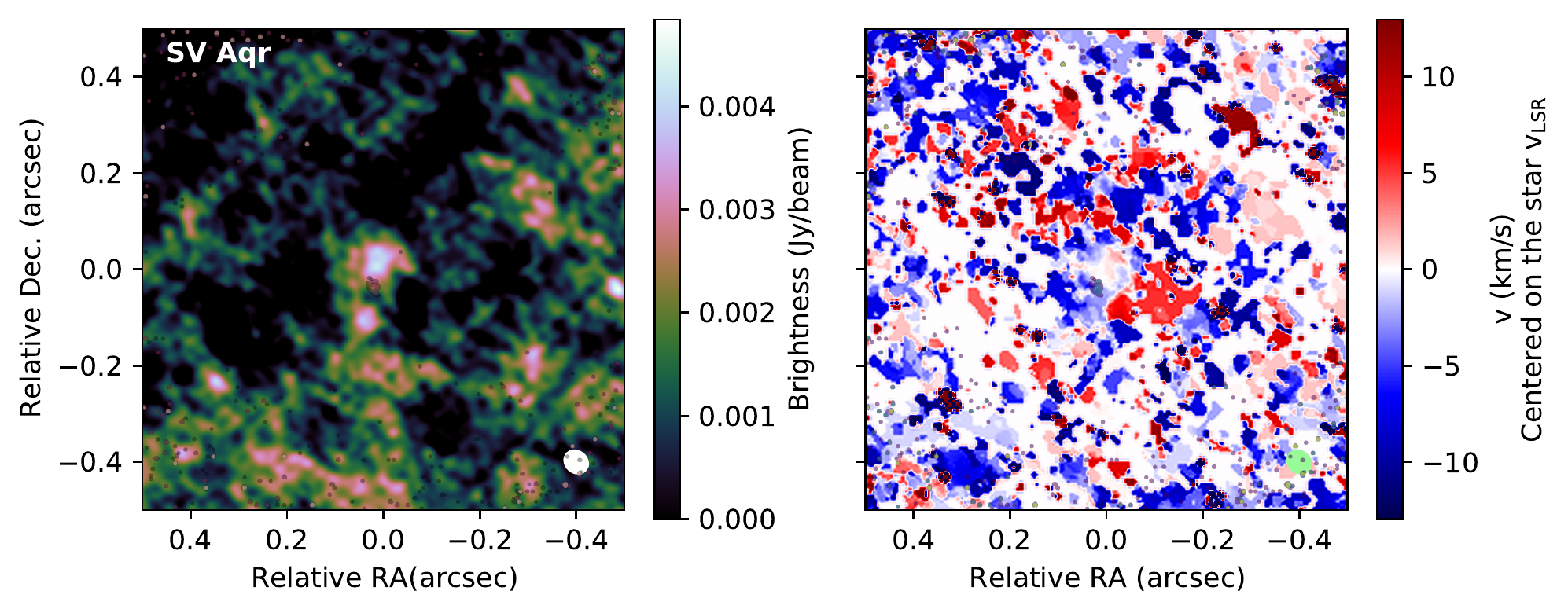}
        
        \includegraphics[width=0.95\columnwidth]{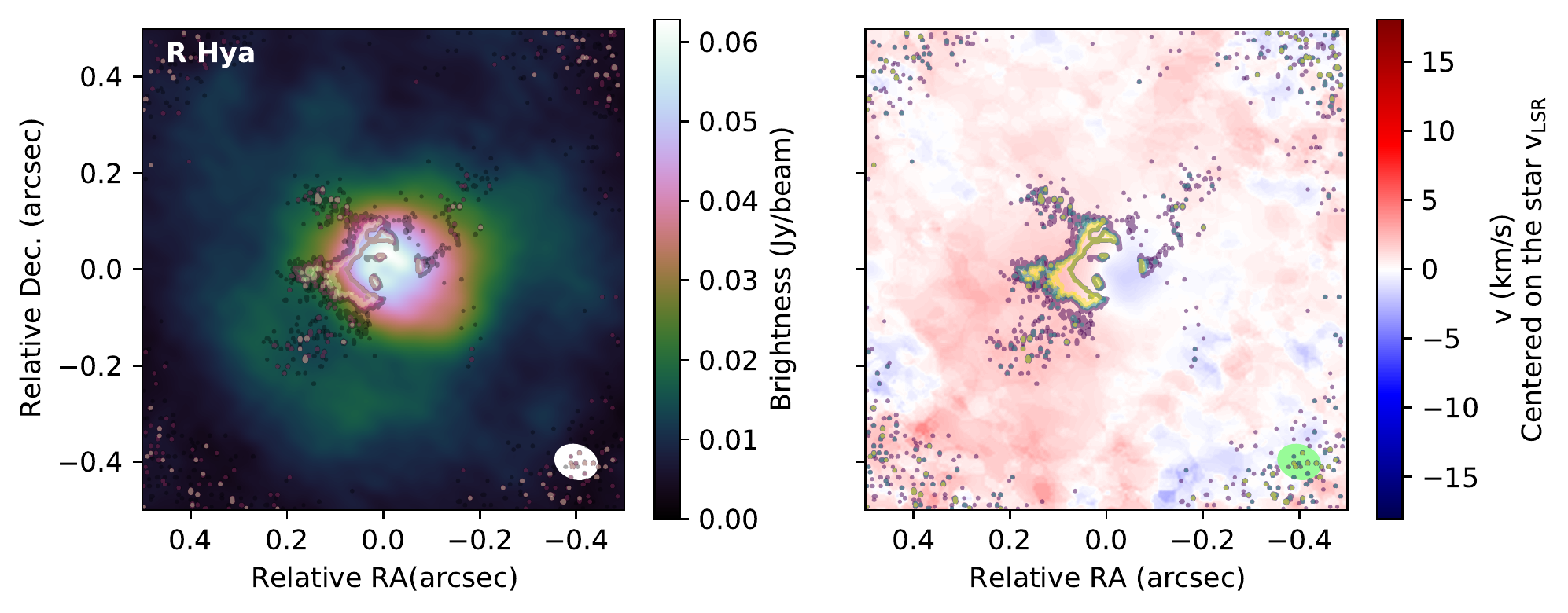}\hfill
        \includegraphics[width=0.95\columnwidth]{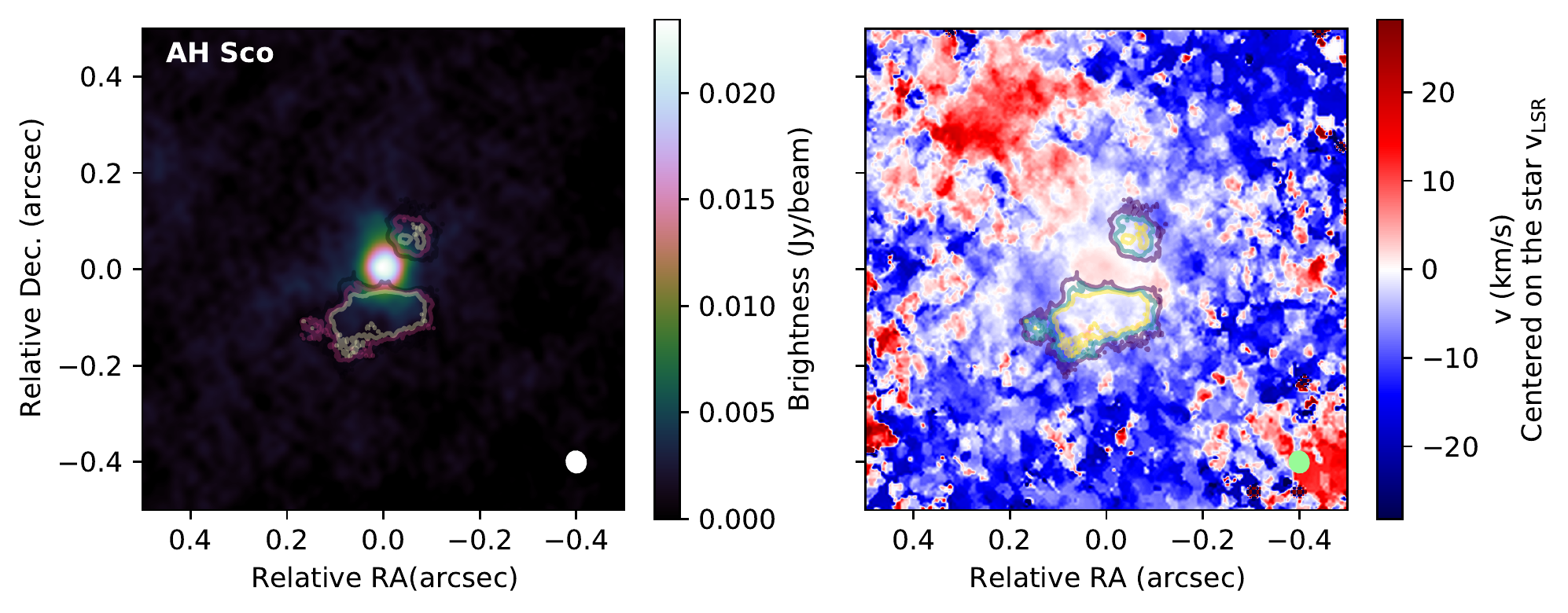}
        
        \includegraphics[width=0.95\columnwidth]{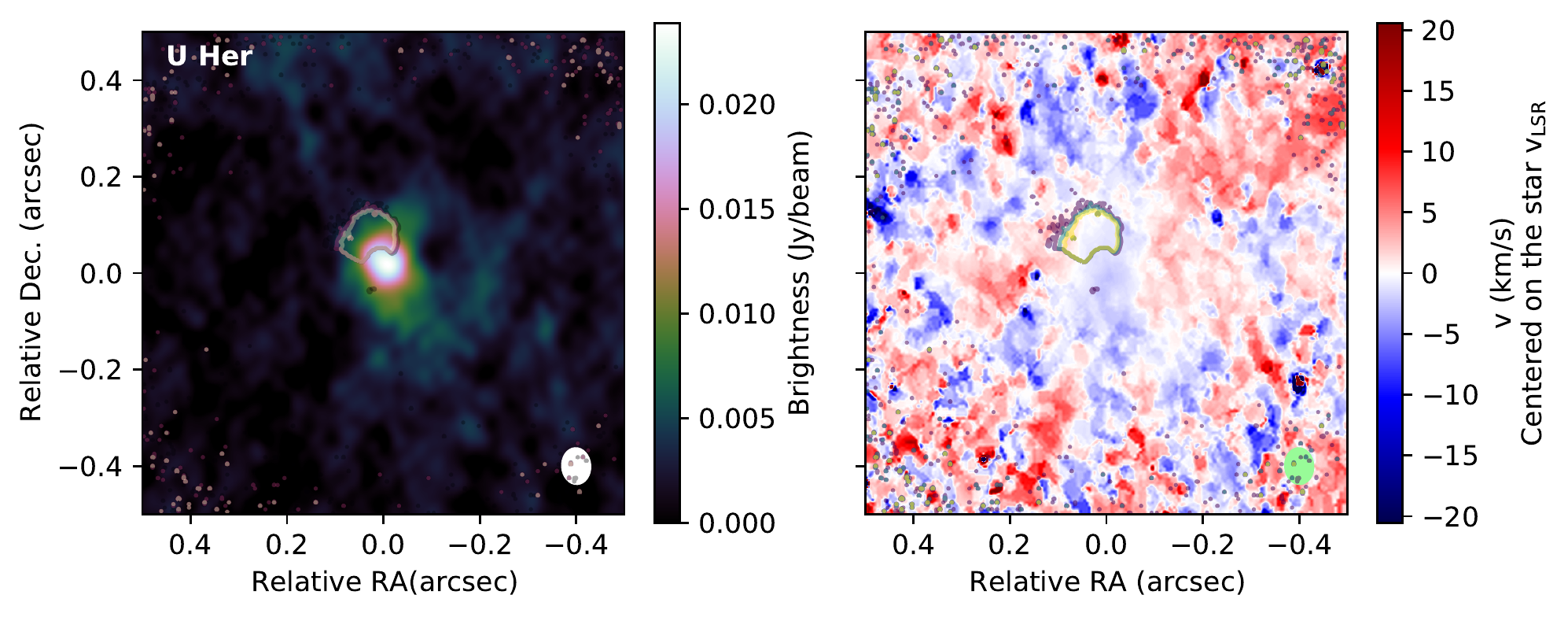}\hfill
        \includegraphics[width=0.95\columnwidth]{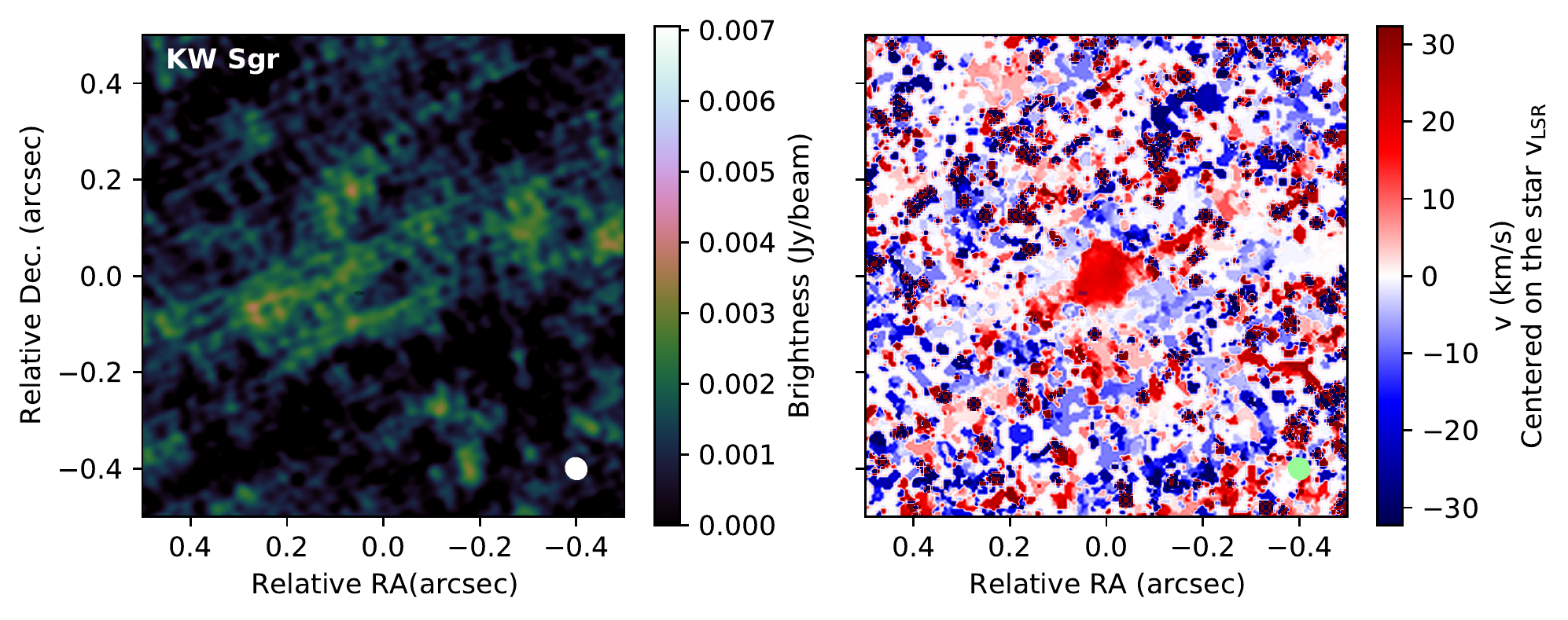}
        
        \includegraphics[width=0.95\columnwidth]{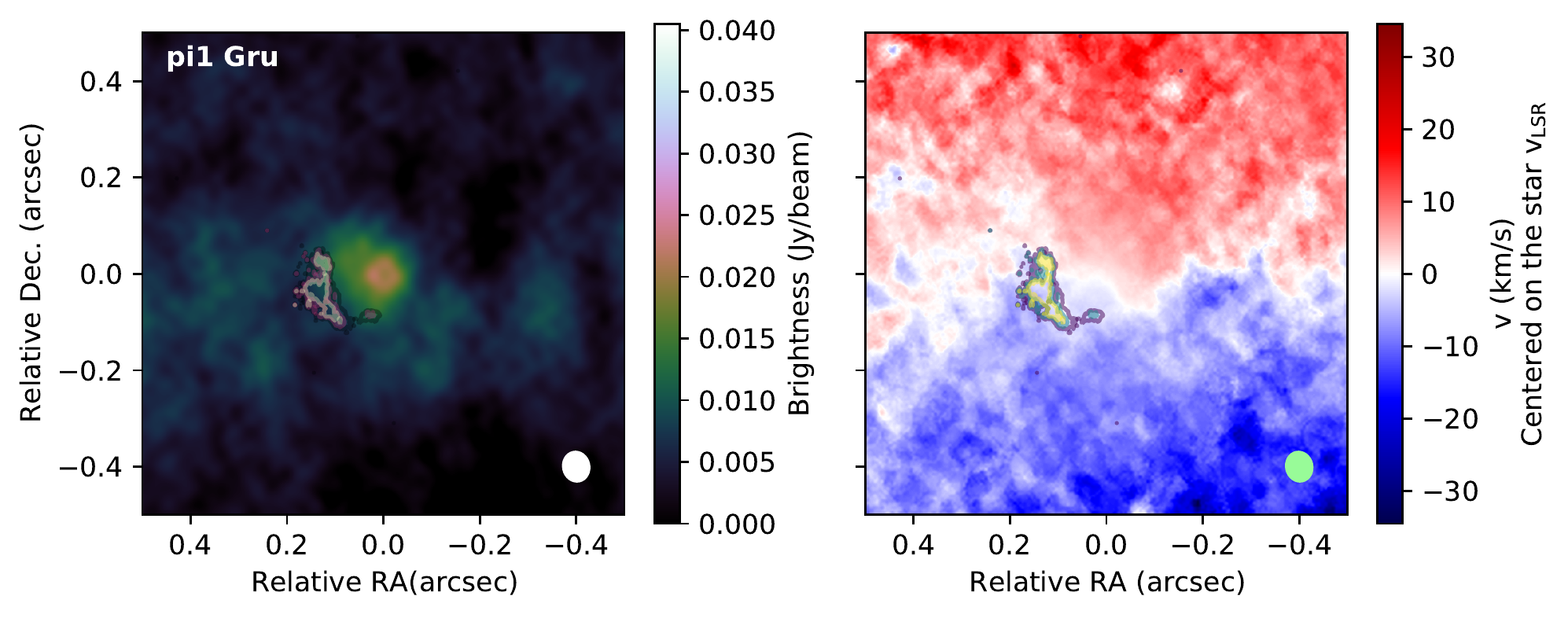}\hfill
        \includegraphics[width=0.95\columnwidth]{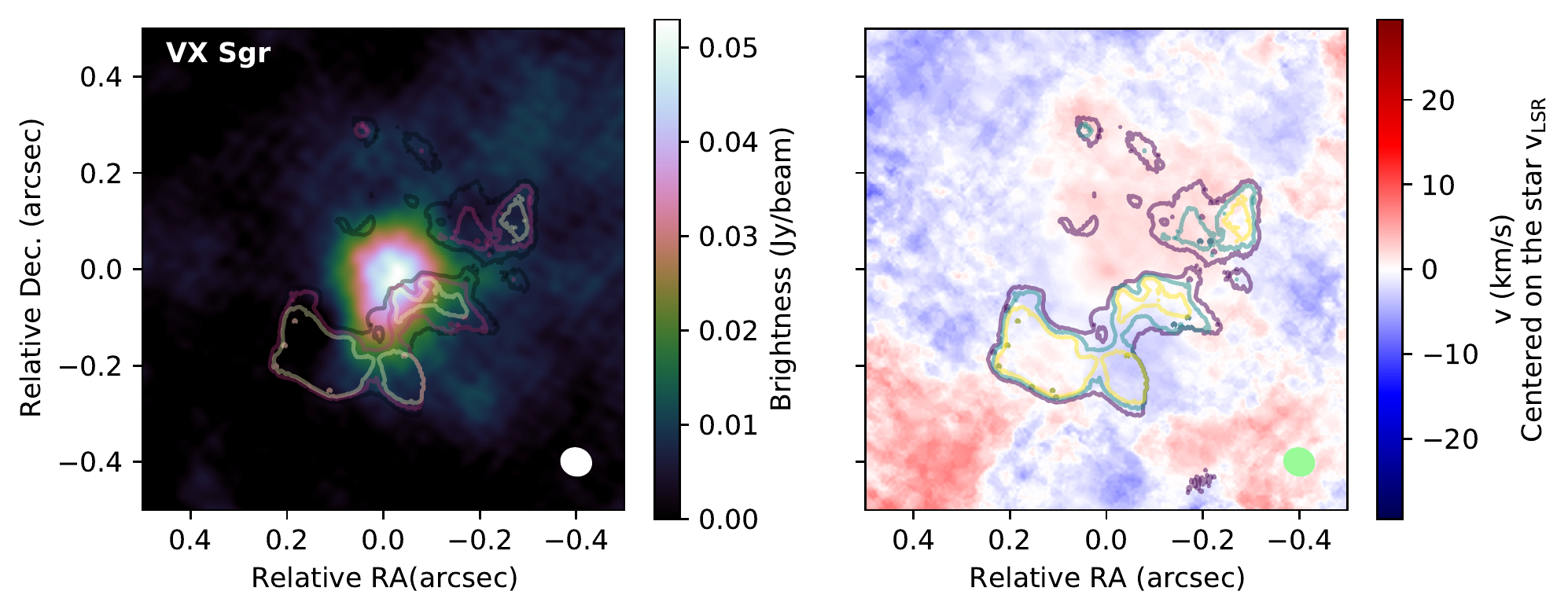}
        
        \caption{Comparison between the ZIMPOL DoLP and the ALMA CO $J=2-1$ observations (combined configurations; see \citeads{2022A&A...660A..94G}). The ZIMPOL DoLP contours correspond to 5$\sigma$ (violet or blue), 6$\sigma$ (pink or green), and 7$\sigma$ (yellow). The maps in the first and third columns show the 0 km~s$^{-1}$ channel map in the star reference frame. The second and fourth columns show the moment 1 maps.}\label{Fig:ALMA_SPHERE}
\end{figure*}

\begin{figure*}
        \centering
        \includegraphics[width=0.95\columnwidth]{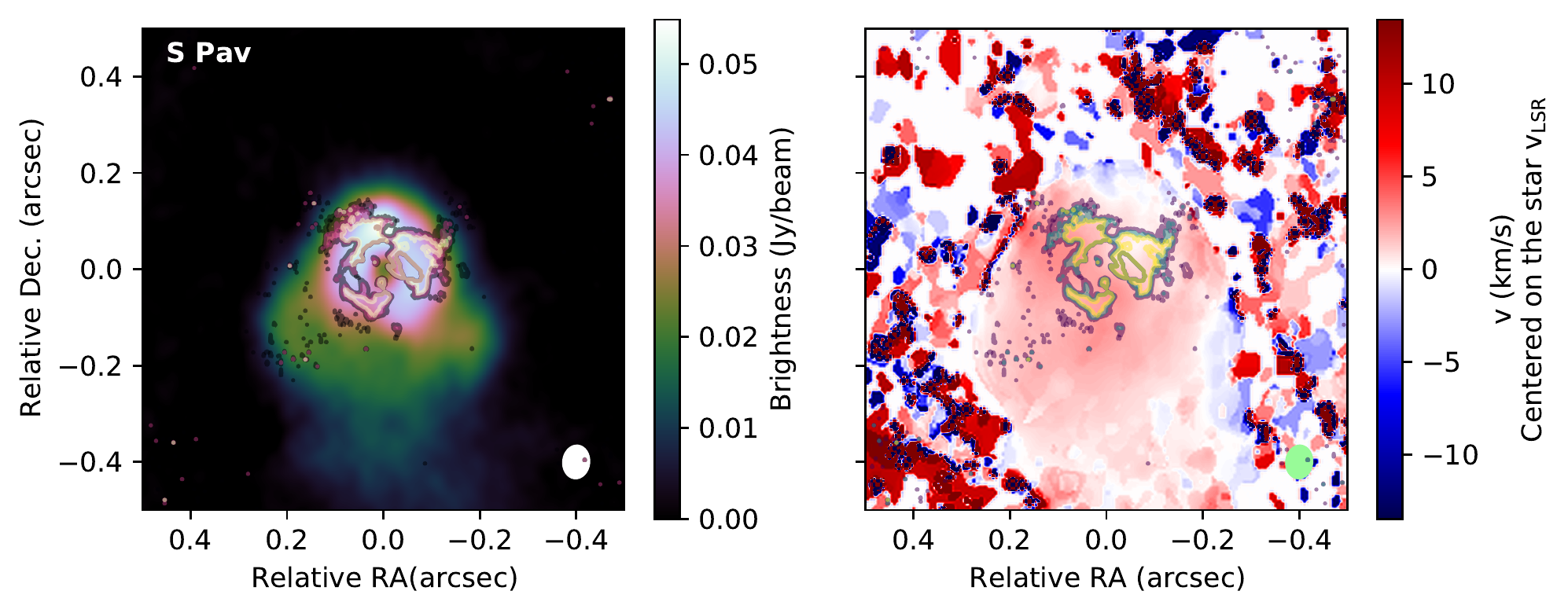} \hfill
        \includegraphics[width=0.95\columnwidth]{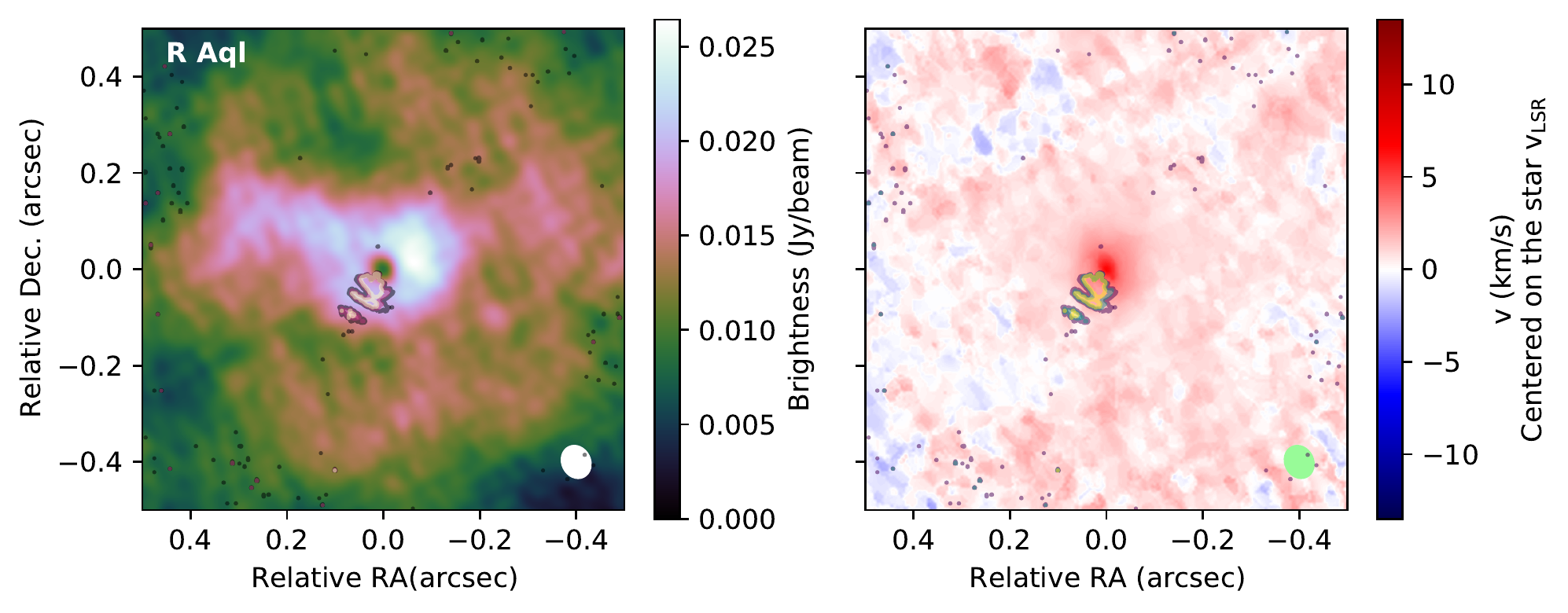}
        
        \includegraphics[width=0.95\columnwidth]{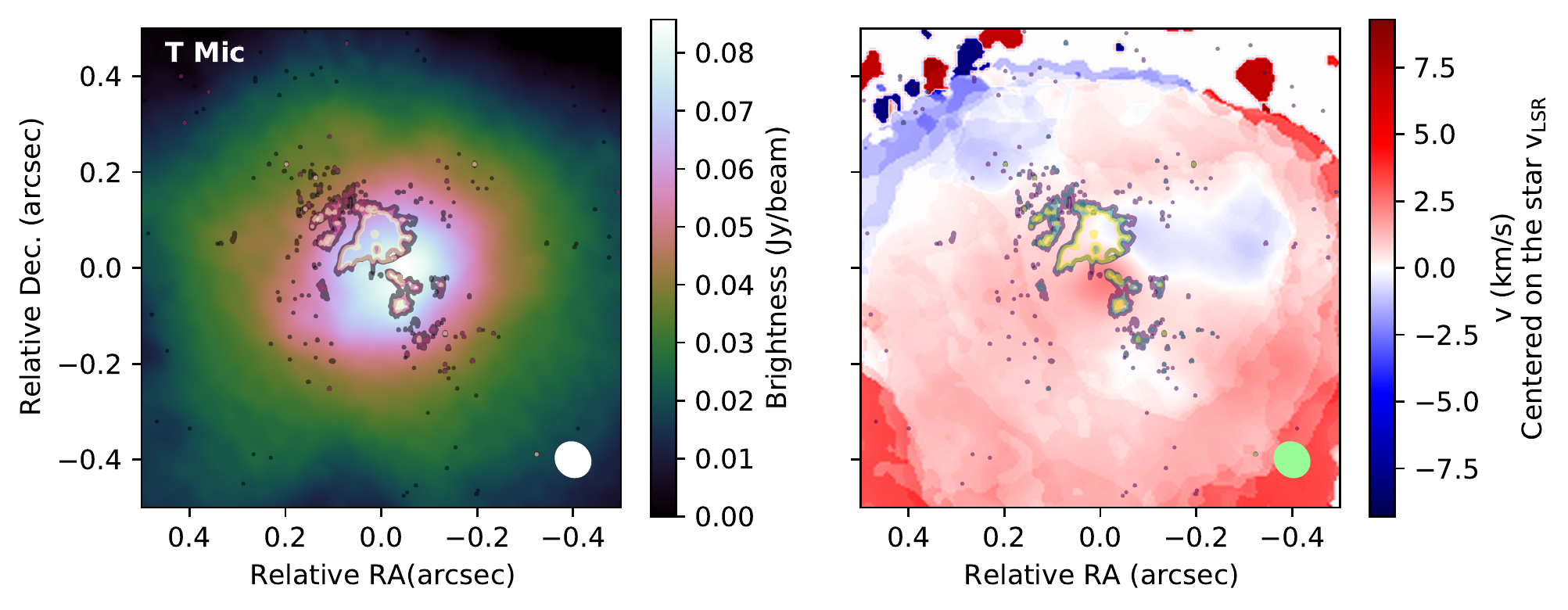}\hfill
        \includegraphics[width=0.95\columnwidth]{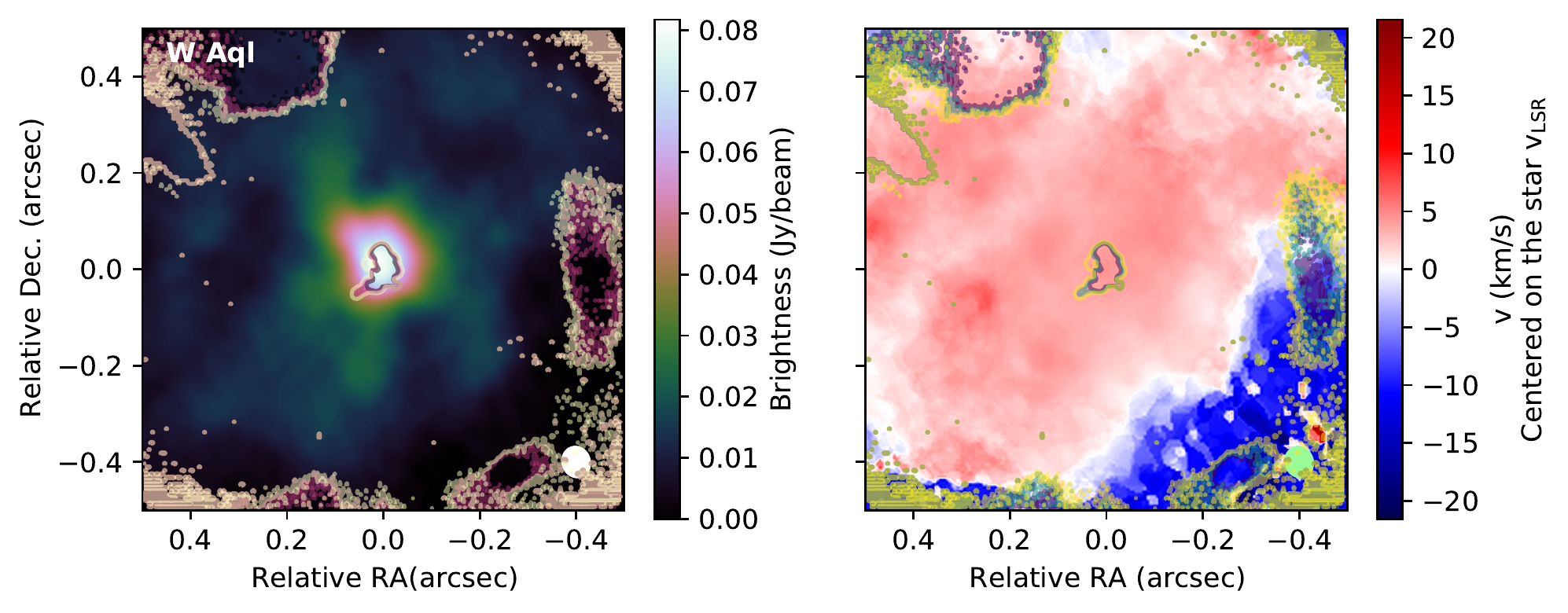}
        
        \includegraphics[width=0.95\columnwidth]{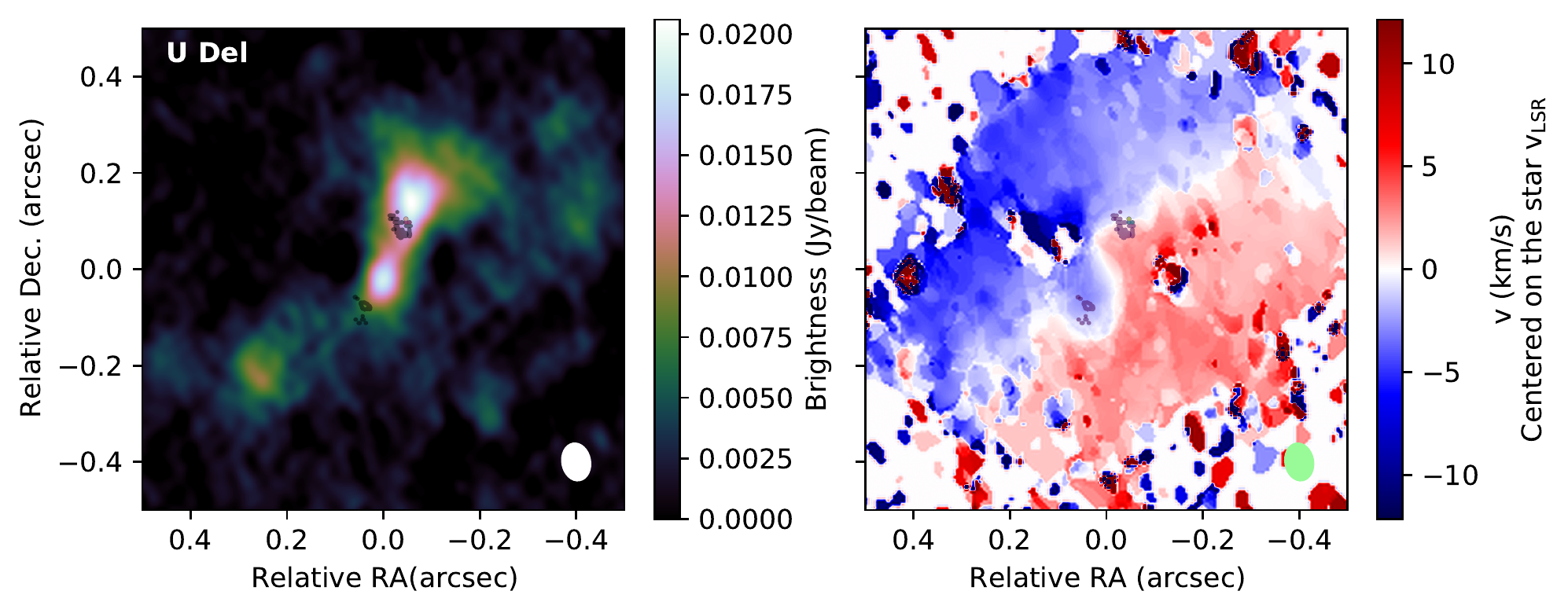}\hfill
        \includegraphics[width=0.95\columnwidth]{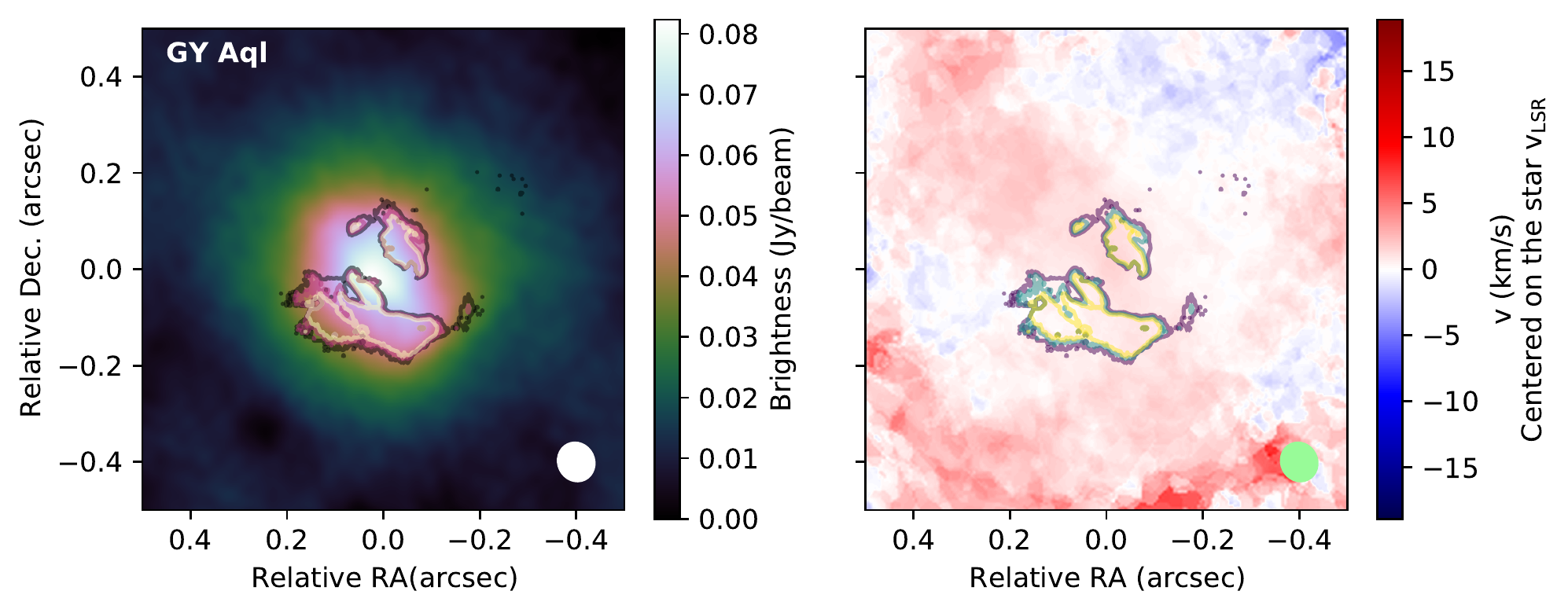}
        
        \includegraphics[width=0.95\columnwidth]{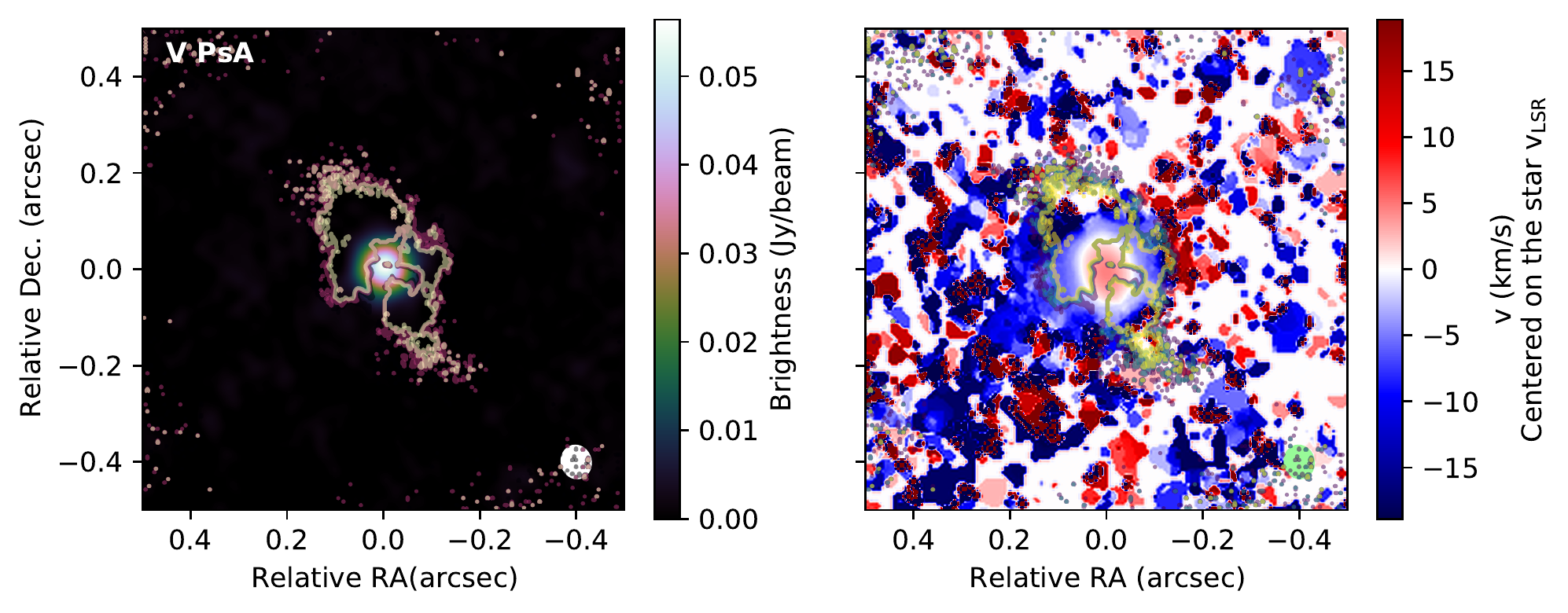}\hfill
        \includegraphics[width=0.95\columnwidth]{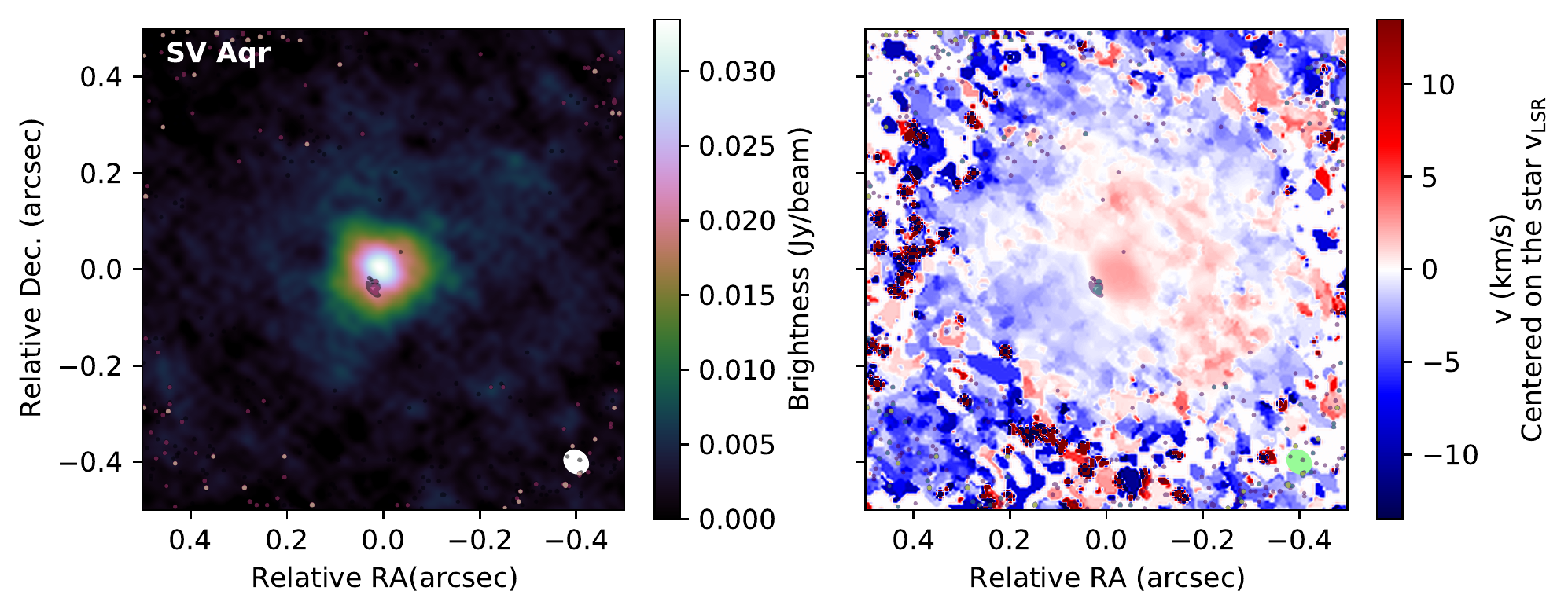}
        
        \includegraphics[width=0.95\columnwidth]{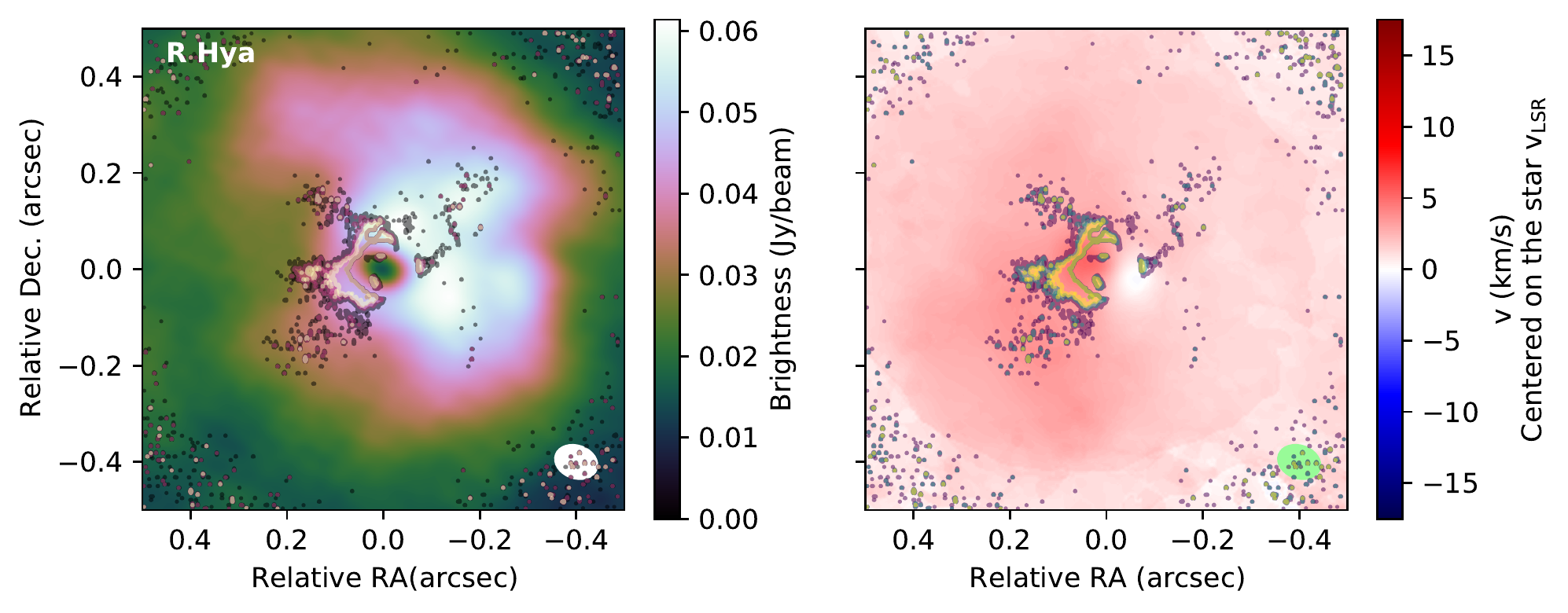}\hfill
        \includegraphics[width=0.95\columnwidth]{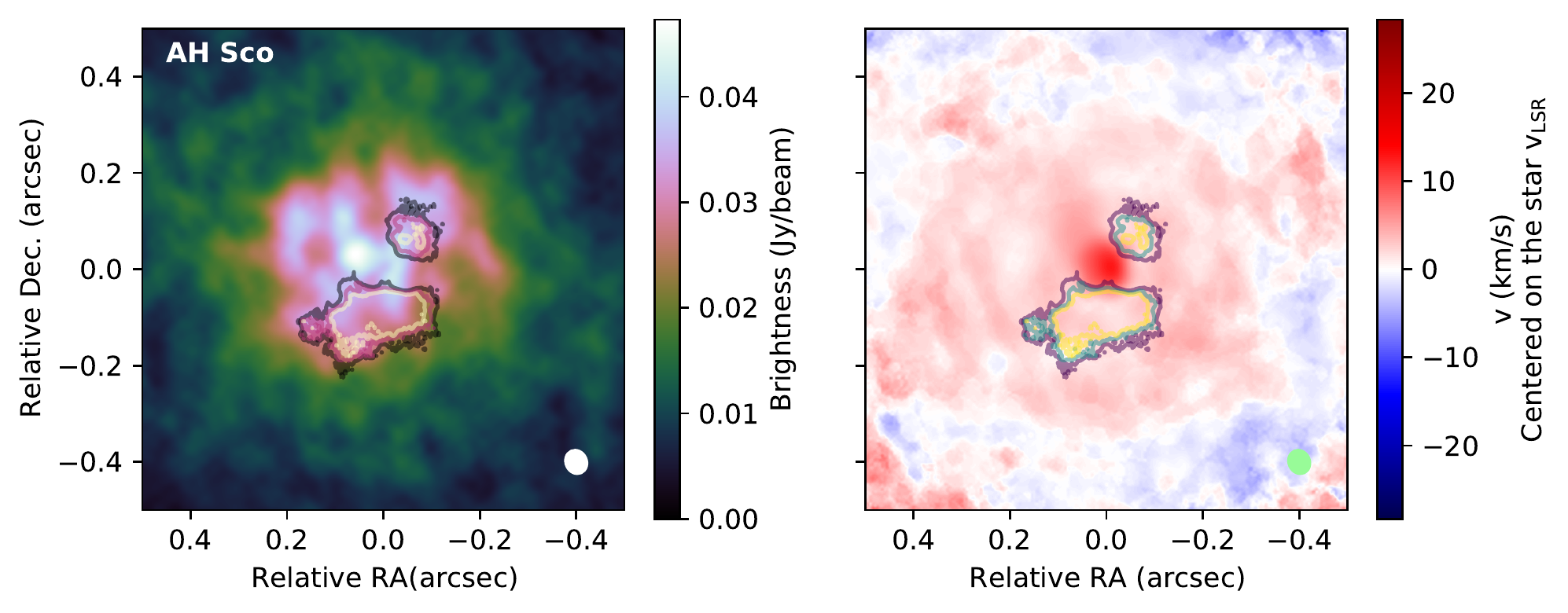}
        
        \includegraphics[width=0.95\columnwidth]{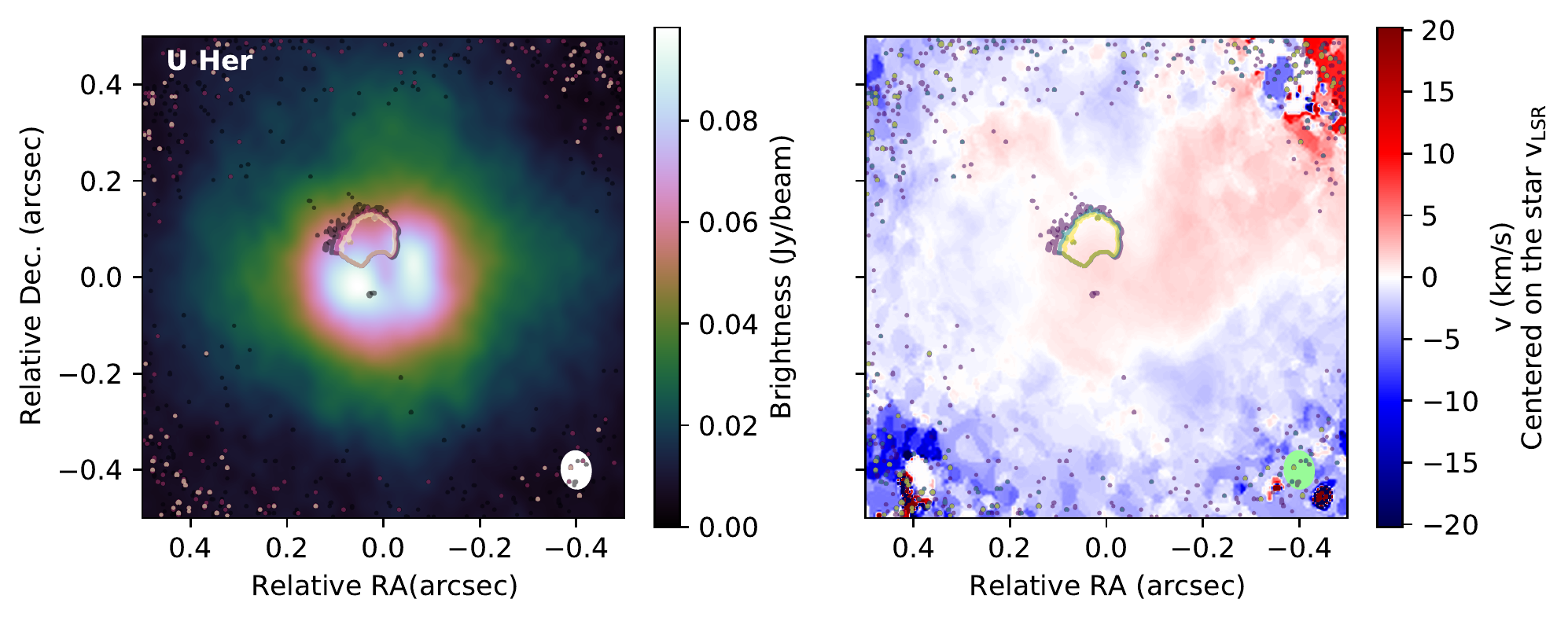}\hfill
        \includegraphics[width=0.95\columnwidth]{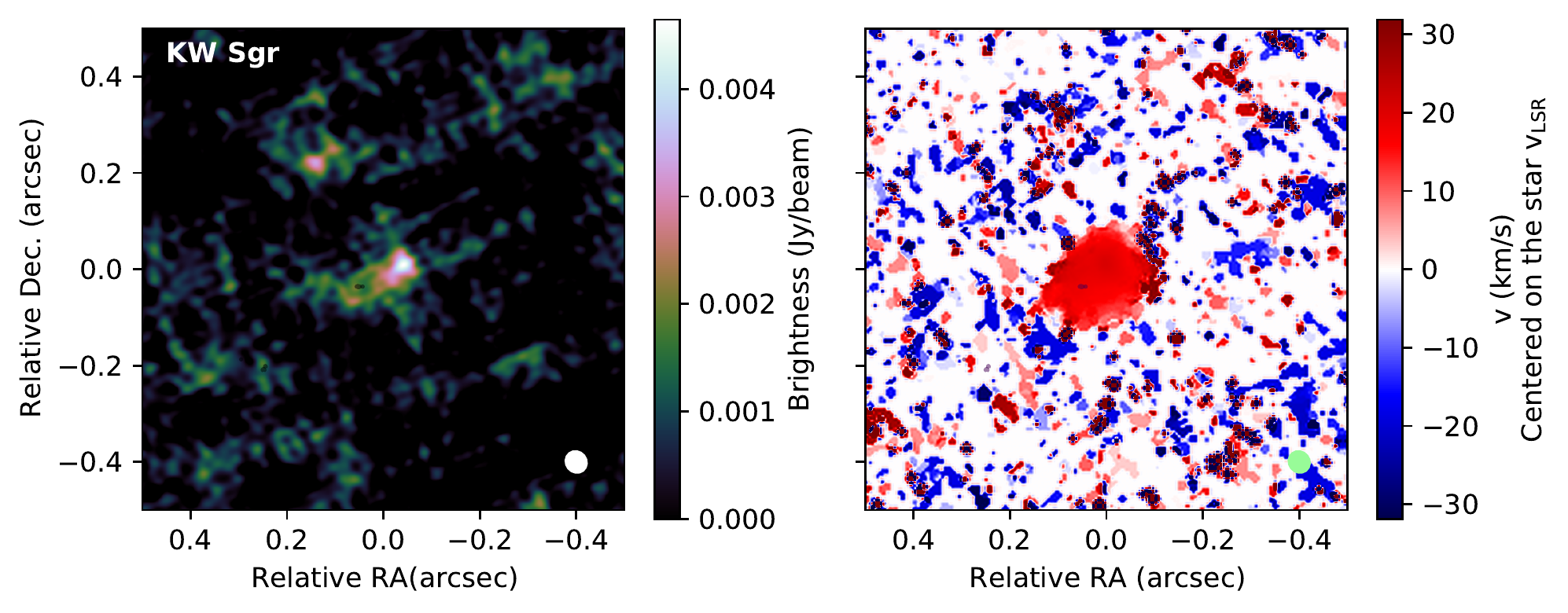}
        
        \includegraphics[width=0.95\columnwidth]{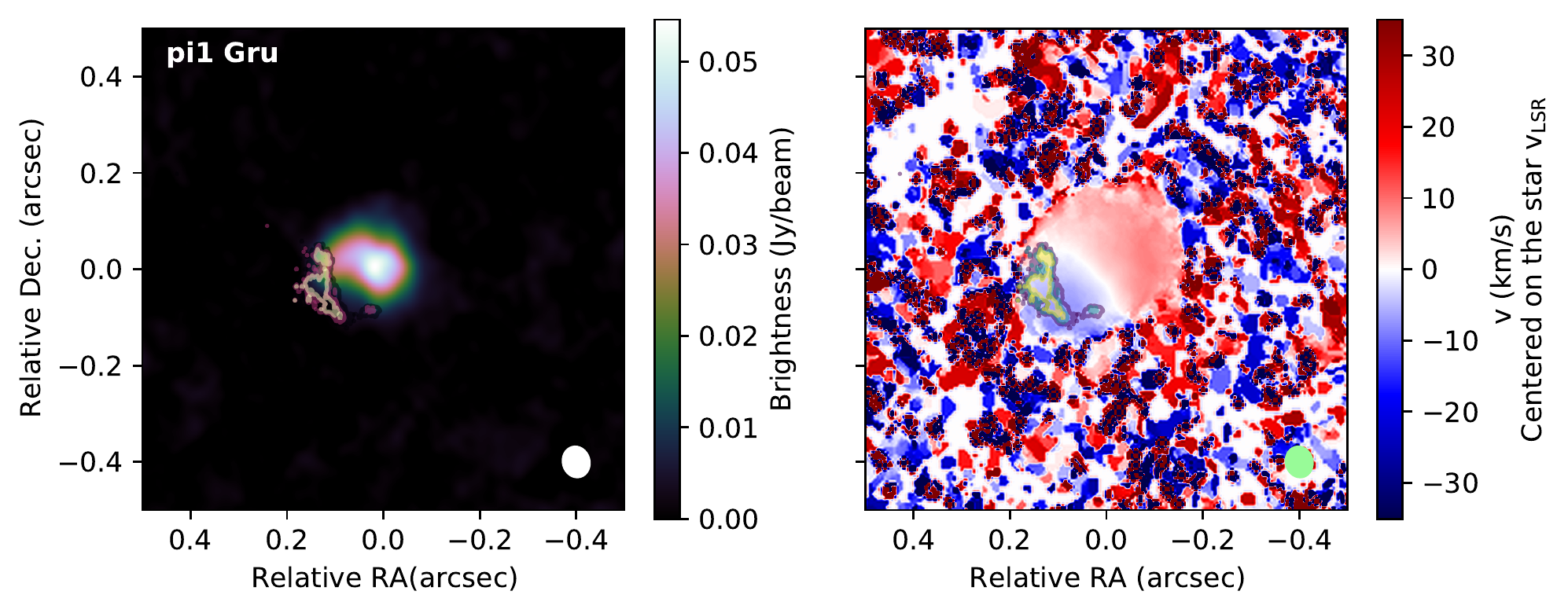}\hfill
        \includegraphics[width=0.95\columnwidth]{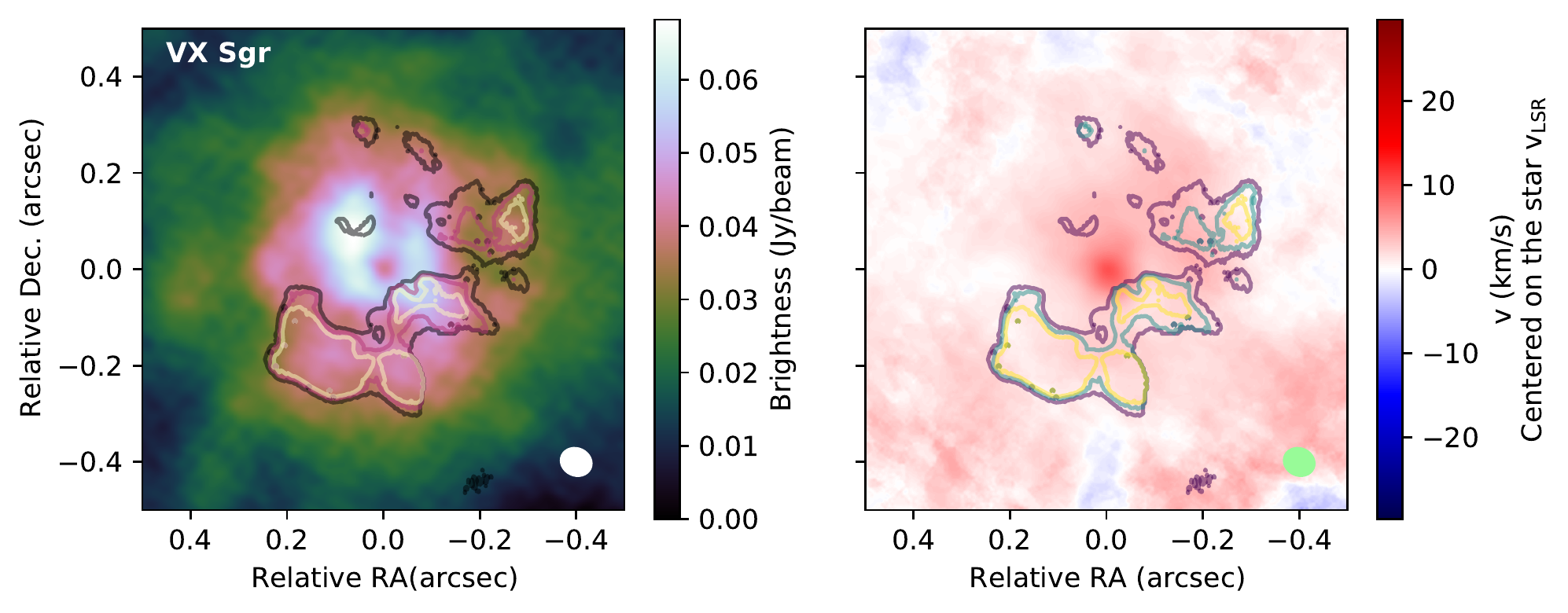}
        
        \caption{Comparison between the ZIMPOL DoLP and the ALMA SiO $\varv=0, J=5-4$ observations (combined configurations; see \citeads{2022A&A...660A..94G}). The ZIMPOL DoLP contours correspond to 5$\sigma$ (violet or blue), 6$\sigma$ (pink or green), and 7$\sigma$ (yellow). The maps in the first and third columns show the 0 km~s$^{-1}$ channel map in the star reference frame. The second and fourth columns show the moment 1 maps.}\label{Fig:ALMA_SPHERE_SiO}
\end{figure*}

Through molecular emission lines, ALMA is sensitive to the gas content of the circumstellar environment. In the continuum, the maps are dominated by the central star but are also sensitive to warm dust close to the star. Except for $\pi^1$~Gru \citepads{2020A&A...644A..61H}, there is no detection of spatially resolved dust features in the \textsc{Atomium} ALMA continuum maps. Here we compare the DoLP detections with ZIMPOL, with the rotational lines observed with ALMA. We chose to focus our discussion on the CO $\varv = 0, \ J=2-1$ and SiO $\varv=0$, $J=5-4$ lines. Owing to its low dipole moment \citepads{2016CPL...663...84C}, CO is mostly insensitive to any radiation field. \citetads{2019A&A...622A.123D} state that infrared pumping of ground or vibrationally excited states should be minimal in the inner regions of AGB envelopes -- therefore the high gas densities close to the star ensure that CO is mainly excited through collisions. Consequently, CO is a good tracer of density (when it is optically thin) and temperature (when it is optically thick). Owing to its higher Einstein coefficient \citepads{2005A&A...432..369S,2013JPCA..11713843M}, the SiO rotational transition is more sensitive to de-excitation effects, which makes it an efficient probe of the wind dynamics close to the photosphere. However, because it participates in the formation of silicate grains, it can be depleted farther out, particularly around O-rich stars. In Sect.~\ref{SubSect:DiscDustPolar} we show that, for a given dust density, composition and distance to the star, the dominant contribution to the DoLP is from the plane of the sky. Although the geometry may be complex, for this particular systematic comparison, we assumed the null velocity channel to trace the gas within the plane of the sky through the star center. This assumption would be totally correct if the bulk of the motion would be radial. Therefore, in Figs.~\ref{Fig:ALMA_SPHERE} and \ref{Fig:ALMA_SPHERE_SiO} we plot the DoLP contours versus the null velocity channel map of each star (where the local standard rest velocity of each star is taken from \citeads{2020Sci...369.1497D}) and the moment 1 map (which shows the average velocity at a given pixel, weighted by the brightness in the different spectral channels), for the CO and SiO transitions, respectively. The ALMA maps were centered using the photocenter of the corresponding continuum maps \citepads{2022A&A...660A..94G}. The SPHERE-ZIMPOL DoLP contours were centered using the intensity maps. We used the maps produced by combining the different ALMA configurations, because they provide the sharpest angular resolution with the best S/N.\\

In most cases, no correlations are observed either in the null velocity or the moment~1 maps. However, there are some notable exceptions: 
\paragraph{S Pav.} The DoLP corresponds to the CO emission in the rest frame of the star, indicating that in the plane of the sky, the dust and gas are colocated in S~Pav. 
\paragraph{GY~Aql.} The two lobes of the DoLP observed with ZIMPOL correspond to the onset of two spiral arms in the moment~1 map of CO. The spirals are detected at larger scales in the images obtained in the mid-configuration in the {\sc Atomium} program \citep{2020Sci...369.1497D}. However, it is difficult to attain a clear interpretation of the correlation between the observations of GY~Aql with ALMA and ZIMPOL, owing to the complexity of the environment in this star. Several spiral arms overlap at various positions along the line of sight, and each arm probably hosts dust and gas. The dominant contribution to the DoLP is in the plane of the sky (Sect.~\ref{SubSect:DiscDustPolar}), but this implies that the correlation of the DoLP with the moment~1 maps is accidental, whereas there is no correlation with the null velocity channel map. 
\paragraph{AH Sco.} The DoLP coincides with a blueshifted zone of the CO line, surrounded by a red shifted area in the moment~1 map. We observe a similar  configuration to GY~Aql, except that the extended emission is not visible in the null velocity map of CO  in AH~Sco, possibly owing to an accidental correlation between the moment~1 map and the DoLP. 
\paragraph{$\pi^1$ Gru.} This is the only star in which there is a remarkable coincidence between the SiO emission and the DoLP.  In $\pi^1$~Gru, the coma-shaped DoLP contour corresponds to the blueshifted area of the SiO moment\,1 map, and appears to arise from the rotating disk around the AGB star described by \citetads{2020A&A...644A..61H}. According to the scenario from \citeauthor{2020A&A...644A..61H}, the DoLP could be a dust tail formed in the wake  of the companion as it moves in our direction, hence the blueshifted SiO area.\\

From these comparisons -- except for S Pav, $\pi^1$ Gru, and perhaps GY~Aql -- it is not possible to state a definite colocation between the dust detected with ZIMPOL, and density enhancements in the gas detected by ALMA. We have seen in Sect.~\ref{SubSect:DiscDustPolar} that the DoLP was mainly sensitive to polarization created by light scattering off small dust grains, in the plane of the sky through the star, with features close to an optical depth of unity. So the conclusion from the comparison with the ALMA maps is that dust regions that meet these criteria do not coincide with the highest gas densities observed with ALMA in the plane of the sky. 

The main conclusion of \citetads{2020Sci...369.1497D} is that the processes leading to the shaping of planetary nebulae are already at work in the wind of AGB stars observed in the \textsc{Atomium} large program. The most probable mechanism invoked is the interaction with (sub)stellar companions. The possibility of shaping the AGB star mass loss are maximized when the semimajor axis of their orbit is in the range $3 - 500$~au. This includes the region where most companions are detected around AGB stars ($\gtrsim 20$~au, \citeads{2017ApJS..230...15M}), and the region examined by the ZIMPOL observations. Table \ref{Table:Companions} shows the result from the proper motion analysis in the \textit{Gaia} EDR3 \citepads{2021A&A...649A...1G} compared to the \textsc{Hipparcos} 2 data release \citepads{2007A&A...474..653V} for the \textsc{Atomium} sources, extracted from \citet{2022A&A...657A...7K}. The S/N of the proper-motion anomaly is an estimator of the probable presence of a detectable companion around the central AGB star (probable if~$ > 3$). We do not discuss the data on the three RSGs (AH~Sco and KW~Sgr because their \textsc{Hipparcos} parallax is negative, and VX~Sgr because its \textit{Gaia} EDR3 parallax is most certainly biased). We note that for the AGB stars the S/N of the proper motion anomaly is larger than 3 only for R~Hya, $\pi^1$~Gru, R~Aql, and GY~Aql. $\pi^1$~Gru's companion is detected by \citetads{2020A&A...644A..61H}, which allows us to identify the coma shape of the DoLP as a dust tail formed in the wake of the companion's motion. For GY~Aql, we saw a correlation between the ALMA maps and the DoLP, which suggests that the dust we detect is at the head of spiral arms, implying that there might be structures that are created by an as-yet undetected companion. For R~Hya and R~Aql, no feature can be conclusively associated with the presence of a potential companion either in the ZIMPOL observations or in the ALMA data. 

\begin{table*}
        \centering
        \caption{Inferred companion presence and mass from the \textit{Gaia} EDR3 proper motion analysis for the \textsc{Atomium} sample \citep{2022A&A...657A...7K}.}\label{Table:Companions}
        \begin{tabular}{llllllll}
                \hline
                \hline
                \noalign{\smallskip}
                Star & \multicolumn{2}{c}{Parallax} & RUWE\tablefootmark{a} & S/N Prop. Mot. & R$_1$\tablefootmark{b} & M$_1$\tablefootmark{c} &  M$_2$ @ 5 au \tablefootmark{d} \\
                & \textsc{Hipparcos} 2 & \textit{Gaia} EDR3 & & anomaly & & & \\
                & (mas) & (mas) & & Hip2 - \textit{Gaia} EDR3  & (R$_\odot$) & (M$_\odot$) & (M$_\odot$)\\
                \hline
                \noalign{\smallskip}
                S Pav                                    &  $5.44 \pm 1.27$ & $5.758 \pm 0.704$ &  4.191 &         1.86 &                                    $710 \pm 36$ &  $1.93 \pm 0.39$ &     $0.12^{+0.06}_{-0.05}$ \\
                \noalign{\smallskip}
                T Mic                                    &  $4.75 \pm 1.01$ & $5.435 \pm 0.283$ &  2.059 &         0.90 &                                    $722 \pm 37$ &  $2.19 \pm 0.44$ &     $0.04^{+0.04}_{-0.03}$ \\
                \noalign{\smallskip}
                U Del                                    &  $2.08 \pm 0.72$ & $3.031 \pm 0.115$ &  1.060 &         2.23 &                                    $449 \pm 23$ &  $1.46 \pm 0.07$ &     $0.05^{+0.03}_{-0.02}$ \\
                \noalign{\smallskip}
                V PsA &    - &     - &    - &          - & - &    - &        - \\
                SV Aqr &    - &     - &    - &          - & - &    - &        - \\
                \noalign{\smallskip}
                R Hya                                    &  $8.05 \pm 0.69$ & $6.761 \pm 0.532$ &  2.913 &         4.60 &                                    $845 \pm 43$ &  $0.73 \pm 0.04$ &     $0.14^{+0.05}_{-0.03}$ \\
                \noalign{\smallskip}
                U Her                                    &  $4.26 \pm 0.85$ & $2.402 \pm 0.088$ &  1.320 &         2.12 &                                   $1111 \pm 59$ &  $2.37 \pm 0.47$ &     $0.08^{+0.03}_{-0.03}$ \\
                \noalign{\smallskip}
                pi1 Gru                                  &  $6.13 \pm 0.76$ & $6.202 \pm 0.512$ &  2.908 &        18.44 &                                    $737 \pm 37$ &  $0.64 \pm 0.03$ &     $0.42^{+0.13}_{-0.05}$ \\
                \noalign{\smallskip}
                R Aql                                    &  $2.37 \pm 0.87$ & $4.323 \pm 0.183$ &  1.439 &         4.16 &                                    $551 \pm 29$ &  $0.65 \pm 0.03$ &     $0.07^{+0.03}_{-0.02}$ \\
                \noalign{\smallskip}
                W Aql &    - &     - &    - &          - & - &    - &        - \\
                \noalign{\smallskip}
                GY Aql                                   &  $4.48 \pm 4.02$ & $1.528 \pm 0.190$ &  1.479 &         4.45 &                                  $1836 \pm 102$ &  $2.07 \pm 0.41$ &     $0.52^{+0.19}_{-0.11}$ \\
                \noalign{\smallskip}
                AH Sco                                   & $-0.09 \pm 0.57$ & $0.608 \pm 0.091$ &  0.961 &         4.08 &                                  $1427 \pm 168$ & $11.29 \pm 2.26$ &     $7.62^{+2.45}_{-0.96}$ \\
                \noalign{\smallskip}
                KW Sgr                                   & $-2.43 \pm 0.94$ & $0.485 \pm 0.083$ &  1.094 &         0.30 &                                  $1166 \pm 158$ & $12.07 \pm 2.41$ &      $0.6^{+0.29}_{-0.24}$ \\
                \noalign{\smallskip}
                VX Sgr                                   &  $3.82 \pm 2.73$ & $0.080 \pm 0.214$ &  2.451 &         6.58 &                                $20772 \pm 2136$ & $23.79 \pm 4.76$ & $256.67^{+82.44}_{-31.62}$ \\
                \hline
        \end{tabular}
	\tablefoot{Note that for VX~Sgr the values are meaningless due to the much-biased value of the \textit{Gaia} EDR3 parallax. We also saw in Sect. \ref{Sect:TargetsObs} that the \textit{Gaia} measurements for U~Her and GY~Aql were not reliable.\\
	\tablefoottext{a}{The RUWE is the \textit{Gaia} EDR3 renormalized unit weight error.}
	\tablefoottext{b}{R$_1$ refer to the radius of the primary star.}
	\tablefoottext{c}{M$_1$ refer to the mass of the primary star.}
	\tablefoottext{d}{M$_2$ refer to the mass of a companion with semimajor axis of the orbit of 5~au.}
}
\end{table*}

\subsection{Individual cases}\label{SubSect:IndivTargets}

In this section we highlight sources of interest for which the DoLP presents features that can be identified or are remarkable. This implies that the dust in their inner and intermediate circumstellar environments has the characteristics to produce a significant polarized signal as seen from Earth. However, other sources may have interesting dust features, only visible through their thermal emission (e.g., in the MIR).

\subsubsection{W Aql}

W~Aql is one of two S-type stars of the \textsc{Atomium} sample. The \textit{Gaia} DR3 \citep{GaiaDR3} puts it at $ 374 \pm 9$~pc. We saw that the main component, the AGB star, is orbited by a companion (Fig.~\ref{Fig:W_Aql_intensity}) that was previously observed \citepads{1965VeBam..27..164H,2011A&A...531A.148R,2015A&A...574A..23D}. As we saw in Sect~\ref{SubSect:DiscDustPolar}, W~Aql exhibits one of the strongest MIR excess and the strongest DoLP (both in terms of the averaged value, and the spatial extent), implying that W~Aql produces dust grains that are the most efficient in our sample at producing a polarized signal owing to grain sizes and/or composition and/or dust location. Because $\pi^1$~Gru, the second S star in the sample, has a much weaker signal, we deduce that the peculiar DoLP characteristics of W~Aql do not arise solely because it is an S-type star.

\citetads{2011A&A...531A.148R} have already imaged the inner circumstellar environment of W~Aql through polarization in the R band (see their Fig. 6 for images through their coronagraph). They report a northeast to southwest elongation of the polarized signal that we do not observe in our own observations. This suggests a change in the dust distribution between their observations with PolCor in 2008 and ours with SPHERE-ZIMPOL in 2019. 

\subsubsection{$\pi^1$ Gru}

 The other S-type star in the sample, $\pi^1$~Gru, is located at $162 \pm 12$~pc \citep{GaiaDR3}. In addition to the main AGB star, it is known to have a secondary G0V companion (separated by 2.8"), creating a spiral arm observed with the Photodetector Array Camera and Spectrometer (PACS) of the \textit{Herschel} mission by \citetads{2014A&A...570A.113M}. These authors, and both \citetads{2006ApJ...645..605C} and \citetads{2017A&A...605A..28D}, claim that there is a second, fainter and closer companion. This is suggested by: (1) the tilting of the inner circumstellar disk that does not coincide with the inclination of the orbit of the G0V companion; (2) the closure phase measured by the Astronomical Multi-BEam combineR (AMBER) of the VLT Interferometer (VLTI) that differs from 0$^\circ$ or 180$^\circ$; and (3) a photocenter displacement measured by \textsc{Hipparcos}. The companion is imaged in the \textsc{Atomium} project by \citetads{2020A&A...644A..61H} through its ALMA continuum emission. It is located at $\sim 40$~mas from the central AGB star ($\sim 6$~au), which corresponds to the inner DoLP detection in our maps (Fig.~\ref{Fig:ZIMPOL_DoLP_all}). We propose that this dust could be formed from the interaction between the companion identified by \citetads{2020A&A...644A..61H} and the AGB outflowing wind, which will be discussed in details in a forthcoming paper (Montargès et al. in prep.)

\subsubsection{GY Aql}

GY~Aql is an O-rich (C/O<1) AGB star, located at $340 \pm 30$~pc \citepads{2022arXiv220903906A}. This source shows a potential correlation between the DoLP and the moment map 1 of the CO $J=2-1$ line (Sect.~\ref{SubSect:ALMA}), indicating that we may be able to examine the dust nucleation or initial growth of the dust grains within the spiral. In this particular case, the ALMA channel maps \citepads{2020Sci...369.1497D} will be of great help in determining the morphology of the dust distribution, on the assumption that the gas and the dust are colocated. This would eliminate one of the most constraining uncertainties in trying to reproduce the polarized signal through radiative transfer simulations, which results from the confinement of the DoLP to the region near the plane of the sky (Sect.~\ref{SubSect:RADMC3D_ParamStudies}).

%%%%%%%%%%%%%%%%%%%%%%%%%%%%%%%%
%        CONCLUSION
%%%%%%%%%%%%%%%%%%%%%%%%%%%%%%%%

\section{Conclusion}\label{Sect:Conclusion}

We presented the observations of 14 stars (out of the 17) of the \textsc{Atomium} sample with VLT/SPHERE-ZIMPOL. Owing to the AO system of SPHERE, we were able to obtain polarization maps of the inner and intermediate circumstellar environments of these cool evolved stars in the visible. The observations were performed a few days after the same sources had been observed with the extended configuration of ALMA. This allowed us to compare observations of the gas (ALMA) and of the dust (ZIMPOL) at the same spatial scale and within the same time frame.

The DoLP observed with ZIMPOL reveals dusty circumstellar environments that are mainly clumpy, with isolated dust features, although there are a few exceptions: in GY~Aql and $\pi^1$~Gru, there is a hint of larger-scale organization (spirals). In the RSG VX~Sgr, a nearly complete shell is observed. The S star W~Aql prominently stands out with a strong DoLP in the full field of view.

Pilot \texttt{RADMC3D} simulations (Sect.~\ref{Sect:Simulations}) allow us to draw several conclusions: 

\begin{enumerate}
        \item Dust features producing the maximum DoLP in the ZIMPOL images most likely: (a) are located near or in the plane of the sky (as polarization is a strong function of the scattering angle, peaking at 90$^{\circ}$); (b) have an optical depth close to unity (i.e., optically thick dust regions do not produce a clear maximum DoLP); and (c) are located just outside the PSF halo (between 85 and 125~mas). 
        
        \item For three sources ($\pi^{1}$\,Gru, W\,Aql, and AH Sco), the maximum DoLP is too high to be reproduced in this way, which may indicate that the chemical composition of the grains is different. Melilite, enstatite, and forsterite have stronger polarization properties and could potentially yield the high maximum DoLP that is seen. Of these possibilities, only forsterite and enstatite are composed of the most abundant elements (Mg and Si; see \citeads{2009ARA&A..47..481A}). They are more likely to dominate over the others. 
        
        \item The dependence of the polarization on wavelength is a probe of the size distribution of the grains, with small, intermediate-sized, and large grains respectively showing a negative, flat, and positive slope in the wavelength interval studied here (from 645 to 817\,nm at best). In the part of the sample for which multiwavelength data are available, the grains seem either small ($0.01-0.1\,\mu$m) or possibly intermediate sized ($0.01-1\,\mu$m), but likely not large ($0.1-1\,\mu$m). However, in order to place firmer constraints on the size range, a wider wavelength baseline for comparison would be useful, for example toward the infrared with the InfraRed Dual-band Imager and Spectrograph (IRDIS) subunit of SPHERE.
\end{enumerate}

We compared the spatial structure of the dust, which gives rise to the polarized radiation observed with SPHERE-ZIMPOL, to the spatial structure of the gas by observing the CO $\varv = 0, J = 2 - 1$ and SiO $\varv = 0, J = 5 - 4$ lines with ALMA in the {\sc Atomium} project. The images of the CO and SiO emission obtained in the combined configuration with ALMA allowed us to explore the inner circumstellar environment with a high S/N and the same spatial scale observed with SPHERE-ZIMPOL. The confinement of the DoLP to regions near the plane of the sky facilitates the comparison with the null-velocity channel maps (in the rest frame of the star) and the moment 1 maps obtained with ALMA. There are few correlations between the dust and the gas distributions observed in the work here. Except for $\pi^1$\,Gru (and perhaps GY~Aql), we do not detect interactions between potential companions (see Table~\ref{Table:Companions} and \citeads{2020Sci...369.1497D}) and the circumstellar material, even though we probe the area where such interactions are anticipated to occur. However, this does not imply that there is no colocation of the two phases of the circumstellar environment, because only part of the dust is detectable from the polarization measurements -- nor does it imply that there are no wind--companion interactions.

ZIMPOL reveals the 3D structure of the inner environment where dust nucleation takes place, circular shells or envelopes are absent, and clumps and arcs are dominant; and the DoLP confirms that significant regions are isolated. As a result, it is evident that dust nucleation is not spherically symmetric and instead occurs in isolated regions where the conditions are favorable. This in turn implies that estimates of the mass-loss rate and the gas-to-dust ratio in the inner circumstellar region need to account for the 3D distribution of the material. In future studies, there should be contemporaneous observations in the MIR of the thermal emission of the dust at different spatial scales (e.g., with VLT/VISIR for the outer regions and VLTI/MATISSE for the inner ones) and parallel observations of the gaseous environment with ALMA.

\begin{acknowledgements}
        We would like to thank Dr. Julien Milli for useful discussions in handling the polarized signal from ZIMPOL.
        Based on observations collected at the European Southern Observatory under ESO programme 0103.D-0772(A).
        We thank the anonymous referee whose kind comments helped improve the present paper.
        This paper makes use of the following ALMA data: ADS/JAO.ALMA\#2018.1.00659.L. ALMA is a partnership of ESO (representing its member states), NSF (USA) and NINS (Japan), together with NRC (Canada), MOST and ASIAA (Taiwan), and KASI (Republic of Korea), in cooperation with the Republic of Chile. The Joint ALMA Observatory is operated by ESO, AUI/NRAO and NAOJ.
        This work has made use of data from the European Space Agency (ESA) mission {\it Gaia} (\url{https://www.cosmos.esa.int/gaia}), processed by the {\it Gaia} Data Processing and Analysis Consortium (DPAC, \url{https://www.cosmos.esa.int/web/gaia/dpac/consortium}). Funding for the DPAC has been provided by national institutions, in particular the institutions participating in the {\it Gaia} Multilateral Agreement.
    This project has received funding from the European Union's Horizon 2020 research and innovation program under the Marie Sk\l{}odowska-Curie Grant agreement No. 665501 with the research Foundation Flanders (FWO) ([PEGASUS]$^2$ Marie Curie fellowship 12U2717N awarded to M.M.). 
    EC acknowledges funding from the KU Leuven C1 grant MAESTRO C16/17/007.
    LD and MM acknowledge support from the ERC consolidator grant 646758 AEROSOL.
    T.D. and S.H.J.W. acknowledge support from the Research Foundation Flanders (FWO) through grants 12N9920N and 1285221N, respectively.
    D.G. was funded by the project grant "The Origin and Fate of Dust in Our Universe" from Knut and Alice Wallenberg foundation.
    IEM has received funding from the European Research Council (ERC) under the European Union’s Horizon 2020 research and innovation programme (grant agreement No 863412).
    JMCP was supported by the UK's STFC (grant number ST/T000287/1).
    This work has made use of the the SPHERE Data Centre, jointly operated by OSUG/IPAG
    (Grenoble), PYTHEAS/LAM/CESAM (Marseille), OCA/Lagrange (Nice), Observatoire de Paris/LESIA
    (Paris), and Observatoire de Lyon.
    We used the SIMBAD and VIZIER databases at the CDS, Strasbourg (France)\footnote{Available at \url{http://cdsweb.u-strasbg.fr/}}, and NASA's Astrophysics Data System Bibliographic Services.
    This research made use IPython \citep{PER-GRA:2007}, Numpy \citep{5725236}, Matplotlib \citep{Hunter:2007}, SciPy \citep{2020SciPy-NMeth}, Pandas \citep{reback2020pandas,mckinney-proc-scipy-2010}, Astropy\footnote{Available at \url{http://www.astropy.org/}}, a community-developed core Python package for Astronomy \citepads{2013A&A...558A..33A}, and  Uncertainties\footnote{Available at \url{http://pythonhosted.org/uncertainties/}}: a Python package for calculations with uncertainties.\\
\end{acknowledgements}


\bibliographystyle{aa}
\bibliography{biblio} 

\begin{appendix}
        
\section{Observation log}

\begin{table}[ht!]
        \centering
        \caption{Observation log.}\label{Tab:ObsLog}
        \begin{tabular}{lllllll}
                \hline
                \hline
                \noalign{\smallskip}
                Date &  Time &    Object & ND & Filter 1 & Filter 2 &  Seeing \\
                &  (UT) &           &    &          &          & (arcsec)\\
                \hline
                \noalign{\smallskip}
                2019-06-07 & 09:01 &    SV Aqr & ND\_1.0 &      VBB &      VBB &    0.60 \\
                & 09:54 & HD 220340 &   OPEN &      VBB &      VBB &    0.69 \\
                2019-07-08 & 02:19 &     U Her & ND\_1.0 &      VBB &      VBB &    0.83 \\
                & 03:12 & HD 153898 & ND\_1.0 &      VBB &      VBB &    0.75 \\
                & 03:31 &    AH Sco &   OPEN &      N\_R &      N\_R &    0.78 \\
                & 03:44 & HD 152473 &   OPEN &      N\_R &      N\_R &    0.94 \\
                & 04:01 &    AH Sco &   OPEN &   Cnt820 &   Cnt748 &    0.98 \\
                & 04:33 & HD 152473 &   OPEN &   Cnt820 &   Cnt748 &    0.70 \\
                & 04:56 &    KW Sgr & ND\_1.0 &      VBB &      VBB &    0.48 \\
                & 05:46 & HD 163105 & ND\_1.0 &      VBB &      VBB &    0.83 \\
                & 06:09 &    VX Sgr &   OPEN &      VBB &      VBB &    1.10 \\
                & 07:00 & HD 166031 &   OPEN &      VBB &      VBB &    0.62 \\
                & 07:20 &     T Mic &   OPEN &      N\_R &      N\_R &    0.60 \\
                & 07:47 & HD 193244 &   OPEN &      N\_R &      N\_R &    0.49 \\
                & 08:07 &     T Mic & ND\_1.0 &   Cnt820 &   Cnt748 &    0.37 \\
                & 08:32 & HD 193244 &   OPEN &   Cnt820 &   Cnt748 &    0.43 \\
                & 08:47 &   $\pi^1$ Gru &   OPEN &    CntHa &    CntHa &    0.37 \\
                & 09:03 & HD 214987 &   OPEN &    CntHa &    CntHa &    0.43 \\
                & 09:20 &   $\pi^1$ Gru & ND\_1.0 &   Cnt820 &   Cnt748 &    0.49 \\
                & 09:59 & HD 214987 & ND\_1.0 &   Cnt820 &   Cnt748 &    0.43 \\
                2019-07-09 & 04:53 &     W Aql &   OPEN &      VBB &      VBB &    0.67 \\
                & 05:44 & HD 180459 &   OPEN &      VBB &      VBB &    0.47 \\
                & 06:08 &     S Pav &   OPEN &      N\_R &      N\_R &    0.40 \\
                & 06:35 & HD 187807 &   OPEN &      N\_R &      N\_R &    0.47 \\
                & 06:55 &     S Pav & ND\_1.0 &   Cnt820 &   Cnt748 &    0.54 \\
                & 07:36 & HD 187807 & ND\_1.0 &   Cnt820 &   Cnt748 &    0.37 \\
                & 07:57 &     V PsA & ND\_1.0 &      N\_I &      N\_I &    0.39 \\
                & 08:35 & HD 216556 & ND\_1.0 &      N\_I &      N\_I &    0.37 \\
                & 08:51 &     V PsA &   OPEN &      N\_R &      N\_R &    0.35 \\
                & 09:48 & HD 216556 &   OPEN &      N\_R &      N\_R &    0.42 \\
                2019-07-27 & 01:24 &     R Hya & ND\_1.0 &   Cnt820 &   Cnt748 &    0.41 \\
                & 01:35 &     R Hya &   OPEN &    CntHa &    CntHa &    0.36 \\
                & 02:03 & HD 121758 &   OPEN &   Cnt820 &   Cnt748 &    0.32 \\
                & 02:13 & HD 121758 &   OPEN &    CntHa &    CntHa &    0.35 \\
                2019-07-30 & 00:53 &     R Aql &   OPEN &      N\_R &      N\_R &    0.73 \\
                & 01:35 &     R Aql & ND\_1.0 &   Cnt820 &   Cnt748 &    0.72 \\
                & 01:57 & HD 174350 & ND\_1.0 &   Cnt820 &   Cnt748 &    0.75 \\
                & 02:08 & HD 174350 &   OPEN &      N\_R &      N\_R &    0.64 \\
                & 03:11 &    GY Aql & ND\_1.0 &      VBB &      VBB &    0.61 \\
                & 04:05 & HD 184413 &   OPEN &      VBB &      VBB &    1.22 \\
                & 05:04 &     U Del &   OPEN &    CntHa &    CntHa &    0.62 \\
                & 05:57 & HD 195835 &   OPEN &    CntHa &    CntHa &    0.57 \\
                2019-08-05 & 05:56 &     U Del & ND\_1.0 &   Cnt820 &   Cnt748 &    0.52 \\
                & 06:22 & HD 201298 & ND\_1.0 &   Cnt820 &   Cnt748 &    0.56 \\
                \hline
        \end{tabular}
	\tablefoot{The \textsc{Atomium} sources are designated by their name while the PSF  calibrators are designated through their HD number. A PSF reference star is always observed after its scientific source. We also indicate the neutral density (ND) filter that was used as well as the scientific filters in the two arms of ZIMPOL.}
\end{table}

\section{Additional figures of the degree of linear polarization}

\begin{figure*}[ht!]
        \centering
        \includegraphics[width=1.9\columnwidth]{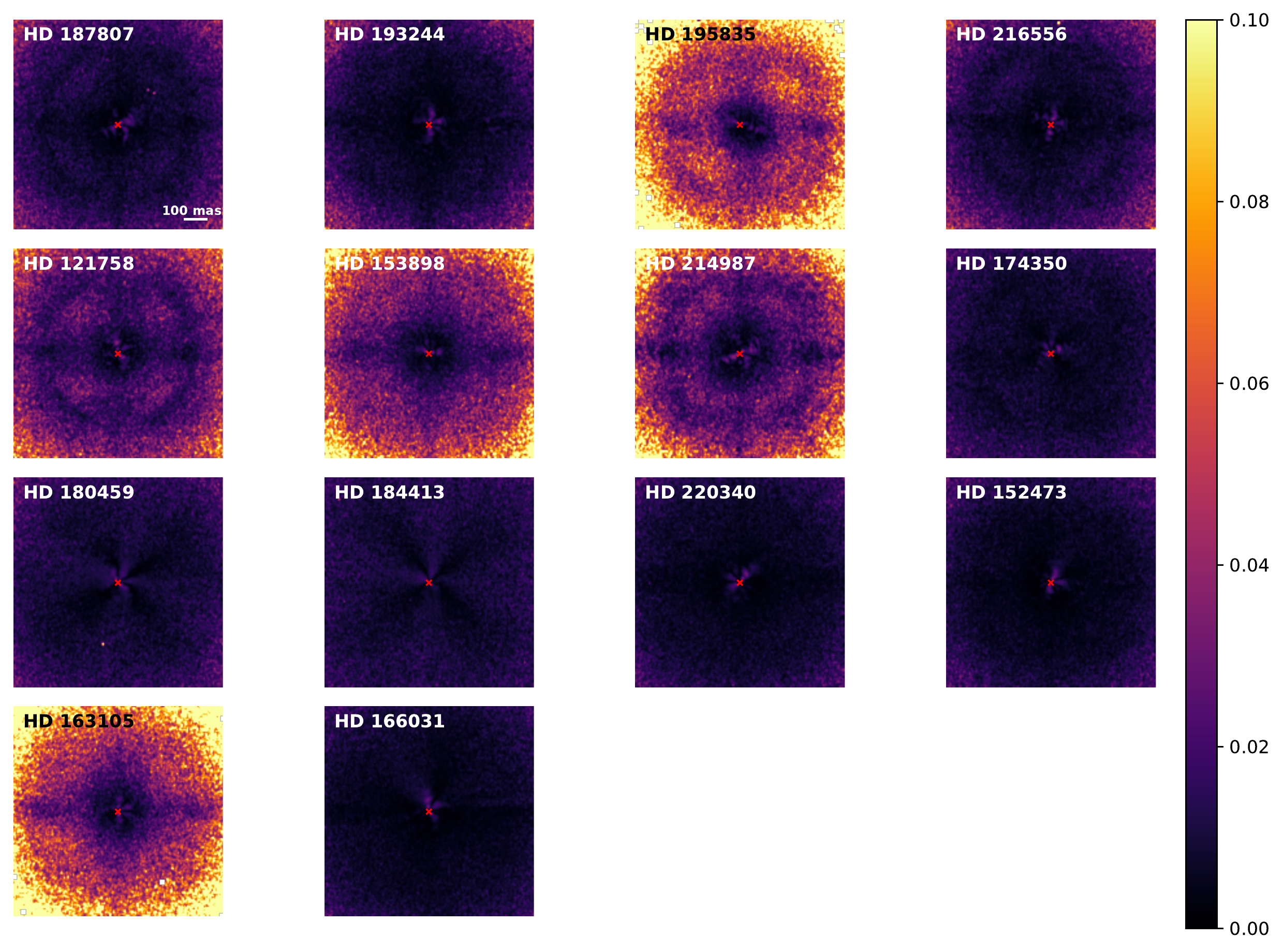}
        \caption{DoLP observed with VLT/SPHERE-ZIMPOL on the PSF sources. The field of view is 1 arcsec x 1 arcsec for each source. The red cross indicates the star position. North is up, and east is to the left.}
        \label{Fig:ZIMPOL_DoLP_PSF}
\end{figure*}

\begin{figure*}
        \centering
        \includegraphics[width=2\columnwidth]{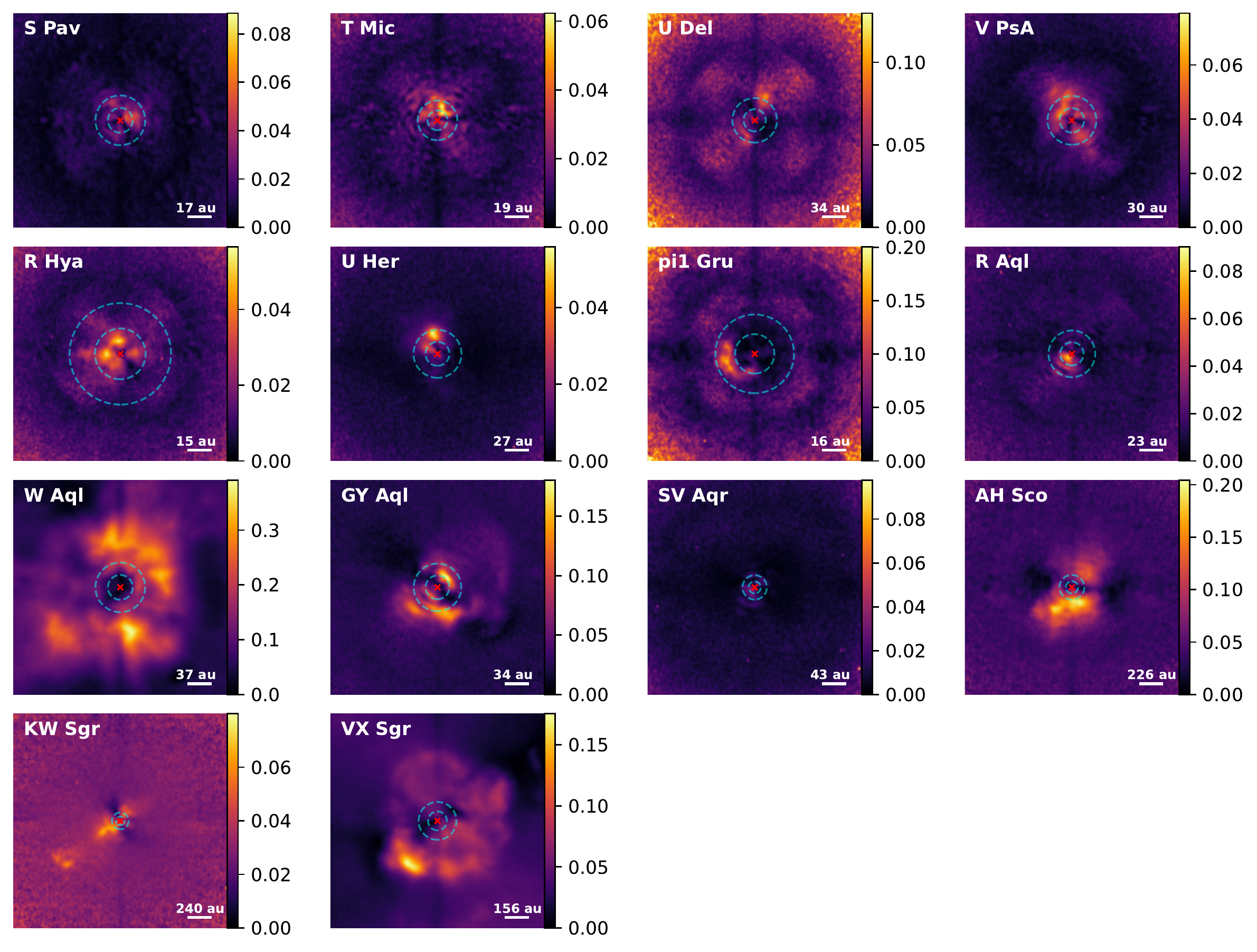}
        \caption{DoLP observed with VLT/SPHERE-ZIMPOL around the \textsc{Atomium} sources, with the original computation of the polarized flux from Stokes U and Q. The image setup is similar to that of Fig.~\ref{Fig:ZIMPOL_DoLP_all}.}
        \label{Fig:ZIMPOL_DoLP_original}
\end{figure*}

\begin{figure*}
        \includegraphics[width=2\columnwidth]{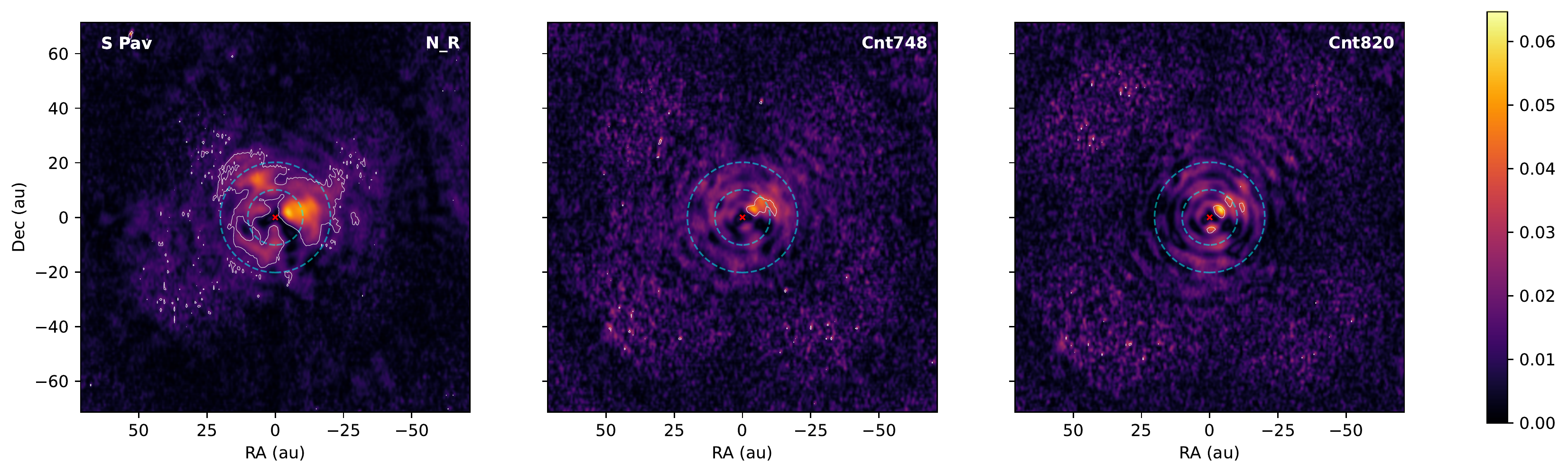}\\
        \includegraphics[width=2\columnwidth]{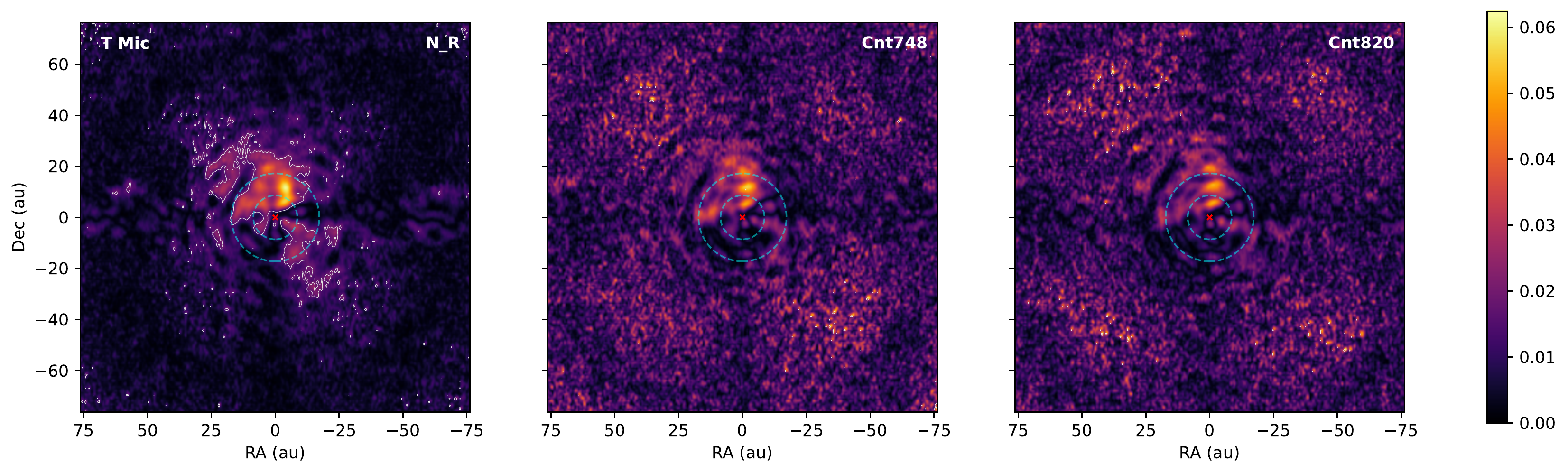}\\
        \includegraphics[width=2\columnwidth]{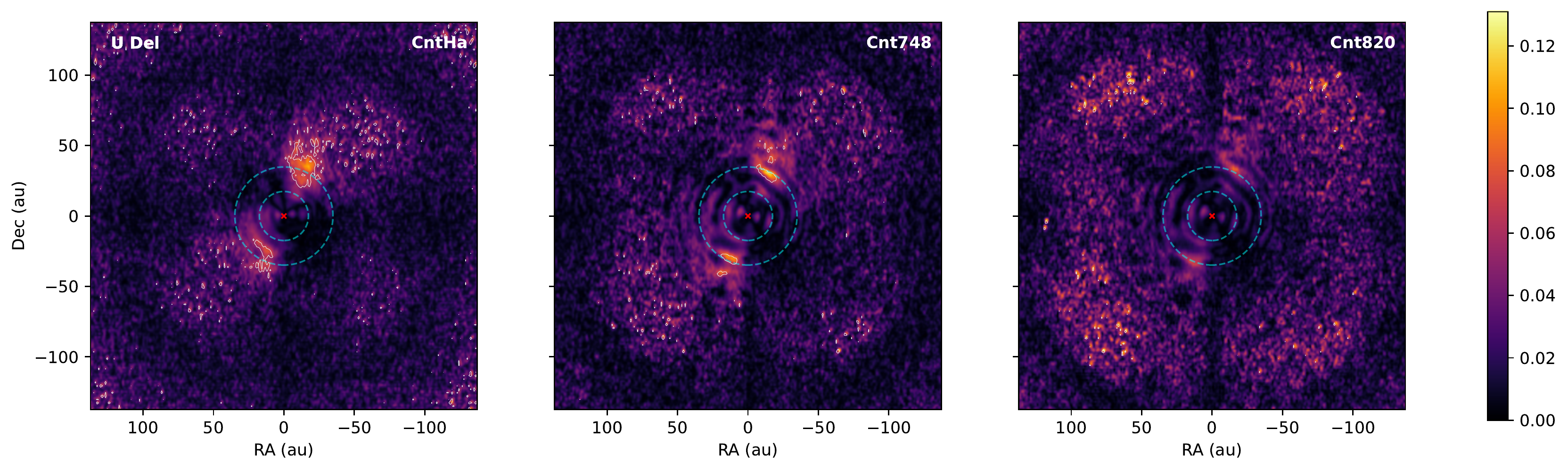}\\
        \includegraphics[width=1.33\columnwidth]{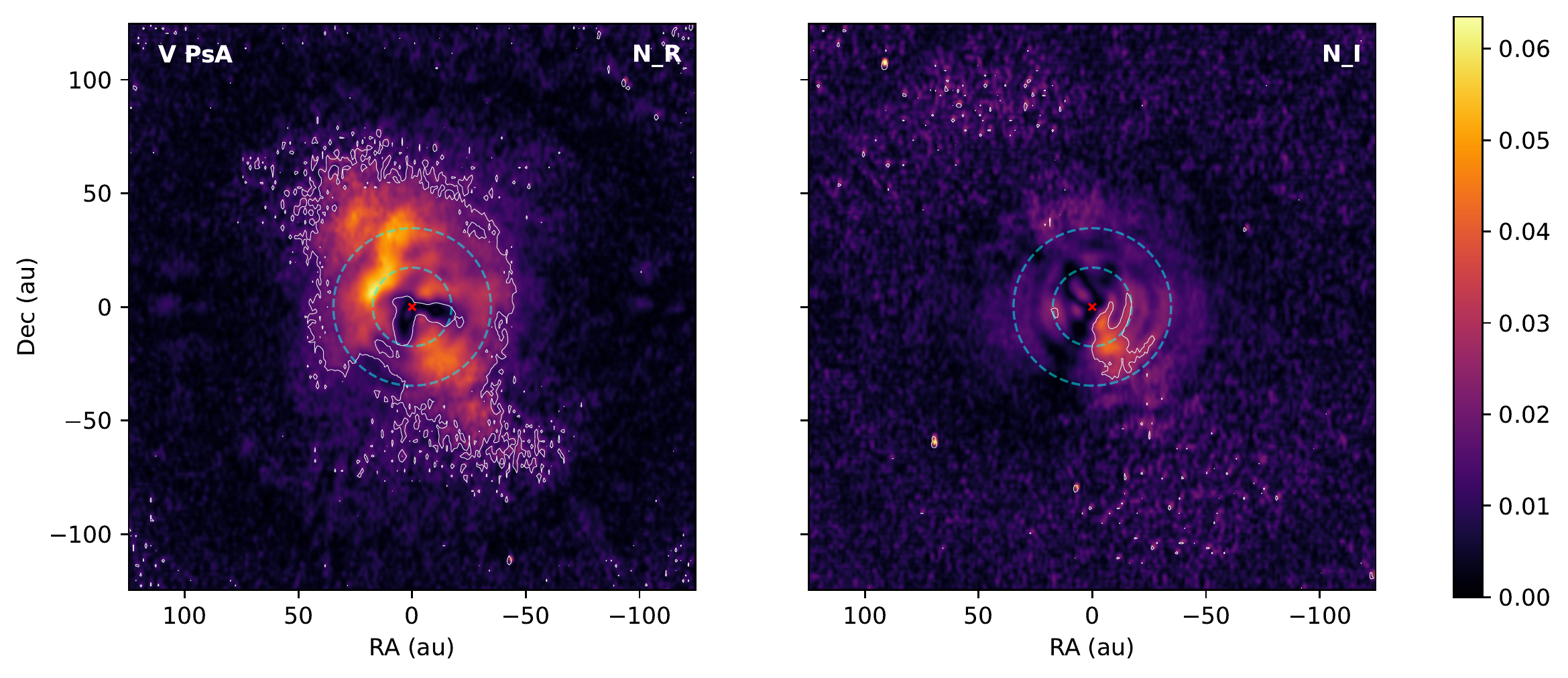}
        \caption{DoLP observed with VLT/SPHERE-ZIMPOL around the \textsc{Atomium} sources when several filters have been observed. The white contours correspond to the $5\sigma$ level, and the dashed cyan circles correspond to distances of 10 and 20 R$_\star$ from the star center. The red cross indicates the star position. The field of view is 1 arcsec x 1 arcsec for each source. North is up, and east is to the left. Note that the color scale has been adapted to each source to highlight the DoLP.}\label{Fig:DoLP_all_filters}
\end{figure*}

\setcounter{figure}{2}

\begin{figure*}
        \includegraphics[width=2\columnwidth]{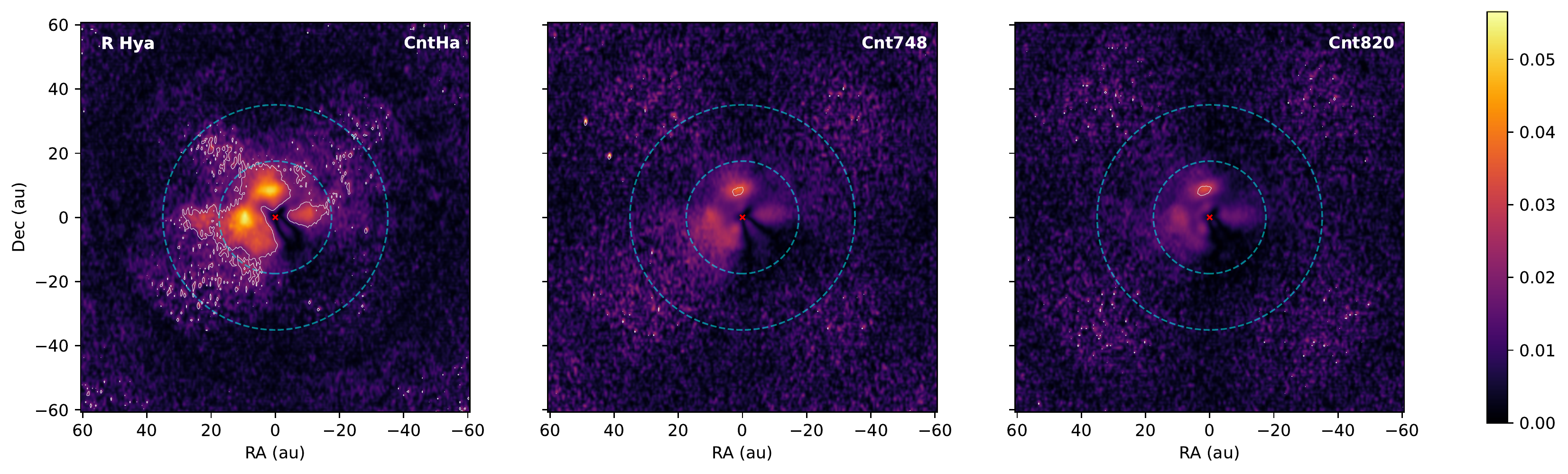}\\
        \includegraphics[width=2\columnwidth]{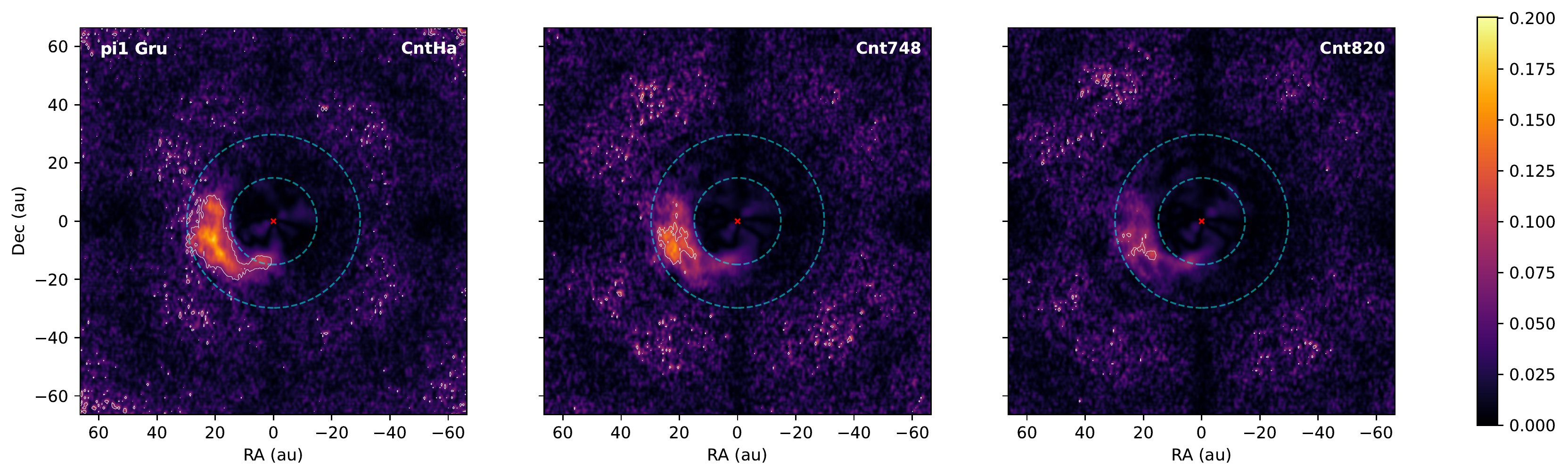}\\
        \includegraphics[width=2\columnwidth]{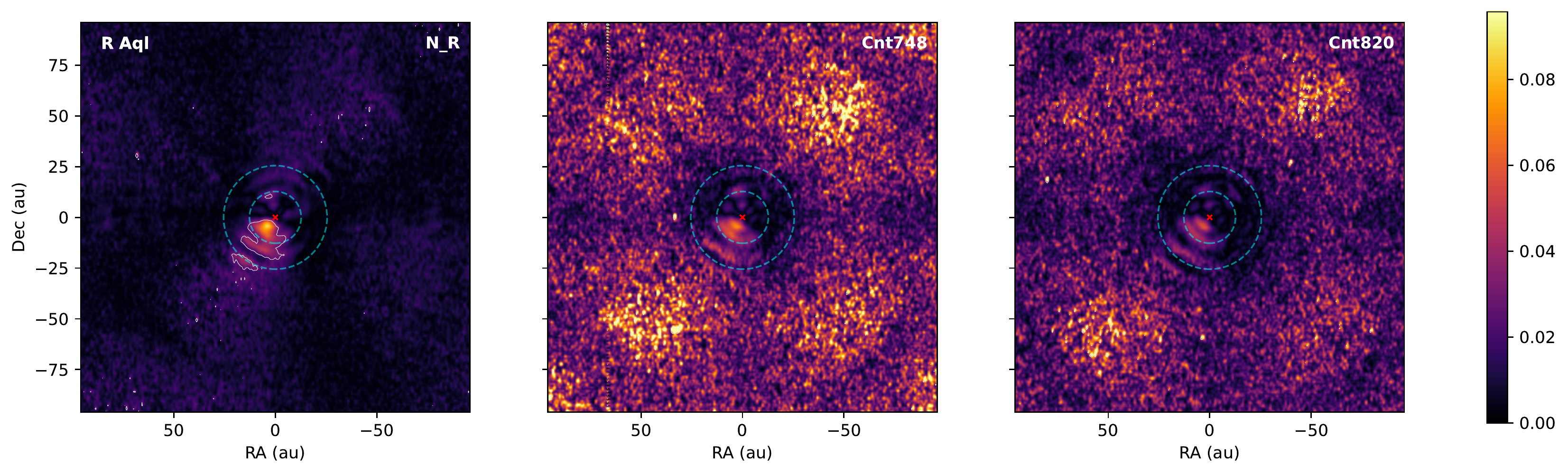}\\
        \includegraphics[width=2\columnwidth]{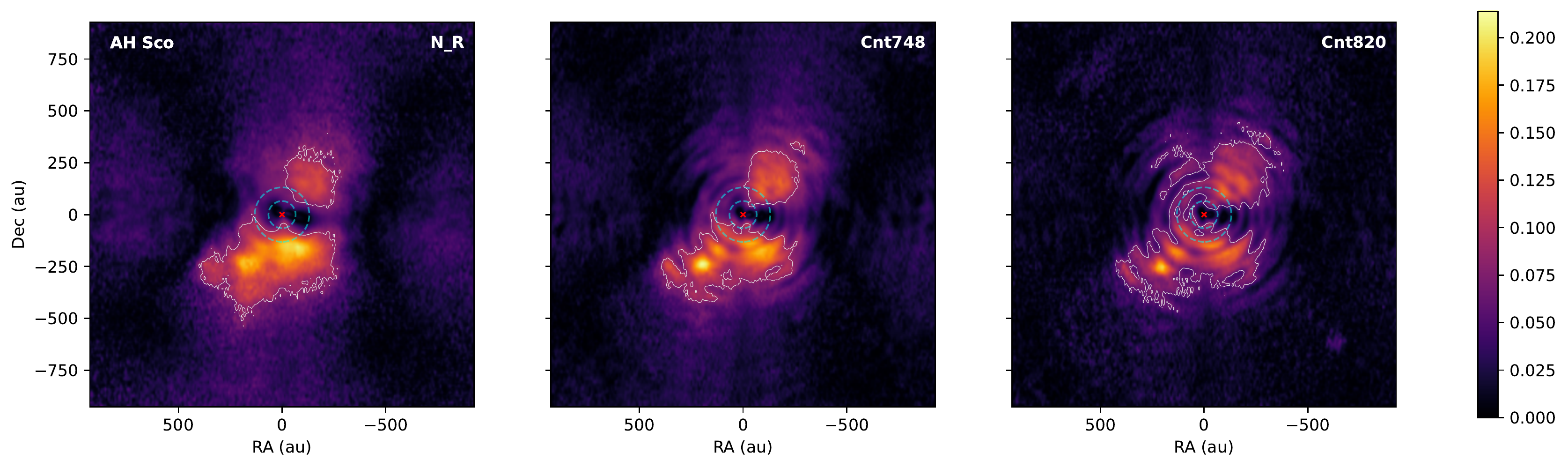}
        \caption{continued}
\end{figure*}

\end{appendix}

\end{document}